\documentclass[twocolumn, tighten]{aastex631}
\usepackage[utf8]{inputenc}
\usepackage{aas_macros,natbib,graphics,
graphicx,amsmath,amssymb,threeparttable,xcolor}

\newcommand\astrosun{$\odot$}
\newcommand\msun{\ensuremath{\mathrm{M}_{\textnormal{\astrosun}}}}
\newcommand\lsun{\ensuremath{\mathrm{L}_{\textnormal{\astrosun}}}}

\shortauthors{McKay et al.}

\begin{document}

\title{Dust Properties of 870 Micron Selected Galaxies in the GOODS-S}

\correspondingauthor{Stephen J. McKay}
\email{sjmckay3@wisc.edu}

\author[0000-0003-4248-6128]{S.~J.~McKay}
\affiliation{Department of Physics, University of Wisconsin--Madison, 1150 University Avenue,
Madison, WI 53706, USA}

\author[0000-0002-3306-1606]{A.~J.~Barger}
\affiliation{Department of Astronomy, University of Wisconsin--Madison, 475 N. Charter Street, Madison, WI 53706, USA}
\affiliation{Department of Physics and Astronomy, University of Hawaii, 2505 Correa Road, Honolulu, HI 96822, USA}
\affiliation{Institute for Astronomy, University of Hawaii, 2680 Woodlawn Drive, Honolulu, HI 96822, USA}

\author[0000-0002-6319-1575]{L.~L.~Cowie}
\affiliation{Institute for Astronomy, University of Hawaii, 2680 Woodlawn Drive, Honolulu, HI 96822, USA}

\author[0000-0002-8686-8737]{F.~E.~Bauer}
\affiliation{Instituto de Astrof\'isica and Centro de Astroingenier\'ia, Facultad de F\'isica, 
Pontificia Universidad Cat\'olica de Chile, Casilla 306, Santiago 22, Chile}
\affiliation{Millennium Institute of Astrophysics (MAS), Nuncio Monse{\~{n}}or S{\'{o}}tero 
Sanz 100, Providencia, Santiago, Chile} 
\affiliation{Space Science Institute,
4750 Walnut Street, Suite 205, Boulder, Colorado 80301, USA} 

\author[0000-0003-3910-6446]{M.~J.~Nicandro Rosenthal}
\affiliation{Department of Astronomy, University of Wisconsin--Madison, 475 N. Charter Street, Madison, WI 53706, USA}

\submitjournal{\apj}

\accepted{03 May 2023}

\begin{abstract}
We analyze the dust properties of 57 870~$\mu$m selected dusty star-forming galaxies in the GOODS-S using new deep ALMA 1.2~mm, 2~mm, and 3~mm continuum imaging together with other far-infrared through millimeter data. 
We fit the spectral energy distributions (SEDs) with optically thin modified blackbodies to constrain the emissivity indices and effective dust temperatures, finding a median emissivity index of $\beta = 1.78^{+0.43}_{-0.25}$ and a median temperature of $T_d = 33.6^{+12.1}_{-5.4}$~K. We observe a negative correlation between $\beta$ and $T_d$. By testing several SED models, we determine that the derived emissivity indices can be influenced by opacity assumptions. Our temperature measurements are consistent with no evolution in dust temperature with redshift. 

\end{abstract}



\section{\label{sec:intro}Introduction}

Over the last several decades, dusty star-forming galaxies (DSFGs) have emerged as a critical population at redshifts $z\gtrsim1$. First discovered with the Submillimeter Common User Bolometer Array (SCUBA) on the single-dish James Clerk Maxwell Telescope (JCMT) \citep{sib97,bcs+98, hsd+98,elg+99}, DSFGs 
boast some of the highest star formation rates (SFRs) in the universe (up to several thousand \msun~yr$^{-1}$) and may be responsible for 25\% to 80\% of the star formation rate density between redshifts of $z=6$ to $z=$~2--2.5, respectively \citep{zcm+21}.

Surveys of distant DSFGs have been performed on single-dish facilities, both from the ground (JCMT, IRAM, the South Pole Telescope, and the Large Millimeter Telescope (LMT))
and from space (the Herschel Space Observatory).
Single-dish observations sample large numbers of sources near the peaks of their far-infrared (FIR) spectral energy distributions (SEDs). However, they do not allow for accurate position measurements, and they are affected by source blending due to poor spatial resolution \citep[e.g.,][]{bii+11,bwc+12}. 
The new TolTEC camera on the LMT, with its fast mapping speeds and high sensitivity
\citep{waa+20}, may mitigate some of these issues. 

In contrast to single-dish imaging surveys,
submillimeter/millimeter interferometric surveys using NOEMA, the Submillimeter Array, and the Atacama Large Millimeter/Submillimeter Array (ALMA) have the sensitivity to detect faint sources \citep[e.g.,][]{ook+14, adw+16, ciz+23}. In addition, they provide accurate positions, are not affected by source blending, and can resolve extended emission \citep{hdc20}.
The main drawback is that interferometric surveys are observationally expensive. 
Currently, the most efficient strategy for studying large numbers of DSFGs is to conduct single-dish surveys to identify sources and then to follow them up with interferometry \citep[e.g.,][]{hks+13, cgb+18, sds+19}. 

Constraints on the dust and gas masses of DSFGs are important for determining what is responsible for their high SFRs. The dust masses of DSFGs are the highest of any known galaxy population ($\gtrsim10^8~\msun)$ \citep[e.g.,][]{sss+14a,dws+15, dss+20}. They have been used to estimate the gas masses available to form stars through an assumed calibration \citep[e.g.,][]{sas+14,ssa+16,sok+21}. However, dust mass measurements depend sensitively on the choice of dust emissivity spectral index, $\beta$, which parameterizes how the dust emission varies with wavelength \citep{bsi+02,cnc+14}. 
Since $\beta$ pertains to the intrinsic makeup of the dust, it can vary across a galaxy \citep[e.g.,][]{PC+14}. 

In studies of the local universe, $\beta$ has been found to be between 1.5 and 2.0 \citep{de11, ccc+11, cpf+18}. This range is in line with theoretical predictions of $\beta$ between 1.0 and 2.5 \citep{dl84}. In the absence of direct measurements, values of $\beta$ from the Milky Way and local galaxies are often assumed for high-redshift sources, with a common choice of $\beta = 1.8$ for optically thin SED fits \citep[e.g.,][]{ssa+16,sss+17,dss+21}.
If this assumption is not valid, then it would systematically impact measured dust and gas masses for the DSFG population.

Unfortunately, measuring $\beta$ directly for high-redshift galaxies is difficult, requiring multiple observations spanning the submillimeter/millimeter regime to break the degeneracy between the dust temperature, $T_d$, and $\beta$. Most surveys provide data at one or two wavelengths---often from low-resolution single-dish observations---and only a few recent studies use ALMA data. 
In one such study, \citet{dhc+21} used an optically thin modified blackbody and found a median $\beta = 1.9 \pm 0.4$ for a sample of 27 DSFGs from the ALESS survey \citep{hks+13,ksh+13} with ALMA 870~$\mu$m and 2~mm data. In another study, \citet{ccz+22} used a combined general opacity blackbody and power law fit and found a median $\beta = 2.4 \pm 0.3$ for a sample of 39 DSFGs in the SSA22 field (850~$\mu$m flux $>5.55$~mJy) with SCUBA-2 850~$\mu$m, AzTEC 1.1~mm, and ALMA 2~mm data. 

In this paper, we determine $\beta$ and $T_d$ for a large sample of DSFGs in the GOODS-S that have observations in multiple ALMA bands. Our sample is taken from the catalog of 75 ALMA sources detected at 870~$\mu$m ($>4.5\sigma$) by \citet{cgb+18}, which were originally selected from SCUBA-2 850~$\mu$m imaging.

We structure the paper as follows. In Section~\ref{sec:data}, we discuss the ALMA 870~$\mu$m catalog and our new longer-wavelength ALMA observations, along with ancillary photometry and redshifts. In Section \ref{sec:analysis}, we describe our SED fitting methods and we constrain the dust properties of our sample. In Section~\ref{sec:disc}, we discuss the implications of our results in the context of other studies in the literature. In Section \ref{sec:conc}, we summarize our results.

We assume a flat concordance $\Lambda$CDM cosmology throughout with $\Omega_m= 0.3$, $\Omega_\Lambda = 0.7$, and $H_0 = 70.0$~km~s$^{-1}$~Mpc$^{-1}$.

\section{\label{sec:data}Data} 

\subsection{Total Sample}

Our main sample of galaxies comes from the SUPER GOODS program of \citet{cgb+18} (hereafter, C18). Using ALMA band~7 (central wavelength of 870~$\mu$m), C18 followed up SCUBA-2 850~$\mu$m selected sources in the GOODS-S to obtain accurate positions. We hereafter refer to the resulting 75 galaxies ($>$~4.5$\sigma$; their Table~4) as our \textit{total sample}.

The total sample consists of sources ranging in 870~$\mu$m flux from 0.84~mJy to 8.93~mJy.
Of these 75 sources, 17 (23\%) are at or below the SCUBA-2 4$\sigma$ confusion limit of $\sim$1.6~mJy at 850~$\mu$m \citep[][]{cbh+17}. The optical/NIR counterparts to this sample vary from bright galaxies at lower redshifts to sources that are very faint or undetected in the deep CANDELS HST imaging. 

\subsection{ALMA Band 3, 4, and 6 Observations}
\label{spectralALMA}

In ALMA Project \#2021.1.00024.S (PI: F. Bauer), we made ALMA spectral linescans in band~6 (central wavelength of 1.24~mm),
band~4 (1.98~mm),
and band~3 (3.07~mm)
of 57 sources in the total sample, prioritizing those
with 870~$\mu$m fluxes above 1.8~mJy and lacking a well-established spectroscopic redshift.

The ALMA observation blocks were downloaded and calibrated using {\sc casa} version 6.2.1-7 based on the associated PI scripts. The visibilities from individual spectral setups in a given band were combined using \texttt{concat}. Dirty continuum images were generated using \texttt{tclean}, adopting 0\farcs25 pixels, natural weighting, and a ``common'' restoring beam. Mildly cleaned continuum images were generated adopting multithreshold automasking with standard thresholds, 100 clean iterations with a 0.1 mJy threshold, pixel scales of 0, 5, and 10, and robust=0.5.
The resulting band 3, 4, and 6 images had central frequencies, bandwidths, and beams of 
97.662\,GHz, 27.120\,GHz, and $\theta_{\rm beam}{=}$0\farcs99$\times$0\farcs83;
151.188\,GHz, 23.370\,GHz, and $\theta_{\rm beam}{=}$1\farcs21$\times$1\farcs10; and
241.750\,GHz, 23.250\,GHz, and $\theta_{\rm beam}{=}$1\farcs36$\times$1\farcs08; respectively. 

We measured peak fluxes and errors from the band 3, 4, and 6 cleaned continuum images. C18 found that since the sources were resolved at 870~$\mu$m, the peak fluxes (even those from the tapered images) underestimated the total fluxes. They determined that taking a ratio of aperture flux measurements made over a range of aperture radii gave a similar correction factor to estimates derived by fitting the sources in the uv plane. We adopted the aperture method for the current data as well, finding average correction factors of 1.3 for all three bands (e.g., see Figure 2 of \citealt{cbb23} for the band 4 data).
Thus, we use this correction factor to convert the band 6, 4, and 3 peak fluxes to total fluxes (we hereafter refer to these as 1.2~mm, 2~mm, and 3~mm fluxes, respectively). 

We supplement our ALMA observations with the 1.13~mm observations from the GOODS-ALMA survey \citep{geb+22}. These were also obtained with ALMA in the band~6 frequency window, but their central frequency of 265~GHz is sufficiently different from our central frequency of 242~GHz to make them worth including in our SED fits. Of the 75 sources in the total sample, 39 are also detected in the GOODS-ALMA sample. Although 8 of these 39 sources were not targeted in our spectral program, we at least have the GOODS-ALMA 1.13~mm measurements (hereafter, 1.1~mm).

In Table~1 (Appendix~\ref{appA}), we list our ALMA 1.2~mm, 2~mm, and 3~mm peak fluxes and errors in a table comprised of the 75 sources in the total sample. In addition, we give the 1.1~mm total fluxes for the 39 sources that also appear in the GOODS-ALMA catalog.

\subsection{Ancillary Photometry and Redshifts}
\label{sec:ancillary}

The GOODS-S is one of the most well-observed fields in the sky, with comprehensive multiwavelength coverage from the X-ray to the radio regime. In this section, we summarize the additional photometric and redshift data that we use in our analysis.

\citet{bcb+22} presented deep SCUBA-2 450~$\mu$m observations of the GOODS-S,
including counterparts to some of the 870~$\mu$m sources described here. Since that time, we have continued to deepen our SCUBA-2 450~$\mu$m images of the GOODS-S (see \citealt{cbb23}). 
Our latest maps reach a central rms noise of 1.67~mJy, which is about 10\% deeper than the image used in \citet{bcb+22}.
For our purposes here, the deeper images provide more robust SCUBA-2 450~$\mu$m fluxes for our sample. These fluxes were obtained sequentially by first measuring the peak SCUBA-2 flux for a given source within $2''$ of the ALMA position, then removing that source from the image to prevent blending in later measurements. However, we note that this is not critical for SCUBA-2 450~$\mu$m data due to its higher resolution and shallower depth than SCUBA-2 850~$\mu$m data. The 1$\sigma$ errors were measured from the local rms noise map.

Spitzer/MIPS 24 and 70~$\mu$m and Herschel/PACS 100 and 160~$\mu$m counterparts were obtained from the catalog of \citet{edh+11} using a 1\farcs5 matching radius. 61 sources from the total sample have a 24~$\mu$m counterpart.

We matched our sample within $4''$ to the HerMES DR3 catalog that used Spitzer/MIPS prior positions for deblending \citep{oba+12} to obtain Herschel/SPIRE data at 250, 350, and 500~$\mu$m. We found 51 galaxies with SPIRE counterparts that appear to be reliable based on the images. For 24 sources without a reasonable counterpart in either the PACS or SPIRE catalogs, plus an additional 13 that did not appear in the PACS catalog, we measured the fluxes ourselves from the images at the ALMA positions and normalized them to the catalog fluxes. However, we do not use the SPIRE 500~$\mu$m fluxes in our SED fits due to their source blending and lower spatial resolution. 

\citet{cbb23} presented for the total sample the redshifts, both photometric (hereafter, photz; these come from \citealt{ssq+16}) and spectroscopic (hereafter, specz), including five that were determined from our ALMA data. In total there are 20 sources with speczs in the total sample (27\%). 

The highest specz in our sample is for ALMA~68 (numbered as in Table~4 of C18) at $z=5.58$, obtained in \citet{obn+23} using JWST NIRCam/grism spectra from the FRESCO survey. A NIRCam F444W, F210M, F182M color image of this source is shown in Figure~\ref{alma68cut}. Although this source was undetected in the CANDELS HST images, it is clearly visible in the NIRCam image. 

\begin{figure}
    \centering
    \includegraphics[width=0.9\linewidth]{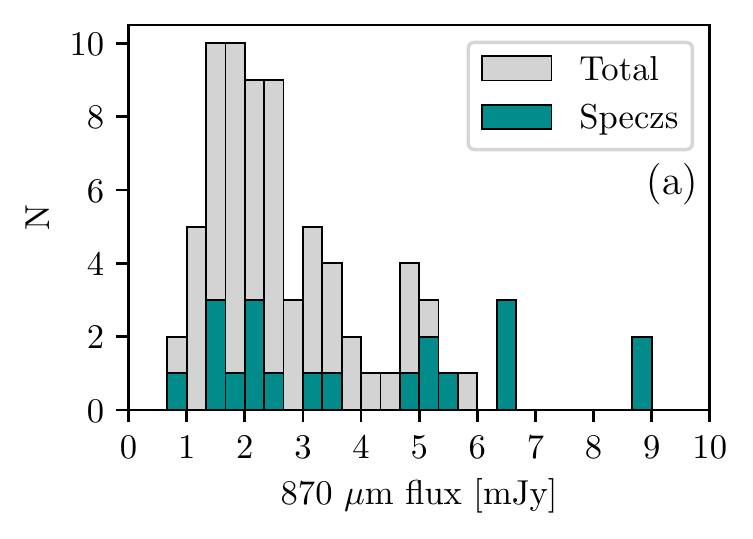}\\\vspace{-8pt}
    \includegraphics[width=0.9\linewidth]{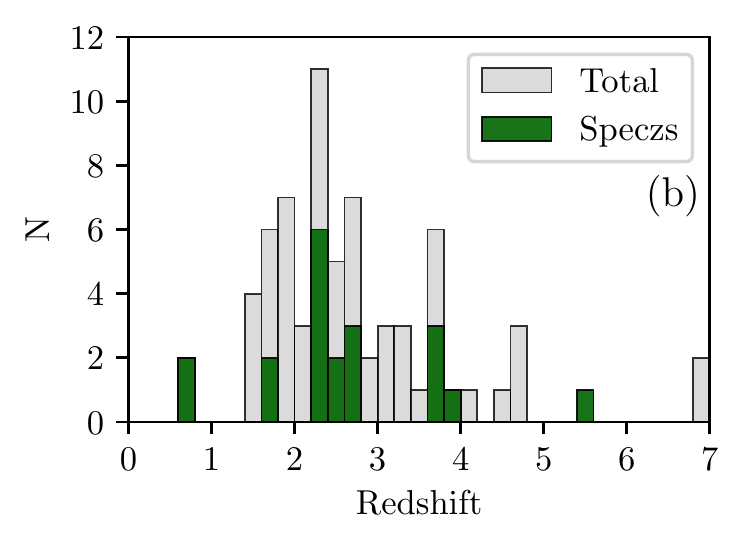}
    \caption{Histograms of the (a) 870~$\mu$m flux densities and (b) redshifts for the total sample (gray). Sources with speczs are shown in color in both plots. The six sources in the total sample without redshifts are not shown in (b), and the two with questionable $z_{\rm phot} > 7$ are shown at a nominal redshift of $z=7$.
}
    \label{flux_z_hist}
\end{figure}

\begin{figure}
\centering
    \includegraphics[width=0.8\linewidth]{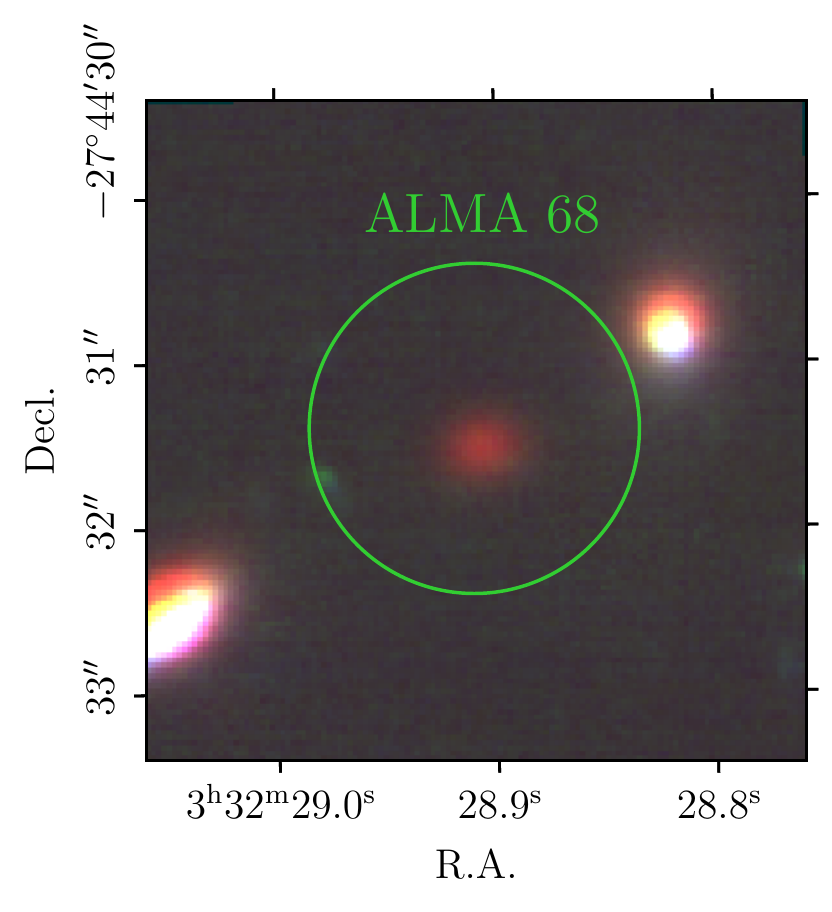}
    \caption{Three-color JWST NIRCam image (red~$=$~F444W, green~$=$~F210M, blue~$=$~F182M) for ALMA~68 from the FRESCO survey \citep{obn+23}. The green circle has a 1$''$ radius and is centered on the ALMA 870~$\mu$m position.
}
    \label{alma68cut}
\end{figure}

The distribution of 870~$\mu$m fluxes and redshifts for our total sample is shown in Figure~\ref{flux_z_hist}. We list all the adopted redshifts in Table~\ref{tab:fluxtable} (Appendix \ref{appA}), including whether they are speczs or photzs (denoted by the number of decimal places).
For the six sources in the table with poor-quality photzs (quality flag $Q>3$ in the \citealt{ssq+16} catalog), we put their photzs in brackets. 

%
%

%
%

\subsection{Robust Subset Selection}
\label{sec:subset}

Although, in principle, only three photometric data points are needed to break the degeneracy between $\beta$ and $T_d$, the results are unlikely to be reliable for individual galaxies unless there are well-constrained fluxes sampling both the peak of the dust SED and the Rayleigh-Jeans (RJ) tail. For example, \citet{dhc+21} found for their bright ALESS sample that without both Herschel detections near the peak and ALMA 870~$\mu$m and 2~mm fluxes on the tail, the derived parameters were poorly constrained. 

Thus, to better determine the dust properties of our sample, we restrict much of our analysis to sources with observations in at least two ALMA bands, i.e., in addition to the ALMA 870~$\mu$m measurement, at least one ALMA measurement at wavelengths longer than 1~mm. We also require the sources to have a redshift estimate. This is satisfied by 52 out of the 57 sources with ALMA spectral observations (the other 5 lack redshifts), and all 8 of the additional sources with GOODS-ALMA 1.1~mm measurements (see Section~\ref{spectralALMA}). However, we exclude one source (ALMA~58) that was not detected in the SPIRE images or SCUBA-2 450~$\mu$m image, and we remove two others (ALMA~45 and ALMA~54) with questionable $z_{\rm phot} > 7$ (their FIR colors suggest lower redshift solutions). After excluding these 3 sources, there are 57 sources which we keep in our analysis; we refer to these as the \textit{robust subset}. The 18 sources not in the robust subset are marked with brackets around their source numbers in Table~\ref{tab:fluxtable}.

Within the robust subset, there are 19 sources with speczs (33\%). The median redshift for the robust subset is $z=2.37$, and the 870~$\mu$m flux range is 0.93--8.93~mJy with a median of 2.54~mJy. Each source in the sample has at least two ALMA measurements by our selection criteria, but 44 (77\%) have four or more ALMA measurements (2 have 3 ALMA bands and the remaining 11 have just 2 ALMA 
bands). Thus, the FIR SEDs of our sample are generally better sampled than others in the literature \citep[e.g.,][]{dhc+21,ccz+22}.

\section{\label{sec:analysis} SED Analysis and Dust Properties}

The main goal of this work is to measure the dust properties of our sample of DSFGs as accurately as possible. We do this by fitting the SEDs using simple isothermal models. We fit the SEDs for 69 sources---the 57 in the robust subset plus the remaining 12 in the total sample with redshifts.

We fit the photometry of each source with a single-temperature modified blackbody (MBB), for which the flux density, $S_\nu$, for rest-frame frequency, $\nu$, is given as $S_\nu  \propto \kappa(\nu) B_\nu(T_d)$. Here $B_\nu(T_d)$ is the Planck distribution and $\kappa(\nu)$ is the frequency-dependent dust opacity. 
Although the dust consists of components at a range of different temperatures, an effective $T_d$ description is often used because it serves as a good trade-off between number of model parameters and quality of fit to the FIR SED. 

Additionally, we make the assumption that the emission is optically thin at the wavelengths we are fitting, such that $\kappa(\nu) \propto \nu^\beta$, where $\beta$ is the dust emissivity spectral index. At the resolution of our observations, this $\beta$ represents a galaxy-averaged value; it manifests primarily in the observed slope of the RJ fall-off. 

The optically thin, isothermal MBB has been widely used as a successful model for the FIR/(sub)millimeter dust emission of galaxies \citep[e.g.,][]{kcd+06,mdb+12,jdm+19,dss+20,dhc+21,bcb+22}, so by choosing it, we enable comparisons of our derived parameters with those from the literature. 

%
%
\begin{figure}
\centering
    \includegraphics[width=0.9\linewidth]{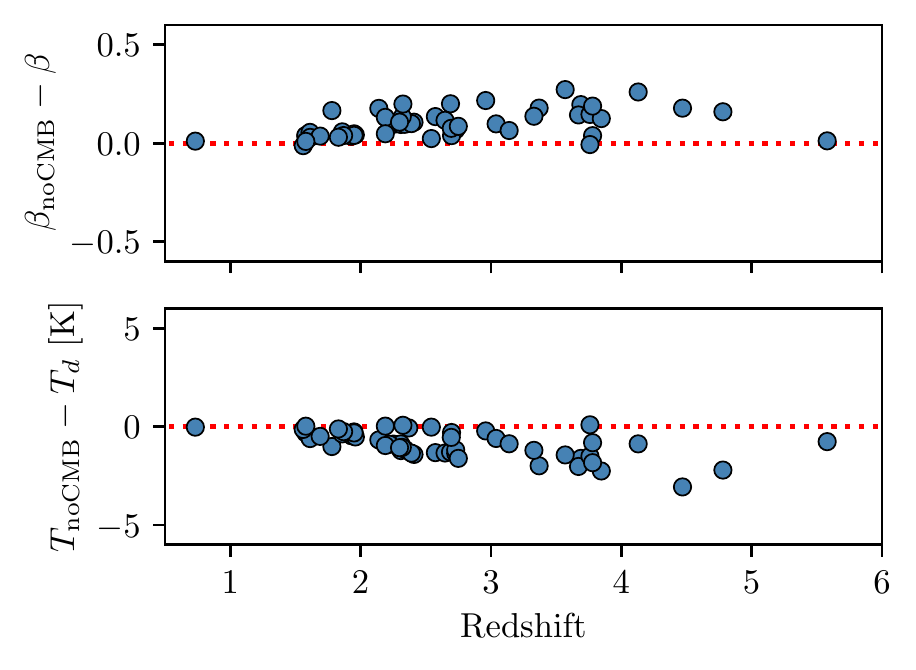}
    \caption{Difference in median likelihood estimates with and without CMB corrections for $\beta$ (top) and for $T_d$ (bottom) vs. redshift, for the robust subset. Models with the CMB included find slightly higher $T_d$ and slightly lower $\beta$ on average, but the effect is minimal.}
    \label{cmb_corr}
\end{figure}

We use the Python-based Markov chain Monte Carlo (MCMC) package \textsc{emcee} \citep{fhl+13} to fit the MBB to our data and recover the posterior likelihood distributions for the parameters.
The free parameters in these fits are the overall normalization, $\beta$, and $T_d$. We use flat priors on $\beta$ between 0.8 and 4.0 and on $T_d$ between 10~K and 90~K.  We choose this Bayesian approach over a simple least-squares fitting algorithm because least-squares fitting has been shown to introduce an artificial correlation between $T_d$ and $\beta$ \citep{sks+09a,sks+09b, kss+12}.

In the fits, we only consider the photometry at rest-frame wavelengths higher than 50~$\mu$m. 
We add a 5\% error in quadrature to the uncertainties to correct for differences in the absolute flux calibration across FIR/(sub)millimeter bands. We perform all of the fits at our adopted redshifts (see Table~\ref{tab:fluxtable}) without allowing redshifts to vary.

We include the corrections to the SED from the CMB as outlined in \citet{dgw+13}. However, these are expected to be small for sources at $z < 5$. We check this by fitting MBBs without the CMB correction to the robust subset and comparing the median likelihood values for $\beta$ and $T_d$ to those which include the CMB effects. The results are shown in Figure~\ref{cmb_corr}. We find small offsets in both $\beta$ and $T_d$, but conclude that the effects of the CMB are fairly negligible for our sample.

Three sources, ALMA~54, ALMA~61, and ALMA~69, none of which is in the robust subset, had 3 or fewer photometric data points included in the fit. Since the number of data points must always be greater than the number of free parameters, for these sources we only fit $T_d$ and not $\beta$.

%
%
\begin{figure*}[ht]
\centering
    \includegraphics[width=0.32\linewidth]{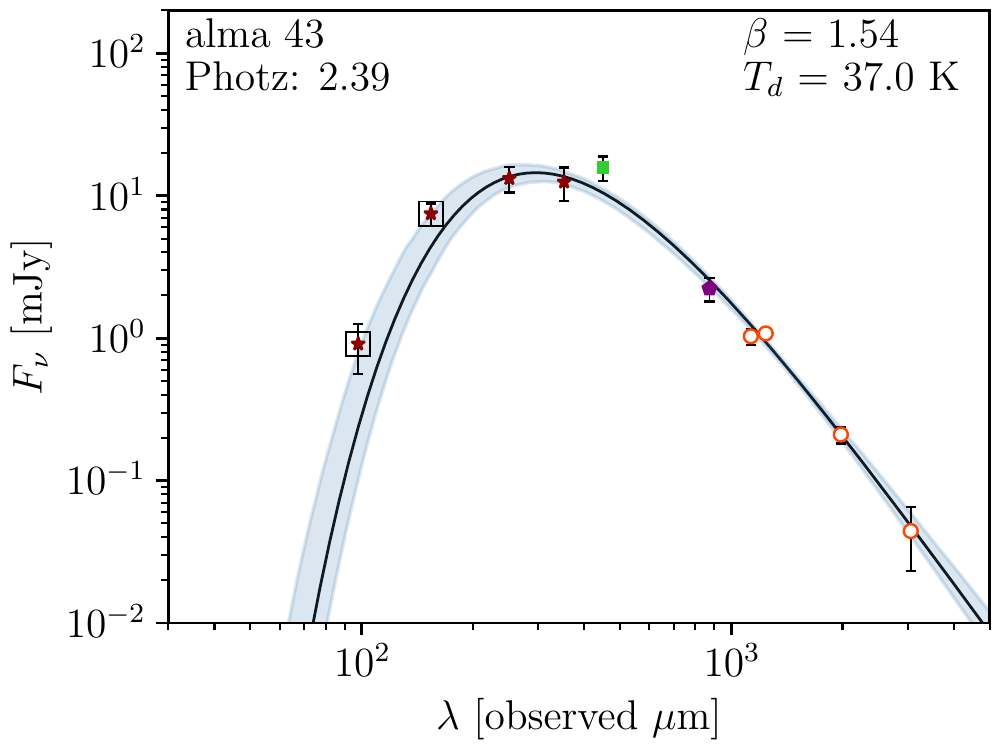}
    \includegraphics[width=0.32\linewidth]{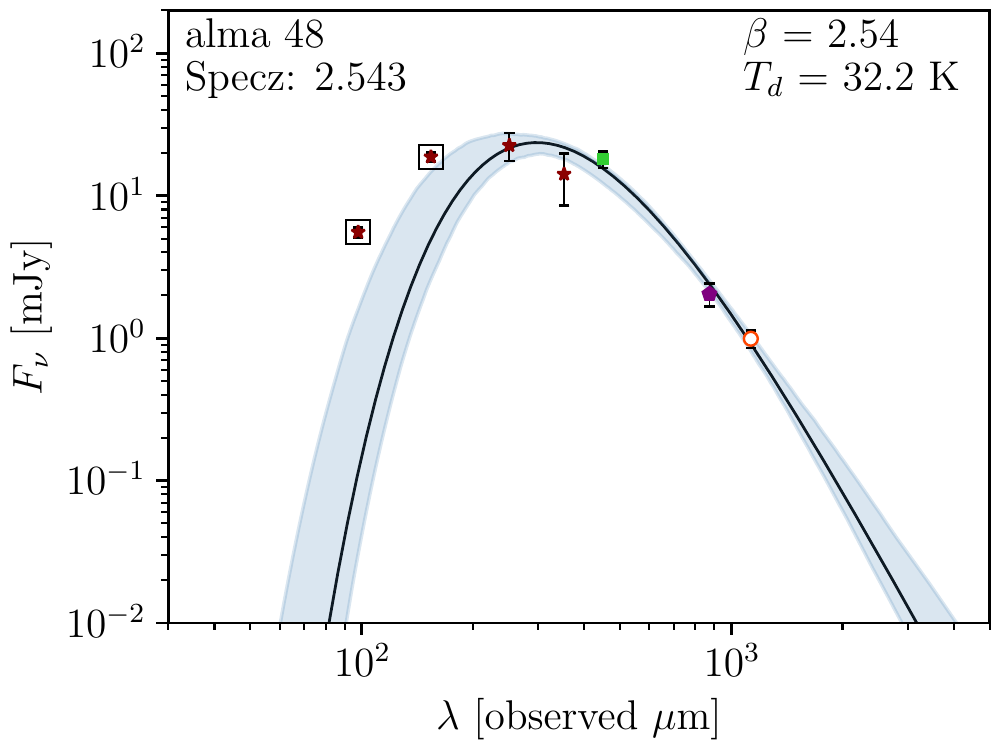}
    \includegraphics[width=0.32\linewidth]{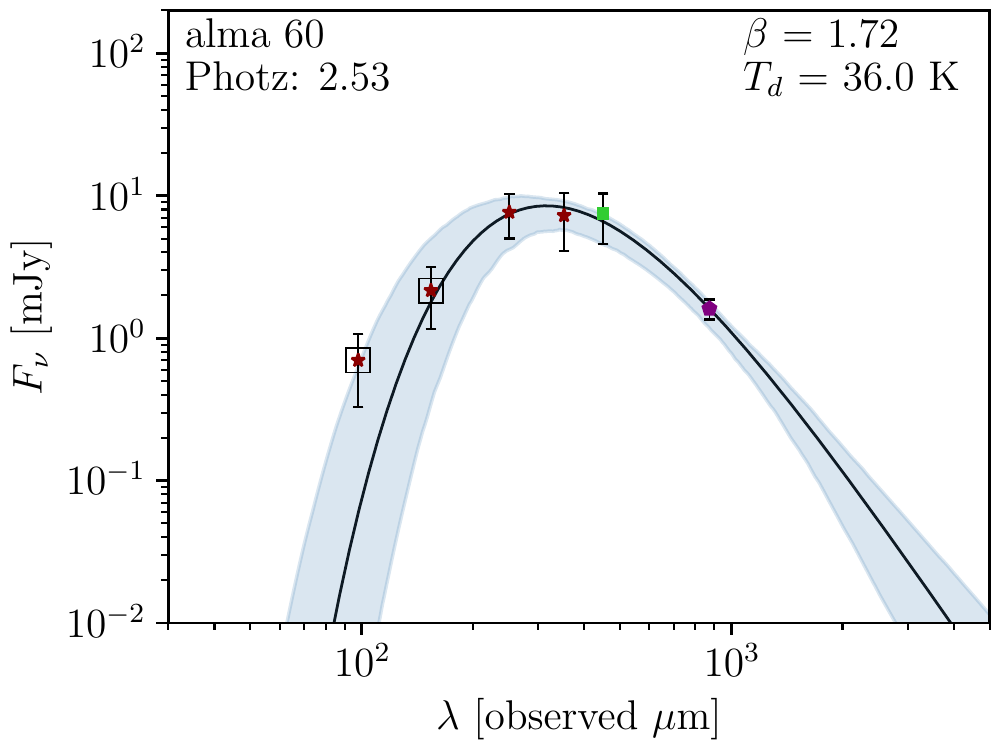}\\
    \includegraphics[width=0.32\linewidth]{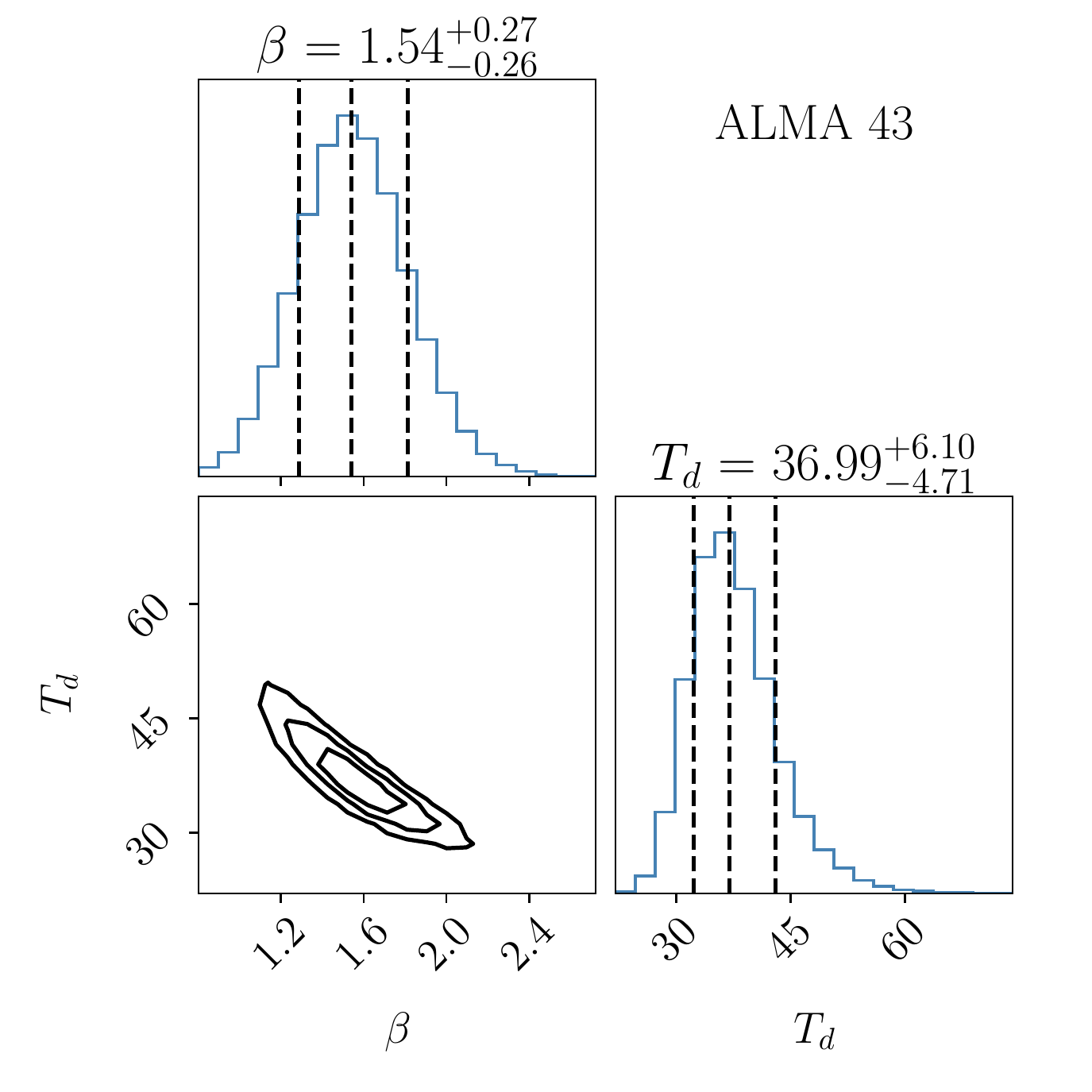}
    \includegraphics[width=0.32\linewidth]{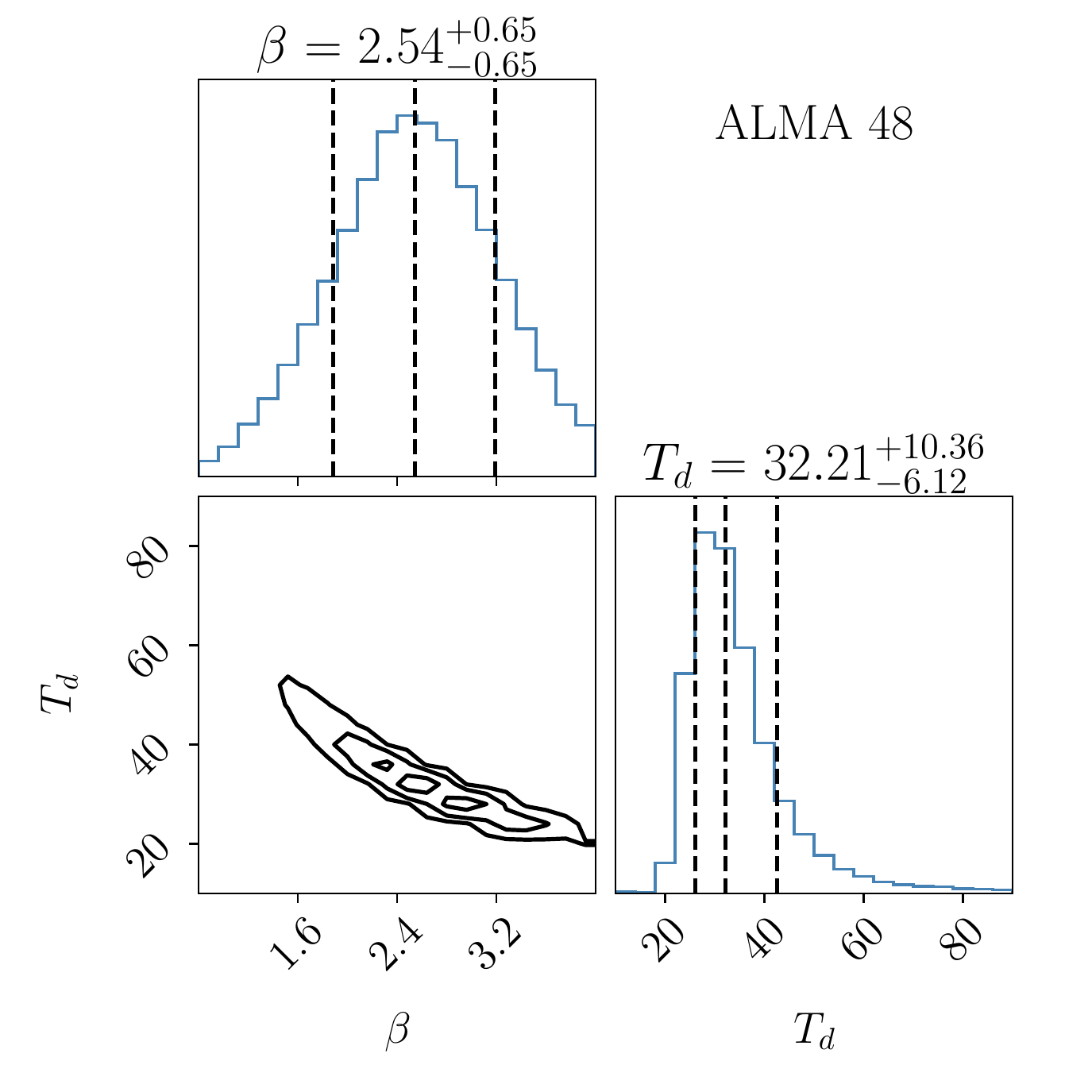}
    \includegraphics[width=0.32\linewidth]{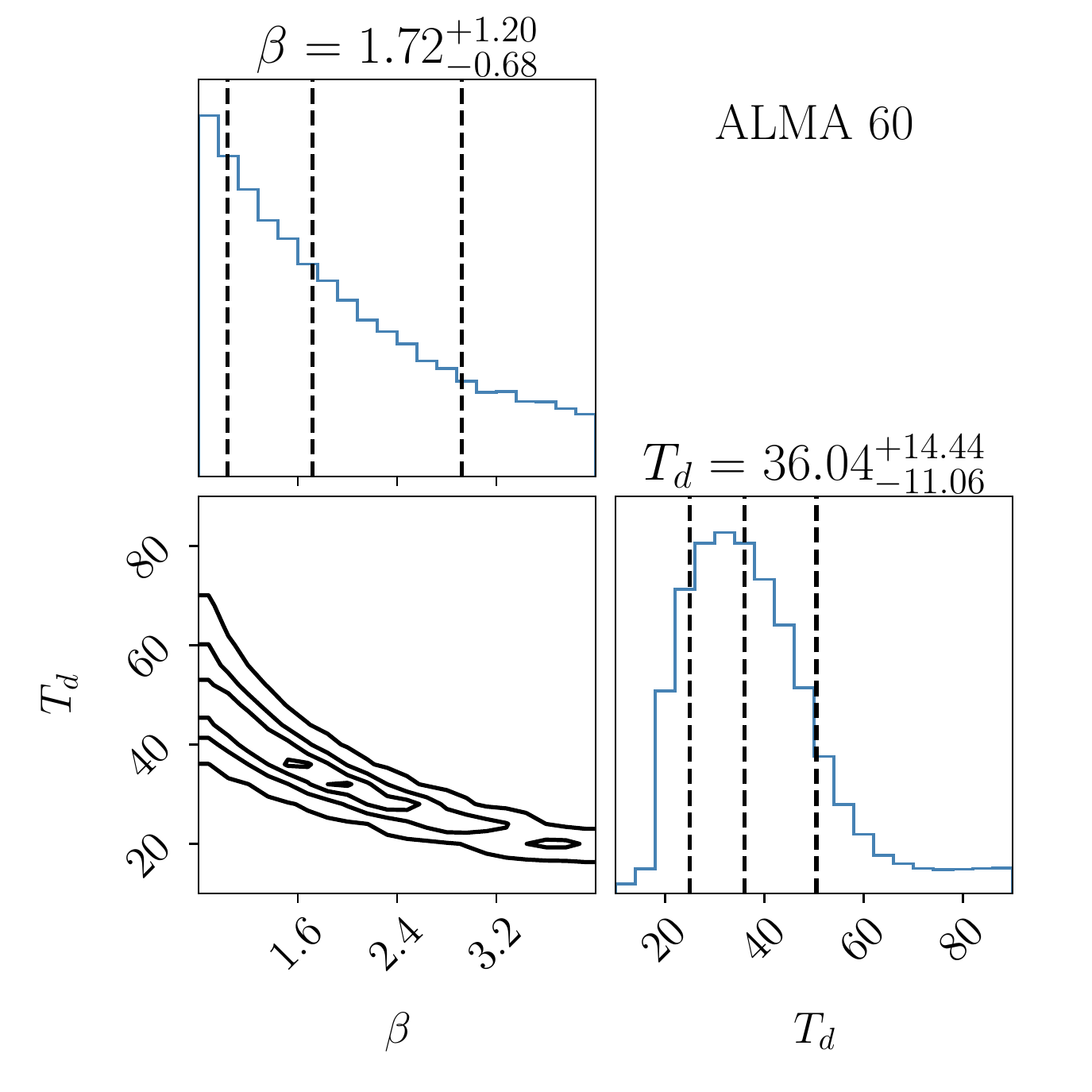}
    \caption{Top row: Best-fit MBB (black curve) and 16th to 84th percentile range from the posterior distribution of the MCMC models (blue shaded region) for ALMA~43 (left), a source in the robust subset with five ALMA measurements; ALMA~48 (center), also in the robust subset but with only two ALMA bands; and ALMA~60 (right), a source not in the robust subset due to the lack of observations at wavelengths longer than 870~$\mu$m. Photometry: Red circles---ALMA 1.1~mm, 1.2~mm, 2~mm, and 3~mm, maroon pentagon---ALMA 870~$\mu$m, green square---SCUBA-2 450~$\mu$m, dark red stars---Herschel/PACS 100 and 160~$\mu$m and SPIRE 250 and 350~$\mu$m. The MBB is only fit to the data at rest-frame wavelengths $\ge$50~$\mu$m (points not included in the fits are marked with black squares). The legend lists the parameters for the MBB fit and the adopted redshift from Table~\ref{tab:fluxtable}. Bottom row: Marginalized likelihood distributions (histograms) and joint likelihood distribution (contours) of $\beta$ and $T_d$ for each of these sources. While the fits to the data are good in all cases, the dust parameters $\beta$ and $T_d$ can only be well constrained for ALMA~43 and ALMA~48, which have millimeter observations.}
    \label{sed_comp}
\end{figure*}

For most of our sources, the breadth of wavelength coverage allows us to constrain the dust parameters tightly (we report the median likelihood estimate of $\beta$ and $T_d$ as the measured value here and throughout the paper). This can be quantified by the uncertainties on the measured parameters, for which we use the 16th to 84th percentile range of the posterior likelihood distribution. For the 44 sources in the robust subsample with 3 or more ALMA measurements, we find a median 16th to 84th percentile range for $\beta$ of 0.45 and a median 16th to 84th percentile range for $T_d$ of 7.0~K. For the 13 sources with only 2 or 3 ALMA bands available, the parameters become more uncertain: the median range on $\beta$ is 1.05 and the median range on $T_d$ is 16.5~K. The errors are generally even higher for sources not in the robust subset. 

We illustrate this for a concrete example in Figure~\ref{sed_comp}, where we compare the fits for ALMA~43, which has 5 ALMA measurements; ALMA~48, which has just 2; and ALMA~60, which is not included in the robust subset due to it only having ALMA 870~$\mu$m and shorter wavelength data. We also show the marginalized likelihood distributions (histograms) and joint likelihood distributions (contour plot) for $\beta$ and $T_d$ for each source.
For ALMA~43, the parameters are tightly constrained, with $\beta = 1.54^{+0.27}_{-0.26}$ and $T_d = 37.0^{+6.1}_{-4.7}$~K. The elongated shape of the joint likelihood distribution reflects the intrinsic degeneracy between $\beta$ and $T_d$, though in this case the parameters are well-constrained regardless. For ALMA~48, we find $\beta=2.54^{+0.65}_{-0.65}$ and $T_d = 32.2^{+10.4}_{-6.1}$~K; the parameters are still well-constrained with slightly higher errors. In contrast, for ALMA~60, though the peak of the SED is sampled by the SPIRE and submillimeter data, the lack of millimeter data means that a much larger range of emissivities are consistent with the available data (clearly apparent in the posterior likelihood distribution for $\beta$). We find $\beta = 1.72^{+1.20}_{-0.68}$ and $T_d=36.0^{+14.4}_{-11.1}$~K for this source. These errors are 2--4$\times$ higher than those for ALMA~43, which is also reflected in the 16th to 84th percentile distribution of models shown in the SED fits and the relatively wide joint likelihood distribution for $\beta$ and $T_d$.

%
%
\begin{figure}
    \centering
    \includegraphics[width=0.9\linewidth]{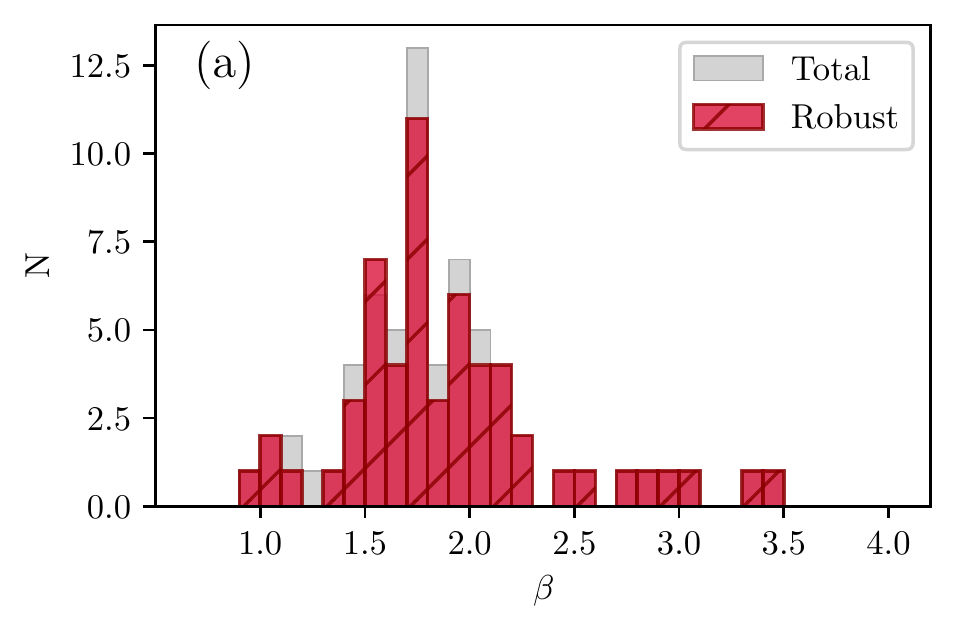}\\
    \includegraphics[width=0.9\linewidth]{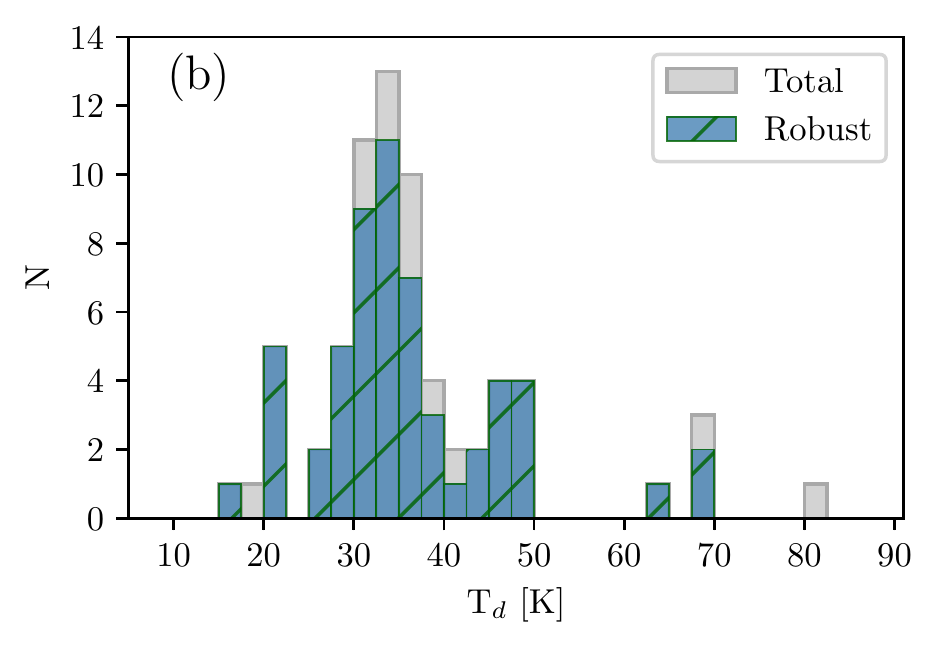}
    \caption{Histograms of the (a) dust emissivity indices and (b) effective dust temperatures measured for the 57 sources in the robust subset (hatched and red in (a); hatched and blue in (b)) and the 12 remaining sources in the total sample with redshifts (gray). Three of the latter sources do not have a measured $\beta$ due to limited photometry and hence are not shown in (a). We only allowed $\beta$ to range between 0.8 and 4.0 and $T_d$ to range between 10~K and 90~K.
}
    \label{param_hist}
\end{figure}

In Figure~\ref{param_hist}, we show histograms of (a) $\beta$ and (b) $T_d$ for the 57 sources in the robust subset (red and hatched) and the remaining 12 sources in the total sample with redshifts (gray). For the robust subset in (a), we find a wide distribution of $\beta$ that peaks in the range $\beta=$ 1.6--2.0 and has a median $\beta = 1.78^{+0.43}_{-0.25}$, where the uncertainties are the 16th to 84th percentile range. The error on the median derived from bootstrapping the sample is $\pm0.06$. Our median $\beta$ is consistent with the frequently-assumed value for high-redshift studies of $\beta=1.8$ and with recent measurements by \citet{dhc+21}, as discussed in the introduction. Considering only the 19 robust subset sources with speczs, we find a median $\beta = 1.74^{+0.28}_{-0.22}$.

\begin{figure}[t!]
\centering
    \includegraphics[width=0.90\linewidth]{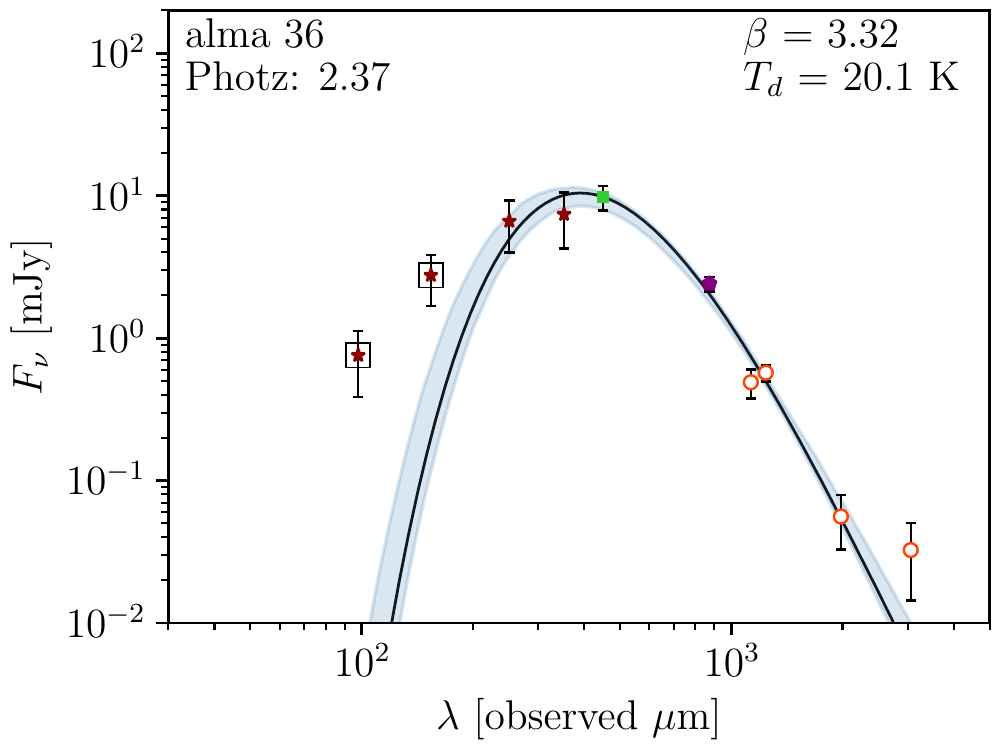}\\
    \includegraphics[width=0.90\linewidth]{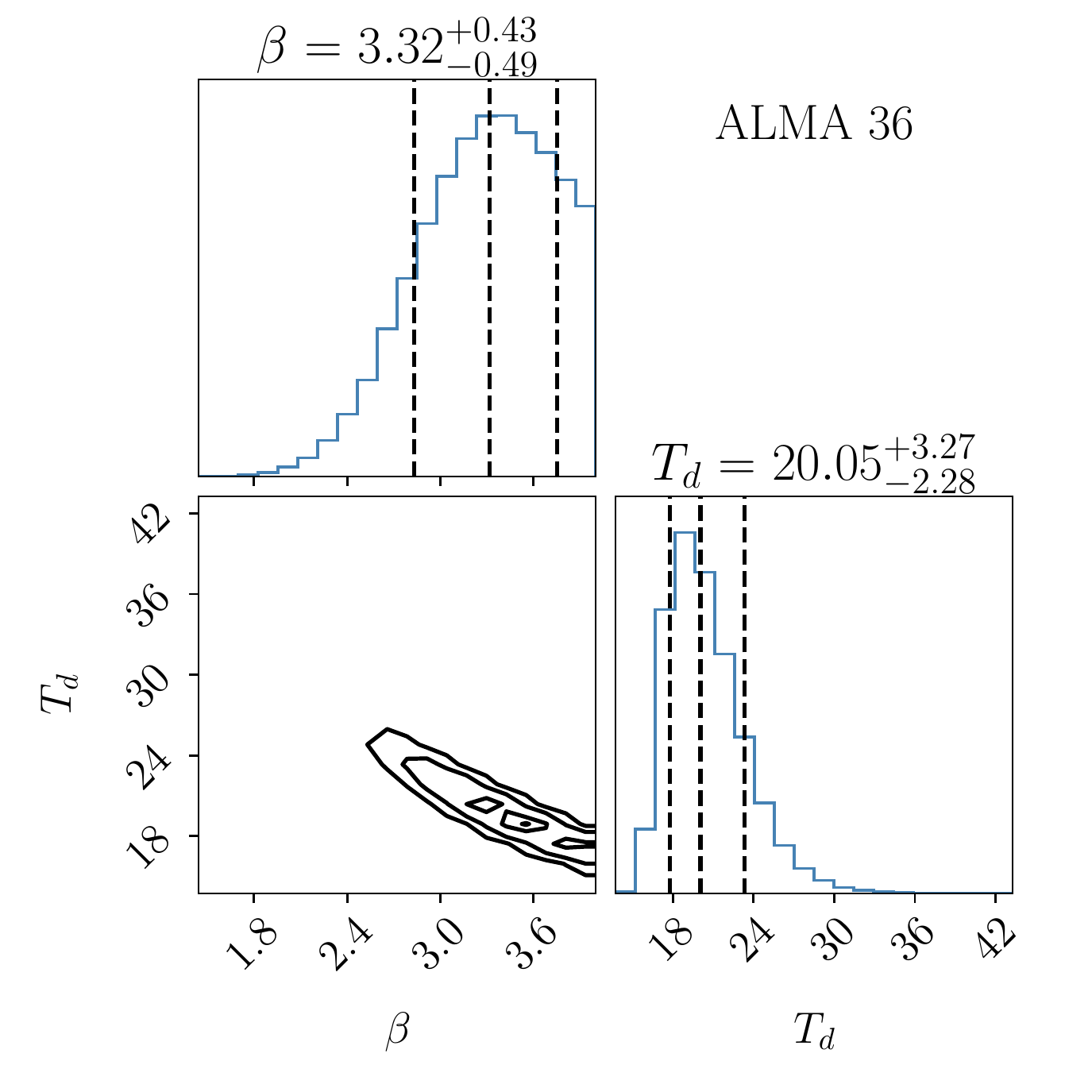}
   \caption{Same as Figure \ref{sed_comp}, but for ALMA 36. }
  \label{highbeta}
\end{figure}

We note that just six sources in the robust subset had measured emissivity indices of $\beta \gtrsim 2.8$. It is possible that these sources do in fact have very steep emissivities, or this could be an artifact of the fitting driven by relatively high uncertainties on the 2~mm and/or 3~mm data. Five of the sources only have photzs, so there is some possibility that an incorrect redshift could affect these fits as well. We show the SED for one of the six sources, ALMA~36, in Figure~\ref{highbeta}, along with the posterior likelihood distribution for $T_d$ and $\beta$. All six of these sources have large errors on $\beta$ that would make them consistent with $\beta \sim 2.8$ or below.

%
%

For the robust subset in Figure~\ref{param_hist}(b) (blue and hatched), we find 15~K~$< T_d <$~70~K with a median $T_d = 33.6^{+12.1}_{-5.4}$~K, where the error refers to the 16th to 84th percentile range of the sample. The error on the median derived from bootstrapping is $\pm 1.0$~K.
This median $T_d$ is similar to the median $T_d = 30.4_{-4.7}^{+6.9}$ K~found by \citet{dss+20} for their ALMA SCUBA-2 UDS sample and to the median $T_d = 30^{+14}_{-8}$~K found by \citet{dhc+21} (again, uncertainties are the 16th to 84th percentile range). Both of these results were derived using similar methods to ours, though \citet{dss+20} assumed a fixed $\beta = 1.8$. Considering only the 19 robust subset sources with speczs, we find a median $T_d = 32.8^{+13.0}_{-1.5}$~K.

We also integrate the best-fit MBB from rest-frame 8~$\mu$m to 1000~$\mu$m for each member of the robust subset to measure the FIR luminosities of our sources and to help in investigating selection biases. We then multiply the result by a correction factor of 1.35 (as was done in, for example, \citealt{jdm+19}) to account for the fact that MBB models for DSFGs typically underestimate the MIR flux that results from warmer dust components. The resulting luminosities range from $L_\text{IR} = 3.9 \times 10^{11}$~\lsun to $L_\text{IR} = 1.4 \times 10^{13}$~\lsun. We discuss the relation between dust temperature and FIR luminosity in Section~\ref{sec:temp_z}.

\begin{figure}
    \centering
    \includegraphics{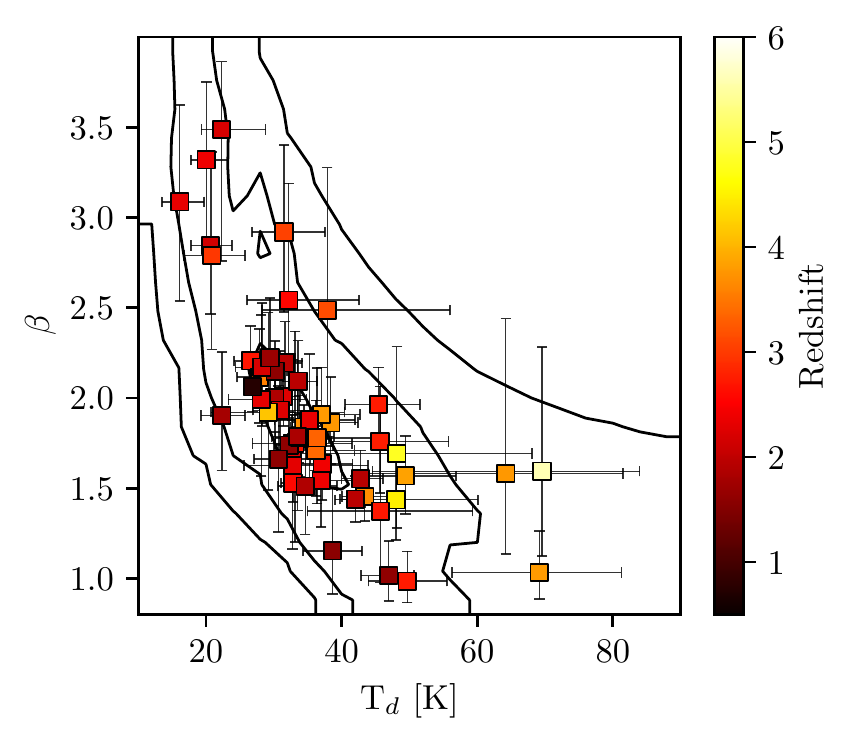}
    \caption{For the robust subset, emissivity index vs. dust temperature, both measured from MBB fits. The data are color-coded by adopted redshift (right-hand scale). Error bars represent 16th to 84th percentile ranges from the likelihood distributions. The stacked joint likelihood distribution for the robusts subset is shown as contours (levels are 39\%, 86.4\%, 98.8\%, i.e., the 1$\sigma$, 2$\sigma$, 3$\sigma$, levels of a 2D Gaussian).}
    \label{beta_temp}
\end{figure}

In Figure~\ref{beta_temp}, we show the relationship between $\beta$ and $T_d$ that we measure for the robust subset, with colors denoting redshifts. We find a general negative correlation, with Pearson coefficient $r = -0.67$, $p$~value $= 1\times10^{-10}$. 
This relationship is expected, both from other studies of high-redshift galaxies \citep[e.g.,][]{dhc+21} and from various observations of molecular clouds within the Galactic plane \citep[e.g.,][]{pvn+10}. Laboratory tests of silicate grains have also shown an intrinsic negative correlation of emissivity with temperature \citep{asj+96,bmn+05, ihc+20}. 

Although the intrinsic degeneracy between $\beta$ and $T_d$ can play a role in a negative correlation \citep[e.g.,][]{ggj18}, our SEDs are very well-sampled and thus we expect that our data are sufficient to break the degeneracy. In Figure~\ref{beta_temp} we overplot the stacked joint likelihood distribution for the robust subset (shown as contours) and find that it traces the median likelihood estimates for $\beta$ and $T_d$ well. This implies that the observed negative correlation is robust. \citet{dhc+21} discuss this correlation and several possible selection effects that could artificially produce this result. However, we find the negative correlation is only minimally affected when considering only sources with $L_{\rm IR} > 2\times 10^{12}$~\lsun and $z<3.5$,  where we expect a greater degree of completeness. This suggests that selection effects are unlikely to bias our result.

\begin{figure}[t]
\centering
    \includegraphics{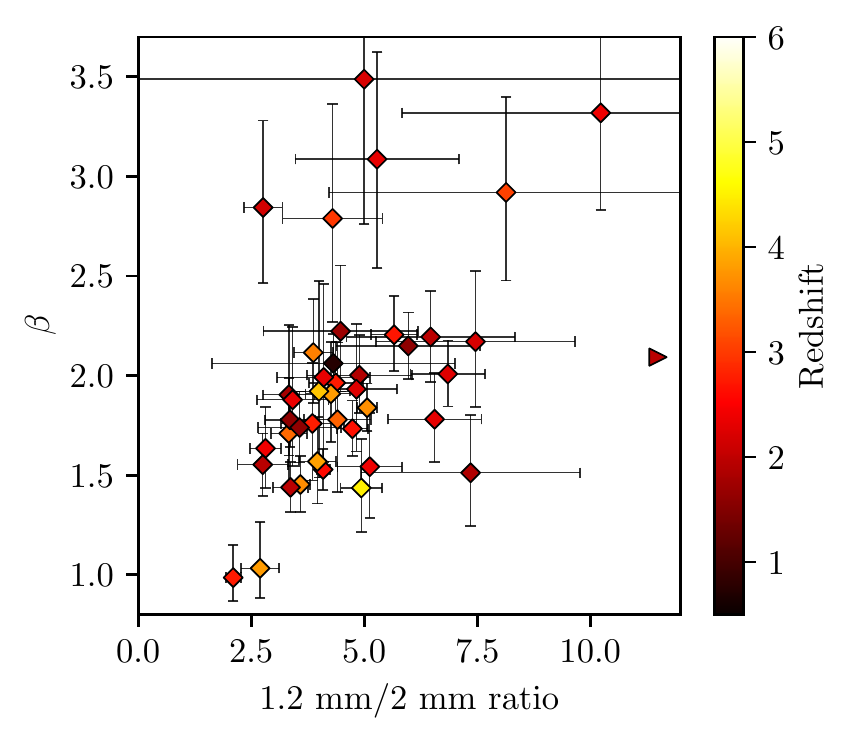}
    \caption{Emissivity index measured from MBB fits vs. the 1.2~mm/2~mm flux ratio, for the robust subset with both ALMA 1.2~mm and 2~mm measurements. The data are color-coded by adopted redshift (right-hand scale). The error bars on $\beta$ are the 16th to 84th percentile range of the likelihood distribution, while the errors on the flux ratio are derived from the respective photometric errors from Table~1. ALMA~34 is shown with a rightward-facing triangle since its flux ratio is off the plot at 26.9, due to its low-significance detection at 2~mm.
    }
    \label{beta_fluxratio}
\end{figure}

In general, one would expect $\beta$ to correlate with a ratio of fluxes on the RJ tail of the blackbody, such as the 1.2~mm/2~mm ratio, although a spread of temperatures and redshifts may introduce scatter into this relation. In Figure~\ref{beta_fluxratio}, we check for this correlation between $\beta$ and the 1.2~mm/2~mm flux ratio for all robust subset sources with a 2~mm measurement. Although there is scatter in the trend as expected, we find a significant correlation ($r = 0.50$, $p = 7\times10^{-4}$. This correlation remains when we only consider sources with 2~mm flux $>$~0.16~mJy, where the sample is expected to be essentially complete \citep{cbb23}. This cut excludes most of the sources with $\beta \gtrsim 2.8$ and the one source with 1.2~mm/2~mm ratio $>20$ (due to a low signal-to-noise detection at 2~mm).

%
%

%
%

In Table~\ref{tab:dustprops}, we summarize the median likelihood estimates for the dust parameters obtained from the MBB fits for the robust subset. In Appendix~\ref{appB}, we show the photometry and best-fit MBB for all 69 sources that we fit.

\section{\label{sec:disc}Discussion}

\subsection{Interpretation of the Measured Emissivity Index}

Our median $\beta = 1.78^{+0.43}_{-0.25}$ is broadly consistent with theoretical predictions for the interstellar medium (ISM), which give $\beta \sim 2$ \citep{dl84}. Measurements of the Milky Way's ISM have yielded $\beta = 1.5$ \citep{pbm09} and $\beta = 1.8$ \citep{PC+11}. These values are within the range spanned by our sample, though our results suggest that $\beta = 1.5$ may not be appropriate as an assumption for high-redshift galaxies as a whole. 

However, as we mentioned in Section~\ref{sec:analysis}, the emissivities that we measure from our SED fits are galaxy-integrated values that characterize the slope of the overall dust emission spectrum on the RJ side. The connection to the intrinsic emissivity of the dust composition of the galaxy is difficult to make and relies on assumptions about the relative proportions of dust grains of different sizes and structures and the ISM geometry, among other factors. Higher values of $\beta$ could suggest a greater proportion of larger crystalline dust grains that are at generally lower temperatures \citep{asj+96}. However, a range of temperatures in the dust components would, in general, flatten the effective RJ slope and hence reduce the measured $\beta$. 

Regardless of the physical interpretation,
we have shown that isothermal MBBs with $\beta \approx$~1.8 can be used to fit the FIR SEDs of high-redshift dusty galaxies spanning a range of submillimeter/millimeter fluxes, redshifts, and observed wavelengths.

\subsection{Comparisons with General Opacity Models}
\label{sec:genop}

While many authors assume an optically thin MBB model as a good trade-off between number of model parameters and quality of fit to the FIR SED, there is no guarantee that the emission is optically thin at shorter wavelengths. Some works (e.g., \citealt{ccv+11,czb+19b}) suggest that the dust may remain optically thick out to rest-frame wavelengths $\lambda_0 \sim$ 200--300 $\mu$m. Accordingly, several recent studies of the dust properties of DSFGs have instead used general opacity factors in their MBB models \citep[e.g.,][]{czm+21,ccz+22}.

%
%
\begin{figure}
    \centering
    \includegraphics[width=0.95\linewidth]{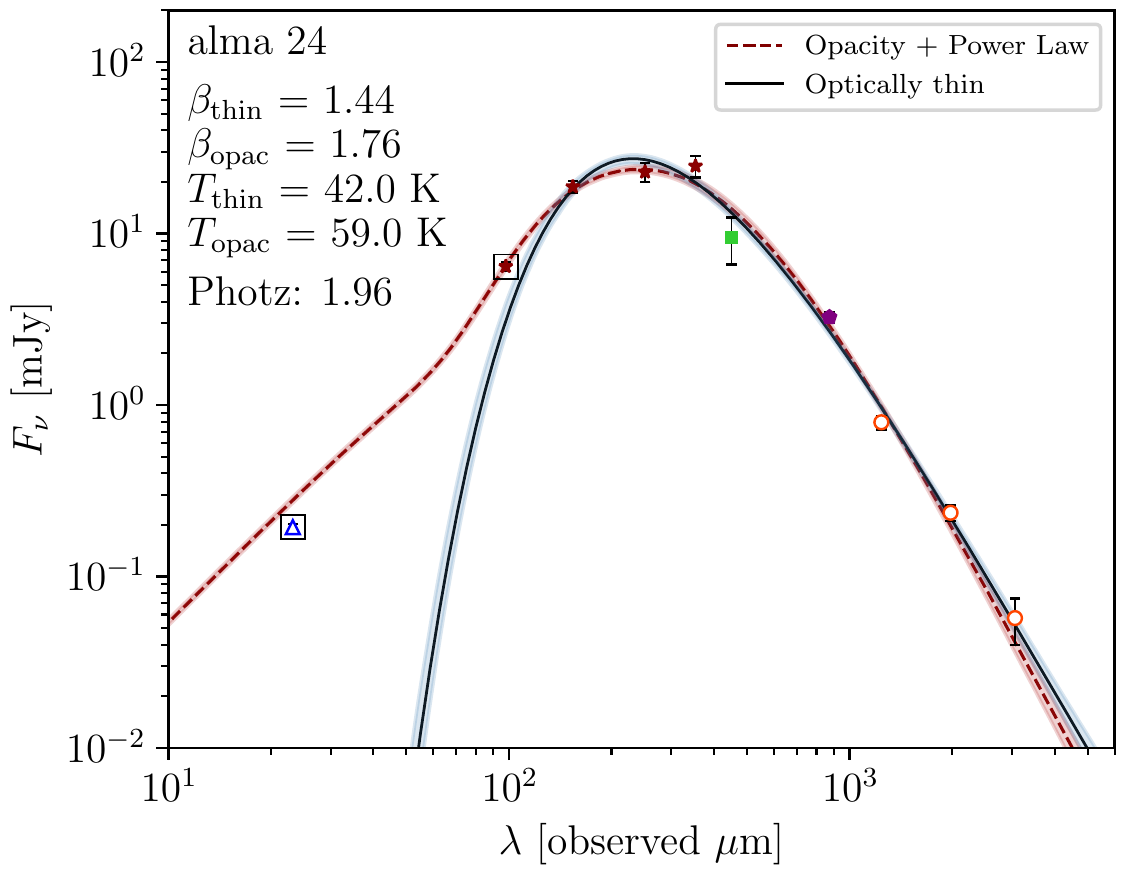}
    \caption{General opacity~+~power law MBB fit (red dot-dashed curve), with MIR power law slope $\alpha=2.0$, compared with optically thin MBB fit (black curve) for ALMA~24. The 16th to 84th percentile range from the posterior distribution of the MCMC models is shaded red for the opacity~+~power law model and shaded blue for the optically thin model. The optically thin model is fit to all points at rest-frame wavelengths greater than 50~$\mu$m (points below this are marked with black squares) while the opacity~+~power law fit is fit to all data.
    The measured dust temperature and emissivity for each are given in the figure legend, showing a slightly steeper $\beta$ and a higher $T_d$ for the opacity~+~power law model.}
    \label{opac_fit_comp}
\end{figure}

To make a direct comparison with these studies, we next fit the SEDs of the robust subset with a general opacity MBB model of the form $S(\nu) \propto (1-e^{-\tau}) B_\nu(T)$, where $\tau$ is the optical depth described by the power law $\tau \propto (\nu/\nu_0)^\beta$. Here $\nu_0$ is the turnover frequency at which the optical depth equals 1. We assume a turnover wavelength $\lambda_0 = 200$~$\mu$m, and we include a MIR power law with $\alpha = 2.0$, following the prescription of \citet{cas12}. 

With the inclusion of the power law on the short-wavelength side of the FIR peak, we fit the photometry down to a rest-frame wavelength of 10 $\mu$m. The MIR power law is included in the SED fits of both \citet{czm+21} and \citet{ccz+22}, though the former vary the slope $\alpha$ in their fits, where possible, and generally find steep slopes ($\alpha \approx $~3.5--7). We choose to leave $\alpha$ fixed at 2.0 so that our model is identical to that of \citet{ccz+22}. 

In Figure~\ref{opac_fit_comp},
we show a comparison of this opacity~+~power law model and the optically thin model fits for ALMA~24. Both fit the data well but with somewhat different median likelihood values. The optically thin fit gives $\beta = 1.44$ and $T_d = 42.0$~K, while the opacity~+~power law fit gives $\beta = 1.76$ and $T_d = 59.0$~K. We also note that for this galaxy, the $\alpha=2.0$ power law slope does a good job of fitting the measured MIPS 70~$\mu$m point, which is not included in our optically thin fit.

%
%
\begin{figure}
    \centering
    \includegraphics[width=0.9\linewidth]{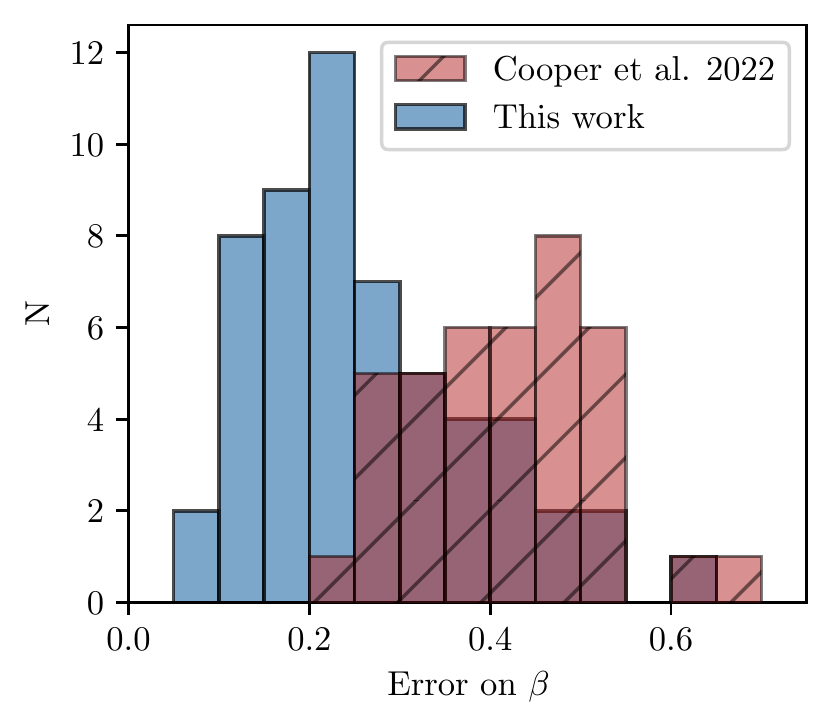}
    \caption{Histogram of errors in our $\beta$ values for the robust subset opacity~+~power law MBB fits (blue) with the errors from \citet{ccz+22} (hatched and red). The errors shown are half of the 16th to 84th percentile range determined for each individual measurement.}
    \label{beta_err}
\end{figure}

We find median $\beta = 2.06^{+0.56}_{-0.37}$ and $T_d = 57.5^{+12.6}_{-12.2} $~K for the robust subset using the opacity~+~power law model. This median $\beta$ is slightly lower than those reported in \citet{czm+21} and \citet{ccz+22}, who find median $\beta = 2.2^{+0.5}_{-0.4}$ and $\beta = 2.4^{+0.3}_{-0.3}$, respectively. The errors here are the 16th to 84th percentile range of each sample and thus reflect the spread in measured $\beta$ values rather than the errors on the individual $\beta$ values. 

Thus, in Figure~\ref{beta_err}, we compare the individual errors on our $\beta$ values with the errors from \citet{ccz+22}, who only had ALMA data in the 2~mm band. Since we and \citet{ccz+22} report both upper and lower errors determined from the likelihood distributions, in the figure we take the average of the upper and lower errors for each source (i.e., half of the 16th to 84th percentile range; for a Gaussian posterior this would be the 1$\sigma$ error) for a rough comparison of errors. We see that the addition of more ALMA bands in our case reduces the errors on the individual $\beta$ measurements.

A Mann-Whitney test gives a 0.05\% probability that our sample comes from the same underlying distribution as that of \citet{ccz+22}. This may be due to their combining single-dish observations with ALMA 2~mm data, or to the brighter flux limit of their sample (SCUBA-2 850~$\mu$m flux~$>$~5.55~mJy). However, the 7 sources in our sample that meet their selection criteria have median $\beta = 1.89$ from the opacity~+~power law fits, suggesting that the discrepancy reflects a problem with mixed single-dish and interferometric measurements.

The median temperature we measure for the opacity~+~power law model is $\sim$24~K lower and the median emissivity is 0.3 higher than the results for the optically thin MBB. For comparison, \citet{dhc+21} found dust temperatures that were $\sim$10~K lower for their optically thin model than for their general opacity model but found that emissivity was robust against different opacity assumptions---though they did not include a MIR power law and they allowed $\lambda_0$ to vary between 60--140~$\mu$m. Meanwhile, for simulated galaxies with optically thick emission out to $\sim$200~$\mu$m, \citet{hkj+11} found that dust temperatures derived from an optically thin model could underpredict its optically thick counterpart by $\sim$20~K, which is closer to the deviation we measure. 

To test whether our discrepancies in $\beta$ and $T_d$ were related to the assumption of $\lambda_0 = 200$~$\mu$m and/or the inclusion of the MIR power law, we fit the SEDs of the robust subset again with three other general opacity models: a general opacity MBB with $\lambda_0$ fixed to 200~$\mu$m with no MIR power law, a general opacity MBB with $\lambda_0$ fixed to 100~$\mu$m with a MIR power law, and a general opacity MBB with $\lambda_0$ fixed to 100~$\mu$m but no power law included (to make a better comparison with the general opacity model used in \citealt{dhc+21}).  

%
%
\begin{figure*}[!ht]
\centering
    \plottwo{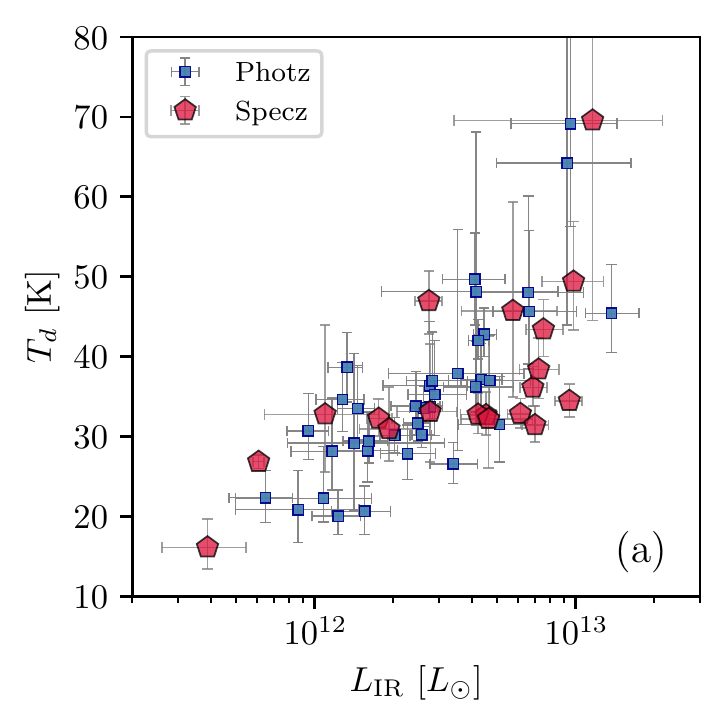}{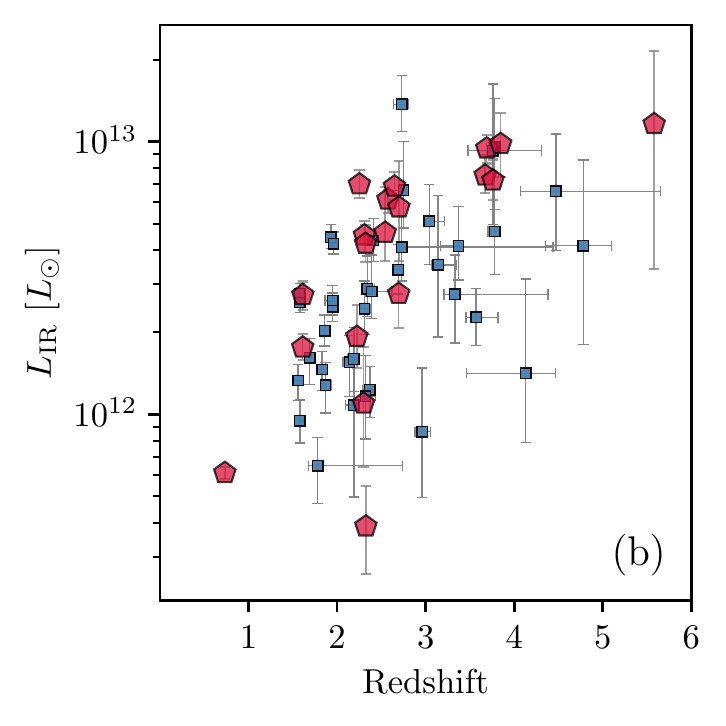}\\
    \plottwo{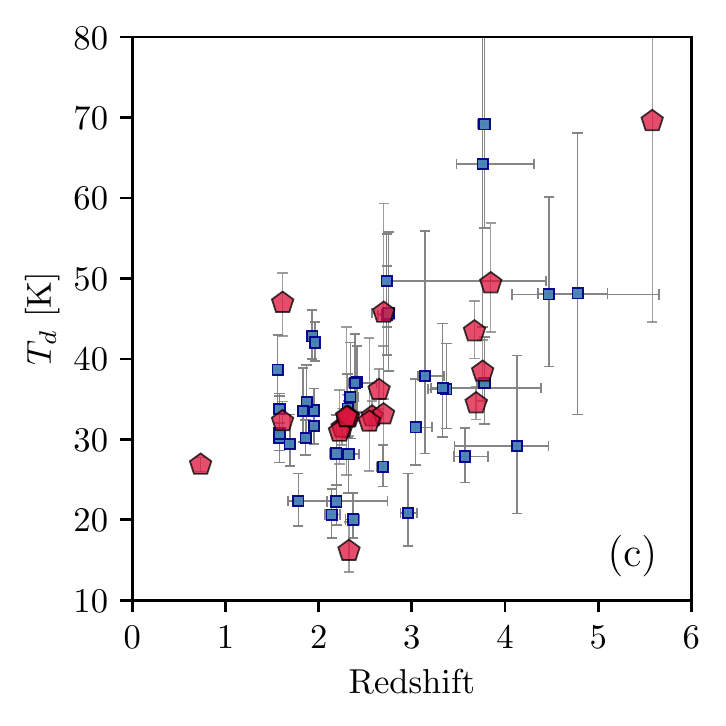}{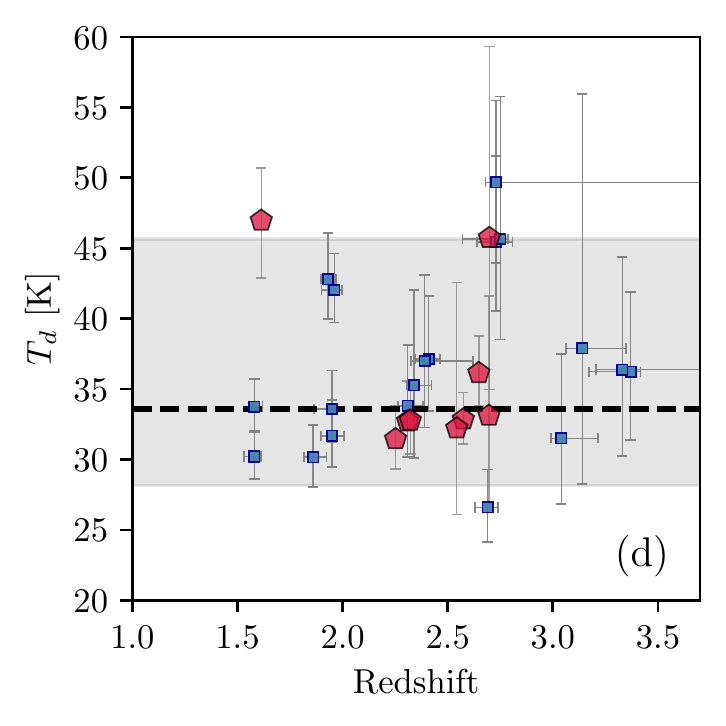}
    \caption{For the robust subset, (a) Dust temperature vs. FIR luminosity, (b) FIR luminosity vs. redshift, (c) Dust temperature vs. redshift, and (d) Same as (c) but only for the sources with a FIR luminosity $>2\times10^{12}$~\lsun and $z < 3.5$, where our sample is more likely to be complete. The dashed line shows the median $T_d = 33.6$~K, and the shaded region denotes the 16th to 84th percentile range of the robust subset. In all panels, sources with speczs are shown as red pentagons while those with photzs are shown as blue squares. 
    }
    \label{temp_lum_z}
\end{figure*}

For the $\lambda_0=100$~$\mu$m MBB model with no power law, we find median $\beta = 1.80$ and median $T_d = 42$~K. The results are only minimally affected by the inclusion of the MIR power law: for the $\lambda_0=100$~$\mu$m MBB model with the power law we find median $\beta = 1.78$ and median $T_d = 45$~K. The median $\beta$ for these models is consistent with the optically thin case, and the median $T_d$ is larger by about $\sim$10~K. These results are consistent with those of \citet{dhc+21}, who used a similar model. However, for the $\lambda_0=200$~$\mu$m MBB model with the power law included, we measure a median $\beta = 2.02$ and median $T_d = 56$~K. Thus, we observe that the assumption of the turnover wavelength has a significant effect on the derived parameters, and the emissivity is not necessarily robust against opacity assumptions if the turnover wavelength is high enough. This may also help to explain why models with $\lambda_0 = 200$~$\mu$m such as that of \citet{ccz+22} find a higher median $\beta$ for their sample than that of \citet{dhc+21}, though as we have already discussed, we find $\beta$ values inconsistent with \citet{ccz+22} even under identical modeling assumptions.

Finally, we note that the general opacity models (with or without a power law) did not necessarily provide a better fit to the FIR data than our optically thin model, aside from being able account for shorter wavelength fluxes in the cases where the power law was included. 

Even if we cannot constrain with certainty the underlying dust properties or opacity of the sources, we can conclude that our range of dust parameters are consistent with theory and with recent studies of the local and high-redshift universe. This is true whether we assume a general opacity or optically thin model, though we caution that---as many authors have pointed out---making direct comparisons between parameters derived using different models is not straightforward. Allowing for differences in the assumed models, we confirm that $\beta$ between 1.8 and 2.0 is appropriate for DSFGs at high redshift.

\subsection{Dust Temperature Variation with Redshift}
\label{sec:temp_z}

Although a number of studies have found that dust temperature increases with redshift \citep[e.g.,][]{mdb+12,mls+14,bdm+15,sep+18,zad+18,sfp+22}, there is debate over whether this is simply a selection bias due to picking out higher luminosity sources at higher redshifts. 
Other recent studies find little to no evidence of temperature evolution with redshift when the luminosity dependence is taken into account \citep{lws+20,dss+20,bcb+22,dc22}. The luminosity ranges of these studies overlap with ours in part or in full, although \citet{dc22} consider a lower-redshift selection of $0 < z < 2$ galaxies with a luminosity range that extends down to $L_\text{IR} \sim 10^{10}$~\lsun.

Our robust subset is also consistent with no evolution.
In Figure \ref{temp_lum_z}(a), we show $T_d$ from our optically thin MBB fits plotted against FIR luminosity. We see an increase in dust temperature with luminosity. From Figure~\ref{temp_lum_z}(b), we also see a general increase in $T_d$ with redshift.
However, when we restrict our analysis to sources with $L_\text{IR} > 2\times10^{12}$~\lsun and $z < 3.5$, where the sample is less likely to suffer from incompleteness, we find the Pearson coefficient for the temperature versus redshift relation is $r=0.10$, with a $p$~value of 0.61, indicating no statistically significant correlation. This is shown in Figure~\ref{temp_lum_z}(c); the lack of visible trend is clear. We note that since we employ photzs for some of our sources and redshift and dust temperature are degenerate in MBB SED fits, incorrect redshifts could affect our results. Thus, it is possible we have underestimated the errors on the dust temperatures of some sources.

\section{\label{sec:conc}Summary}

We analyzed the dust properties of a large sample of GOODS-S galaxies selected at 870~$\mu$m using new ALMA continuum observations at millimeter wavelengths. Compared to other large samples of DSFGs, the ALMA observations, which probe the RJ side of the FIR SED, reach some of the deepest depths at these wavelengths. Here we summarize our main results:

\begin{enumerate}

\item Using optically thin, isothermal MBB fits, we measured a median $T_d = 33.6^{+12.1}_{-5.4}$~K and a median $\beta = 1.78^{+0.43}_{-0.25}$ for our robust subset of 57 sources. 

\item  We observed a negative correlation between $\beta$ and $T_d$ for our robust subset. Since our FIR SEDs are relatively well-sampled (up to nine photometric points from rest-frame 50--1500~$\mu$m) and based on the stacked likelihood distributions, this relationship appears robust. We also confirm that it is unlikely to have been produced by selection effects.

\item We determined that the opacity assumptions used in the MBB fits can affect the measured values for $\beta$ as well as $T_d$. We found that a general opacity MBB with $\lambda_0 = 100$~$\mu$m gave similar values of $\beta$ to optically thin fits, while a general opacity MBB with $\lambda_0 = 200$~$\mu$m gave higher values of $\beta$. In all cases $T_d$ was higher for the general opacity MBB fits. 

\item After restricting to sources in our robust subset with $L_\text{IR} > 2\times10^{12}$~\lsun and $z < 3.5$, we find no evidence for temperature evolution from $z=1$ to $z=3.5$.

\end{enumerate}

This work is one of only a few to directly measure the emissivity index and dust temperature of individual DSFGs using deep ALMA millimeter imaging. We find that the dust emission of DSFGs is well represented by modified blackbodies with $\beta \sim 1.8$.
Future observations of larger and fainter samples of DSFGs using ALMA and TolTEC will confirm whether the dust characteristics of DSFGs differ from local galaxies and help to disentangle competing theories about the origins of their extreme dust masses and star formation rates.

\vspace{6cm}

\textcolor{white}{}

\begin{acknowledgements}
{
We thank the anonymous referee for constructive comments that helped us to improve the manuscript.
We gratefully acknowledge support for this research from 
the William F. Vilas Estate (S.~J.~M.),
a Kellett Mid-Career Award and a WARF Named Professorship from the 
University of Wisconsin-Madison Office of the 
Vice Chancellor for Research and Graduate Education with funding from the 
Wisconsin Alumni Research Foundation (A.~J.~B.),
NASA grant 80NSSC22K0483 (L.~L.~C.), 
the Millennium Science Initiative Program -- ICN12\_009 (F.~E.~B), 
CATA-Basal -- FB210003 (F.~E.~B), and FONDECYT Regular -- 1190818 (F.~E.~B) 
and 1200495 (F.~E.~B).

The National Radio Astronomy Observatory is a facility of the National Science
Foundation operated under cooperative agreement by Associated Universities, Inc.
This paper makes use of the following ALMA data: 
ADS/JAO.ALMA\#2021.1.00024.S.  \\ 
ALMA is a partnership of ESO (representing its member states), NSF (USA), and NINS (Japan), 
together with NRC (Canada), MOST and ASIAA (Taiwan), and KASI (Republic of Korea),
 in cooperation with the Republic of Chile. The Joint ALMA Observatory is operated by 
 ESO, AUI/NRAO, and NAOJ.
 
The James Clerk Maxwell Telescope is operated by the East Asian Observatory on 
behalf of The National Astronomical Observatory of Japan, Academia Sinica Institute 
of Astronomy and Astrophysics, the Korea Astronomy and Space Science Institute, 
the National Astronomical Observatories of China and the Chinese Academy of 
Sciences (grant No.~XDB09000000), with additional funding support from the Science 
and Technology Facilities Council of the United Kingdom and participating universities 
in the United Kingdom and Canada. 

We wish to recognize and acknowledge 
the very significant cultural role and reverence that the summit of Maunakea has always 
had within the indigenous Hawaiian community. We are most fortunate to have the 
opportunity to conduct observations from this mountain.
}
\end{acknowledgements}

\facilities{ALMA, JCMT}
\software{astropy \citep{astropy:2022}, \textsc{casa} \citep{casa}, \textsc{emcee} \citep{fhl+13}}

\appendix
\section{Flux Densities and Dust Properties \label{appA}}

In Table~1, we list the positions and redshifts of the total sample, along with the ALMA 1.2~mm, 2~mm, and 3~mm fluxes from the present work and the 1.1~mm fluxes from \citet{geb+22}. In Table~2, we give the median likelihood $\beta$ and $T_d$ values and the errors from the posterior likelihood distributions for our optically thin MBB fits, as well as the FIR luminosities and errors measured from the fits.

%
%
\startlongtable
\begin{deluxetable*}{cccccccccccc}
\tablewidth{0pt}
\tablecaption{\label{tab:fluxtable}
Total Sample Redshifts and Fluxes}
\tablehead{
\colhead{C18} & & & & \multicolumn{2}{c}{Total}& \multicolumn{2}{c}{Peak}& \multicolumn{2}{c}{Peak}& \multicolumn{2}{c}{Peak} \\
\colhead{No.} & \colhead{R.A.}&\colhead{Decl.}&\colhead{$z$} & \colhead{$f_{1.13\,{\rm mm}}$} & \colhead{Error} & \colhead{$f_{1.24\,{\rm mm}}$} & \colhead{Error} & \colhead{$f_{2\,{\rm mm}}$} & \colhead{Error} & \colhead{$f_{3\,{\rm mm}}$} & \colhead{Error} \\
& \colhead{J2000.0} & \colhead{J2000.0} & & \multicolumn{2}{c}{(mJy)}& \multicolumn{2}{c}{(mJy)} & \multicolumn{2}{c}{(mJy)} & \multicolumn{2}{c}{(mJy)}} 

\colnumbers

\startdata
1 & 53.030373 &   -27.855804 & 2.574  & \nodata & \nodata & 2.48 & 0.08 & 0.61 & 0.02 & 0.146 & 0.019  \\
2 & 53.047211 &   -27.870001 & 3.690  & \nodata & \nodata & 2.71 & 0.09 & 0.54 & 0.02 & 0.119 & 0.019  \\
3 & 53.063877 &   -27.843779  & 2.648  & 2.06 & 0.10 & 1.55 & 0.1 & 0.33 & 0.02 & 0.078 & 0.02  \\
4 & 53.020374 &   -27.779917 & 2.252  & 2.75 & 0.09 & 1.87 & 0.08 & 0.27 & 0.04 & 0.076 & 0.04  \\
5 & 53.118790 &   -27.782888 & 2.309  & 2.15 & 0.11 & \nodata & \nodata & 0.26 & 0.03 & 0.114 & 0.043 \\
{[6]} & 53.195126 &   -27.855804 & \nodata & \nodata & \nodata & 1.41 & 0.08 & 0.41 & 0.02 & 0.039 & 0.021  \\
7 & 53.158371 &   -27.733612 & 3.672  & 2.15 & 0.12 & 1.36 & 0.06 & 0.38 & 0.02 & 0.081 & 0.019  \\
8 & 53.105247 &   -27.875195 & 2.69 (2.62-2.73)  & \nodata & \nodata & 1.41 & 0.1 & 0.25 & 0.02 & 0.05 & 0.02 \\
9 &  53.148876 &   -27.821167  & 2.322  & 2.11 & 0.11 & \nodata & \nodata & 0.32 & 0.04 & 0.052 & 0.018  \\
10 & 53.082085 &   -27.767279 & 2.41 (2.34-2.46)  & 1.34 & 0.11 & 0.73 & 0.1 & 0.26 & 0.02 & 0.032 & 0.022  \\
{[11]} & 53.079376 &   -27.870806  & \nodata  & 2.05 & 0.12 & 0.99 & 0.1 & 0.28 & 0.02 & 0.024 & 0.018 \\
12 & 53.142792 &   -27.827888 & 3.764  & 2.25 & 0.12 & 1.26 & 0.08 & \nodata & \nodata & 0.055 & 0.02  \\
13 & 53.074837 &   -27.875916  & [2.73 (2.68-4.43)]  & 1.67 & 0.10 & 0.7 & 0.06 & 0.33 & 0.02 & 0.033 & 0.018  \\
14 & 53.092335 &   -27.826834 & 2.73 (2.63-2.80)  & 1.07 & 0.09 & 0.8 & 0.1 & 0.18 & 0.02 & 0.012 & 0.019  \\
15 & 53.024292 &   -27.805695 & 2.14 (2.06-2.23)  & \nodata & \nodata & 0.47 & 0.07 & 0.17 & 0.02 & 0.006 & 0.02  \\
16 & 53.082752 &   -27.866585 & 3.37 (3.17-3.41)  & 1.40 & 0.12 & 0.92 & 0.11 & 0.28 & 0.03 &  \nodata & \nodata  \\
17 & 53.146629 &   -27.871029 & 3.57 (3.45-3.81)  & \nodata & \nodata & 0.81 & 0.08 & 0.21 & 0.02 & 0.069 & 0.02  \\
18 & 53.092834 &   -27.801332 & 3.847  & 1.39 & 0.10 & 0.9 & 0.08 & 0.23 & 0.02 & 0.084 & 0.019  \\
19 &  53.108795 &   -27.869028 & [4.47 (4.07-5.65)]  & 1.24 & 0.10 & 1.17 & 0.09 & 0.24 & 0.02 & 0.08 & 0.018  \\
20 & 53.198292 &   -27.747889 & 1.93 (1.89-1.96)  & 0.65 & 0.12 & 0.43 & 0.1 & 0.16 & 0.02 &  \nodata & 0.018  \\
21 & 53.178333 &   -27.870222 & 3.78 (3.70-3.83)   & \nodata & \nodata & 0.78 & 0.09 & 0.18 & 0.02 & 0.06 & 0.016  \\
22 &  53.183460 &   -27.776638  & 2.698  & 1.49 & 0.13 & \nodata & \nodata & \nodata & \nodata &  \nodata & \nodata  \\
23 & 53.157207 &   -27.833500 & 1.58 (1.56-1.61)  & 1.36 & 0.12 & 0.47 & 0.07 & 0.14 & 0.02 & 0.03 & 0.018 \\
24 & 53.102791 &   -27.892860  & 1.96 (1.90-1.99)  & \nodata & \nodata & 0.61 & 0.06 & 0.18 & 0.02 & 0.044 & 0.017  \\
25 & 53.181377 &   -27.777557  & 2.696  & 1.41 & 0.12 & \nodata & \nodata & \nodata & \nodata &  \nodata & \nodata \\
26 &  53.070251 &   -27.845612 & 3.78 (3.70-3.82)  & 0.83 & 0.09 & 0.51 & 0.08 & 0.19 & 0.02 & 0.066 & 0.02  \\
27 & 53.014584 &   -27.844389  & [1.78 (1.67-2.73)]  & \nodata & \nodata & 0.52 & 0.08 & 0.16 & 0.02 & 0.053 & 0.018  \\
28 & 53.139290 &   -27.890722 & [3.33 (3.20-4.38)]  & \nodata & \nodata & 0.56 & 0.08 & 0.13 & 0.02 & 0.05 & 0.02  \\
{[29]} & 53.137127 &   -27.761389 & \nodata  & 0.92 & 0.10 & 0.68 & 0.09 & 0.13 & 0.02 & 0.031 & 0.019  \\
30 & 53.071709 &   -27.843693 & 1.86 (1.81-1.92)  & 0.83 & 0.10 & 0.46 & 0.08 & 0.09 & 0.02 & 0.02 & 0.019 \\
31 & 53.077377 &   -27.859612 & 1.95 (1.89-2.00)  & 0.51 & 0.10 & 0.44 & 0.07 & 0.07 & 0.02 & 0.011 & 0.022  \\
32 & 53.049751 &   -27.770971 & 2.75 (2.57-2.78)  & 0.93 & 0.10 & 0.47 & 0.07 & 0.12 & 0.02 & 0.042 & 0.017  \\
33 & 53.072708 &   -27.834278 & 1.58 (1.53-1.61)  & 0.69 & 0.11 & 0.49 & 0.08 & 0.08 & 0.02 & 0.057 & 0.019  \\
34 & 53.090752 &   -27.782473 & 1.95 (1.86-1.97)  & 0.76 & 0.09 & 0.35 & 0.07 & 0.01 & 0.03 &  \nodata & 0.022  \\
35 & 53.091747 &   -27.712166  & 1.612  & \nodata & \nodata & 0.35 & 0.09 & 0.1 & 0.02 & 0.042 & 0.018\\
36 & 53.086586 &   -27.810249 & 2.37 (2.28-2.42)  & 0.49 & 0.10 & 0.44 & 0.07 & 0.04 & 0.02 & 0.025 & 0.018  \\
37 & 53.146378 &   -27.888807 & 2.96 (2.87-3.05)   & \nodata & \nodata & 0.43 & 0.09 & 0.1 & 0.03 & 0.011 & 0.018  \\
38 & 53.092335 &   -27.803223 & 2.31 (2.26-2.38)  & 0.85 & 0.11 & 0.8 & 0.09 & 0.12 & 0.02 &  \nodata & 0.019  \\
39 & 53.124332 &   -27.882696 & 3.04 (2.99-3.21)   & \nodata & \nodata & 0.35 & 0.08 & 0.04 & 0.02 & 0.032 & 0.018  \\
40 & 53.131123 &   -27.773195 & 2.223  & 0.72 & 0.11 & 0.56 & 0.05 & 0.12 & 0.03 & 0.015 & 0.026  \\
41 & 53.172832 &   -27.858860 & [4.13 (3.45-4.46)]  & \nodata & \nodata & 0.66 & 0.11 & 0.16 & 0.02 & 0.039 & 0.019  \\
42 & 53.091629 &   -27.853390  & 2.34 (2.30-2.42)  & 0.81 & 0.10 & 0.41 & 0.09 & 0.12 & 0.02 & 0.029 & 0.02 \\
43 & 53.068874 &   -27.879723 & 2.39 (2.32-2.62)  & 1.03 & 0.11 & 0.83 & 0.07 & 0.16 & 0.03 & 0.034 & 0.021 \\
{[44]} & 53.087166 &   -27.840195 & \nodata  & 0.85 & 0.11 & 0.73 & 0.09 & 0.18 & 0.02 & 0.07 & 0.017 \\
{[45]} & 53.041084 &   -27.837721 & [7.62 (7.15-7.93)]  & \nodata & \nodata & 0.54 & 0.08 & 0.14 & 0.02 & 0.022 & 0.018  \\
46 & 53.104912 &   -27.705305  & 1.613  & \nodata & \nodata & \nodata & \nodata & \nodata & \nodata & 0.049 & 0.024 \\
47 & 53.163540 &   -27.890556 & 2.19 (2.12-2.22)  & \nodata & \nodata & 0.5 & 0.1 & 0.07 & 0.02 & 0.018 & 0.019  \\
48 & 53.160664 &   -27.776251 & 2.543  & 0.99 & 0.12 & \nodata & \nodata & \nodata & \nodata &  \nodata & \nodata  \\
49 & 53.053669 &   -27.869278 & 1.87 (1.84-1.93)  & \nodata & \nodata & 0.5 & 0.11 & 0.07 & 0.02 &  \nodata & 0.019  \\
50 & 53.089542 &   -27.711666 & 1.69 (1.64-1.70) & \nodata & \nodata & 0.3 & 0.08 & 0.07 & 0.03 & 0.012 & 0.022 \\
51 & 53.067833 &   -27.728889  & 2.32 (2.29-2.43)  & \nodata & \nodata & 0.39 & 0.08 & 0.09 & 0.02 & 0.028 & 0.019 \\
52 &  53.064793 &   -27.862638 & [4.78 (4.35-5.10)]  & 0.54 & 0.10 & \nodata & \nodata & 0.14 & 0.03 & 0.002 & 0.017  \\
53 & 53.198875 &   -27.843945 & 1.56 (1.50-1.60)  & 1.01 & 0.12 & \nodata & \nodata & \nodata & \nodata &  \nodata & \nodata \\
{[54]} & 53.181995 &   -27.814196 & [9.42 (9.35-9.83)]  & \nodata & \nodata & \nodata & \nodata & 0.06 & 0.02 & -0.02 & 0.02 \\
{[55]} & 53.048378 &   -27.770306 & \nodata & 0.71 & 0.11 & 0.24 & 0.1 & 0.06 & 0.02 & 0.01 & 0.019 \\
56 & 53.107044 &   -27.718334 & 2.299  & \nodata & \nodata & \nodata & \nodata & 0.1 & 0.03 &  \nodata & \nodata \\
{[57]} & 53.033127 &   -27.816778 & 3.08 (3.00-3.68)  & \nodata & \nodata & \nodata & \nodata & \nodata & \nodata &  \nodata & \nodata \\
{[58]} & 53.183666 &   -27.836500 & [4.73 (4.39-4.90)]  & 1.23 & 0.12 & \nodata & \nodata & 0.19 & 0.03 & 0.042 & 0.018  \\
59 & 53.094044 &   -27.804195 & 2.325  & \nodata & \nodata & 0.37 & 0.09 & 0.07 & 0.03 & 0.01 & 0.02  \\
{[60]} & 53.124584 &   -27.893305 & 2.53 (2.41-2.60)  & \nodata & \nodata & \nodata & \nodata & \nodata & \nodata &  \nodata & \nodata \\
{[61]} & 53.132751 &   -27.720278 & [4.67 (4.48-5.23)]  & \nodata & \nodata & \nodata & \nodata & \nodata & \nodata &  \nodata & \nodata  \\
{[62]} & 53.080669 &   -27.720861 & 2.94 (2.88-3.03)  & \nodata & \nodata & \nodata & \nodata & \nodata & \nodata &  \nodata & \nodata  \\
63 & 53.120041 &   -27.808277 & 1.83 (1.78-1.88) & 0.55 & 0.12 & \nodata & \nodata & \nodata & \nodata &  \nodata & \nodata  \\
{[64]} & 53.117085 &   -27.874918 & 3.26 (3.20-3.40)  & \nodata & \nodata & \nodata & \nodata & \nodata & \nodata &  \nodata & \nodata \\
65 & 53.131458 &   -27.841389 & 1.58 (1.56-1.62)  & 0.74 & 0.10 & \nodata & \nodata & \nodata & \nodata &  \nodata & \nodata \\
{[66]} & 53.044708 &   -27.802027 & 0.653  & \nodata & \nodata & \nodata & \nodata & \nodata & \nodata &  \nodata & \nodata  \\
{[67]} & 53.072002 &   -27.819000  & 1.69 (1.66-1.84)  & \nodata & \nodata & \nodata & \nodata & \nodata & \nodata &  \nodata & \nodata  \\
68 & 53.120461 &   -27.742083 & 5.58 & 0.95 & 0.12 & \nodata & \nodata & \nodata & \nodata &  \nodata & \nodata \\
{[69]} & 53.113125 &   -27.886639 & 2.55 (2.47-2.64)  & \nodata & \nodata & \nodata & \nodata & \nodata & \nodata &  \nodata & \nodata  \\
70 & 53.141251 &   -27.872860  & 3.14 (3.06-3.34) & \nodata & \nodata & \nodata & 0.06 & 0.04 & 0.02 &  \nodata & 0.019  \\
{[71]} & 53.056873 &   -27.798389  & 1.71 (1.63-1.72)  & \nodata & \nodata & \nodata & \nodata & \nodata & \nodata &  \nodata & \nodata  \\
72 & 53.119957 &   -27.743137  & [3.76 (3.47-4.31)] & 0.71 & 0.12 & \nodata & \nodata & \nodata & \nodata &  \nodata & \nodata  \\
73 & 53.142872 &   -27.874084  & 2.19 (2.09-2.22)  & \nodata & \nodata & 0.04 & 0.08 & 0.01 & 0.02 & -0.01 & 0.015  \\
74 & 53.093666 &   -27.826445  & 0.732  & \nodata & \nodata & 0.19 & 0.1 & 0.04 & 0.03 & 0.017 & 0.018 \\
{[75]} & 53.074837 &   -27.787111 & \nodata & \nodata & \nodata & \nodata & \nodata & \nodata & \nodata &  \nodata & \nodata  \\
\enddata

\tablecomments{Columns:
(1) Source number from Table~4 of C18 (brackets refer to sources not in the robust subset), 
(2) and (3) ALMA 870~$\mu$m R.A. and decl., 
(4) adopted redshift taken from the compilation in \citet{cbb23} (three digits after the decimal point for
spectroscopic redshifts---except for the JWST NIRSpec redshift for source 68 from \citet{obn+23}---and 
two digits after the decimal point for photometric 
redshifts, plus 68\% confidence ranges from \citealt{ssq+16} for photometric redshifts, given in parentheses), 
(5) and (6) total ALMA 1.13~mm flux and error from \citet{geb+22}, 
(7) and (8) measured peak ALMA 1.24~mm flux and error from this work,
(9) and (10) measured peak ALMA 2~mm flux and error from this work,
(11) and (12) measured peak ALMA 3~mm flux and error from this work.
In Column (4), values in brackets refer to photzs which had quality flag $Q>3$ in the catalog of \citet{ssq+16}.}

\end{deluxetable*}

%
%
\startlongtable
\begin{deluxetable}{ccccc}
\tablewidth{0pt}
\tablecaption{\label{tab:dustprops}
Robust Subset Dust Properties}
\tablehead{
\colhead{No.} &\colhead{$z$} &\colhead{$\beta$} &\colhead{$T_d$/K} & \colhead{$\log (L_{\rm IR}/\lsun)$}} 

\colnumbers

\startdata
1 & 2.574 & 1.53$^{+0.1}_{-0.1}$ & 32.8$^{+1.9}_{-1.7}$ & 12.79$^{+0.04}_{-0.05}$\\ 
2 & 3.69 & 1.84$^{+0.12}_{-0.12}$ & 34.5$^{+2.1}_{-2.0}$ & 12.98$^{+0.05}_{-0.06}$\\ 
3 & 2.648 & 1.73$^{+0.14}_{-0.14}$ & 36.1$^{+2.6}_{-2.3}$ & 12.84$^{+0.05}_{-0.05}$\\ 
4 & 2.252 & 2.01$^{+0.17}_{-0.16}$ & 31.4$^{+2.4}_{-2.1}$ & 12.84$^{+0.05}_{-0.05}$\\ 
5 & 2.309 & 1.75$^{+0.16}_{-0.15}$ & 32.7$^{+2.9}_{-2.5}$ & 12.66$^{+0.05}_{-0.06}$\\ 
7 & 3.672 & 1.45$^{+0.14}_{-0.14}$ & 43.4$^{+3.7}_{-3.4}$ & 12.88$^{+0.08}_{-0.07}$\\ 
8 & 2.69 & 2.21$^{+0.2}_{-0.18}$ & 26.6$^{+2.7}_{-2.4}$ & 12.53$^{+0.09}_{-0.09}$\\ 
9 & 2.322 & 1.71$^{+0.16}_{-0.16}$ & 32.8$^{+2.7}_{-2.4}$ & 12.63$^{+0.07}_{-0.07}$\\ 
10 & 2.41 & 1.63$^{+0.21}_{-0.2}$ & 37.1$^{+4.5}_{-3.7}$ & 12.64$^{+0.08}_{-0.08}$\\ 
12 & 3.76 & 1.86$^{+0.25}_{-0.23}$ & 38.4$^{+4.0}_{-3.6}$ & 12.86$^{+0.08}_{-0.07}$\\ 
13 & 2.73 & 0.99$^{+0.16}_{-0.12}$ & 49.7$^{+5.8}_{-5.7}$ & 12.61$^{+0.12}_{-0.12}$\\ 
14 & 2.73 & 1.96$^{+0.21}_{-0.2}$ & 45.4$^{+6.1}_{-4.9}$ & 13.14$^{+0.1}_{-0.1}$\\ 
15 & 2.14 & 2.84$^{+0.44}_{-0.38}$ & 20.6$^{+3.2}_{-2.9}$ & 12.19$^{+0.1}_{-0.13}$\\ 
16 & 3.37 & 1.71$^{+0.28}_{-0.25}$ & 36.2$^{+5.7}_{-4.9}$ & 12.62$^{+0.14}_{-0.13}$\\ 
17 & 3.57 & 2.12$^{+0.27}_{-0.25}$ & 27.9$^{+3.6}_{-3.2}$ & 12.36$^{+0.11}_{-0.1}$\\ 
18 & 3.847 & 1.57$^{+0.22}_{-0.21}$ & 49.4$^{+7.5}_{-6.0}$ & 12.99$^{+0.11}_{-0.12}$\\ 
19 & 4.47 & 1.44$^{+0.25}_{-0.22}$ & 48.0$^{+12.1}_{-9.0}$ & 12.82$^{+0.21}_{-0.22}$\\ 
20 & 1.93 & 1.55$^{+0.16}_{-0.16}$ & 42.8$^{+3.3}_{-2.8}$ & 12.65$^{+0.05}_{-0.04}$\\ 
21 & 3.78 & 1.91$^{+0.26}_{-0.24}$ & 37.0$^{+5.7}_{-5.1}$ & 12.67$^{+0.16}_{-0.16}$\\ 
22 & 2.698 & 1.37$^{+0.55}_{-0.4}$ & 45.7$^{+13.6}_{-10.8}$ & 12.76$^{+0.17}_{-0.2}$\\ 
23 & 1.58 & 1.78$^{+0.14}_{-0.14}$ & 33.8$^{+1.9}_{-1.7}$ & 12.44$^{+0.04}_{-0.04}$\\ 
24 & 1.96 & 1.44$^{+0.13}_{-0.12}$ & 42.0$^{+2.6}_{-2.3}$ & 12.63$^{+0.04}_{-0.04}$\\ 
25 & 2.696 & 1.75$^{+0.62}_{-0.55}$ & 33.1$^{+8.5}_{-6.2}$ & 12.44$^{+0.1}_{-0.13}$\\ 
26 & 3.78 & 1.03$^{+0.23}_{-0.15}$ & 69.2$^{+12.1}_{-12.9}$ & 12.98$^{+0.18}_{-0.23}$\\ 
27 & 1.78 & 1.90$^{+0.35}_{-0.31}$ & 22.3$^{+3.4}_{-3.1}$ & 11.81$^{+0.1}_{-0.14}$\\ 
28 & 3.33 & 1.78$^{+0.39}_{-0.36}$ & 36.4$^{+8.0}_{-6.1}$ & 12.44$^{+0.14}_{-0.18}$\\ 
30 & 1.86 & 2.00$^{+0.2}_{-0.19}$ & 30.2$^{+2.3}_{-2.1}$ & 12.31$^{+0.06}_{-0.06}$\\ 
31 & 1.95 & 2.19$^{+0.23}_{-0.23}$ & 31.7$^{+2.5}_{-2.2}$ & 12.40$^{+0.05}_{-0.05}$\\ 
32 & 2.75 & 1.76$^{+0.3}_{-0.29}$ & 45.7$^{+10.1}_{-7.1}$ & 12.82$^{+0.18}_{-0.14}$\\ 
33 & 1.58 & 2.15$^{+0.17}_{-0.17}$ & 30.2$^{+1.8}_{-1.6}$ & 12.41$^{+0.04}_{-0.04}$\\ 
34 & 1.95 & 2.09$^{+0.23}_{-0.23}$ & 33.6$^{+2.7}_{-2.3}$ & 12.42$^{+0.05}_{-0.05}$\\ 
35 & 1.612 & 1.74$^{+0.18}_{-0.17}$ & 32.3$^{+2.5}_{-2.2}$ & 12.25$^{+0.05}_{-0.05}$\\ 
36 & 2.37 & 3.32$^{+0.43}_{-0.49}$ & 20.1$^{+3.3}_{-2.3}$ & 12.09$^{+0.09}_{-0.1}$\\ 
37 & 2.96 & 2.79$^{+0.58}_{-0.52}$ & 20.9$^{+4.9}_{-4.1}$ & 11.94$^{+0.23}_{-0.24}$\\ 
38 & 2.31 & 1.78$^{+0.23}_{-0.22}$ & 33.8$^{+4.3}_{-3.6}$ & 12.39$^{+0.1}_{-0.1}$\\ 
39 & 3.04 & 2.92$^{+0.48}_{-0.44}$ & 31.5$^{+6.0}_{-4.7}$ & 12.71$^{+0.14}_{-0.16}$\\ 
40 & 2.223 & 1.93$^{+0.33}_{-0.31}$ & 30.9$^{+5.2}_{-4.0}$ & 12.28$^{+0.12}_{-0.11}$\\ 
41 & 4.13 & 1.92$^{+0.55}_{-0.43}$ & 29.2$^{+11.2}_{-8.4}$ & 12.15$^{+0.35}_{-0.25}$\\ 
42 & 2.34 & 1.88$^{+0.37}_{-0.33}$ & 35.3$^{+6.8}_{-5.2}$ & 12.46$^{+0.12}_{-0.1}$\\ 
43 & 2.39 & 1.54$^{+0.27}_{-0.26}$ & 37.0$^{+6.1}_{-4.7}$ & 12.45$^{+0.13}_{-0.1}$\\ 
46 & 1.613 & 1.02$^{+0.19}_{-0.14}$ & 47.0$^{+3.7}_{-4.1}$ & 12.44$^{+0.05}_{-0.05}$\\ 
47 & 2.19 & 2.17$^{+0.35}_{-0.33}$ & 28.3$^{+4.8}_{-3.9}$ & 12.20$^{+0.12}_{-0.12}$\\ 
48 & 2.543 & 2.54$^{+0.65}_{-0.65}$ & 32.2$^{+10.4}_{-6.1}$ & 12.67$^{+0.17}_{-0.11}$\\ 
49 & 1.87 & 1.51$^{+0.29}_{-0.27}$ & 34.6$^{+4.6}_{-4.0}$ & 12.11$^{+0.08}_{-0.1}$\\ 
50 & 1.69 & 2.22$^{+0.33}_{-0.3}$ & 29.4$^{+3.6}_{-2.8}$ & 12.21$^{+0.07}_{-0.1}$\\ 
51 & 2.32 & 1.99$^{+0.47}_{-0.42}$ & 28.2$^{+6.6}_{-4.9}$ & 12.07$^{+0.15}_{-0.16}$\\ 
52 & 4.78 & 1.69$^{+0.59}_{-0.41}$ & 48.1$^{+20.0}_{-15.1}$ & 12.62$^{+0.31}_{-0.36}$\\ 
53 & 1.56 & 1.15$^{+0.35}_{-0.24}$ & 38.7$^{+4.3}_{-4.3}$ & 12.12$^{+0.06}_{-0.07}$\\ 
56 & 2.299 & 1.62$^{+0.59}_{-0.46}$ & 32.8$^{+11.1}_{-7.2}$ & 12.04$^{+0.21}_{-0.23}$\\ 
59 & 2.325 & 3.09$^{+0.54}_{-0.55}$ & 16.1$^{+3.6}_{-2.7}$ & 11.59$^{+0.15}_{-0.18}$\\ 
63 & 1.83 & 1.79$^{+0.42}_{-0.41}$ & 33.5$^{+5.4}_{-3.9}$ & 12.16$^{+0.07}_{-0.08}$\\ 
65 & 1.58 & 1.66$^{+0.41}_{-0.4}$ & 30.7$^{+4.7}_{-3.6}$ & 11.98$^{+0.08}_{-0.08}$\\ 
68 & 5.58 & 1.59$^{+0.69}_{-0.47}$ & 69.6$^{+14.4}_{-25.0}$ & 13.07$^{+0.27}_{-0.53}$\\ 
70 & 3.14 & 2.49$^{+0.79}_{-0.71}$ & 37.9$^{+18.0}_{-9.6}$ & 12.55$^{+0.25}_{-0.26}$\\ 
72 & 3.76 & 1.58$^{+0.86}_{-0.44}$ & 64.2$^{+17.3}_{-20.2}$ & 12.97$^{+0.24}_{-0.27}$\\ 
73 & 2.19 & 3.49$^{+0.38}_{-0.73}$ & 22.3$^{+6.5}_{-2.9}$ & 12.03$^{+0.18}_{-0.34}$\\ 
74 & 0.7323 & 2.06$^{+0.15}_{-0.14}$ & 26.9$^{+0.9}_{-0.9}$ & 11.79$^{+0.02}_{-0.02}$\\ 
\enddata

\tablecomments{Columns:
(1) Source number from Table~4 of C18,
(2) adopted redshift taken from the compilation in \citet{cbb23} and used in MBB fits, 
(3) median likelihood emissivity index and error (16th to 84th percentile range from the likelihood distribution) from the optically thin MBB fits, 
(4) median likelihood dust temperature and error (16th to 84th percentile range from the likelihood distribution) from the optically thin MBB fits,
(3) FIR luminosity measured from the median likelihood model and error (16th to 84th percentile range from the distribution of accepted MCMC models) from the optically thin MBB fits.
Speczs in Column (4) are listed with 3 decimal places---except for the JWST NIRSpec redshift for source 68 from \citet{obn+23}---and photzs with 2 decimal places; values in brackets refer to photzs which had quality flag $Q>3$ in the catalog of \citet{ssq+16}.}

\end{deluxetable}

\section{SED Fits \label{appB}}

\begin{figure}
\figurenum{12}
\includegraphics[width=\linewidth]{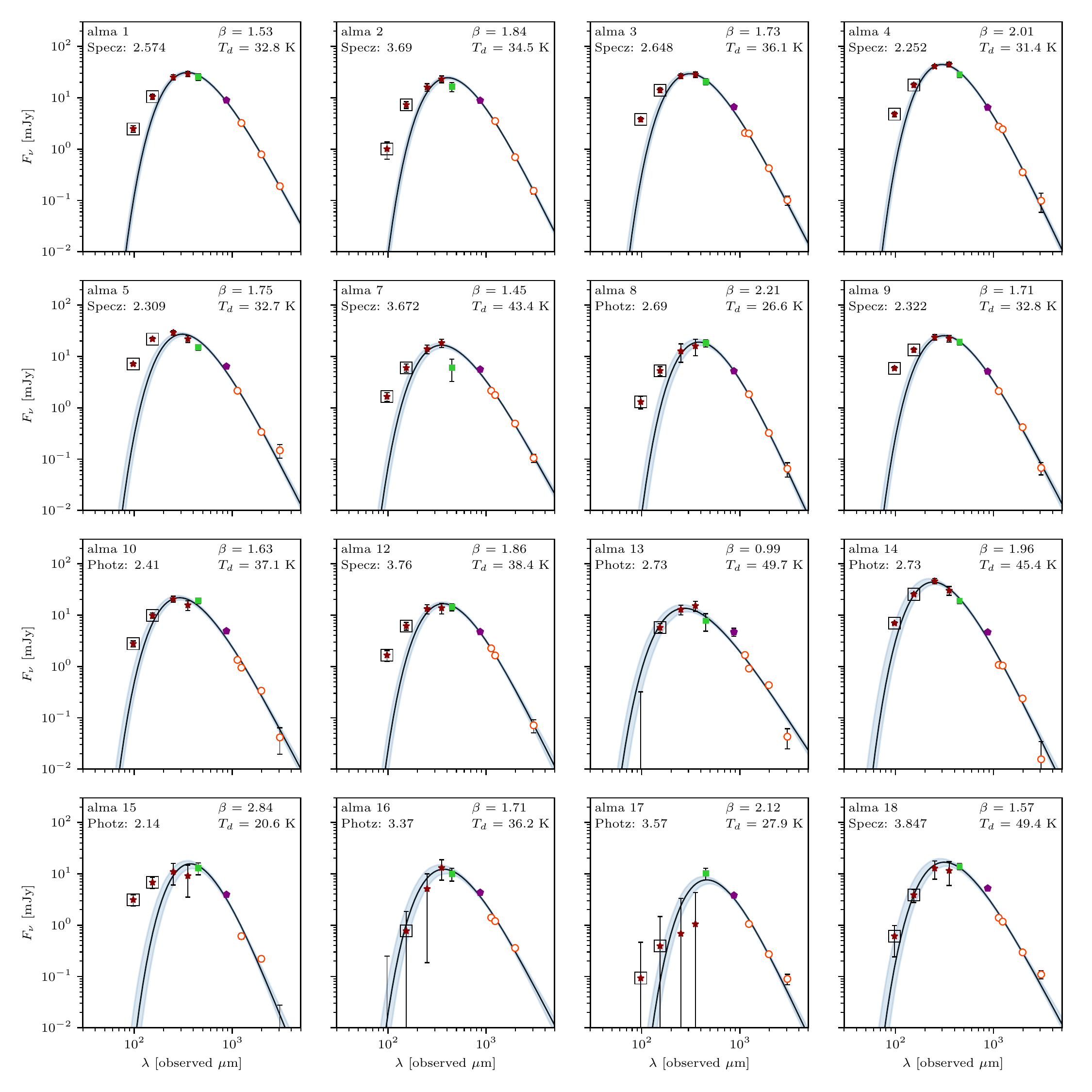}
\caption{Optically thin MBB SED fits (black curves) and 16th to 84th percentile ranges of the accepted MCMC models (blue shaded regions) for the 69 sources with redshifts in Table~1. Photometry: Red circles---ALMA 1.1~mm, 1.2~mm, 2~mm, and 3~mm, maroon pentagon---ALMA 870~$\mu$m, green square---SCUBA-2 450~$\mu$m, dark red stars---Herschel/PACS 100 and 160~$\mu$m and SPIRE 250 and 350~$\mu$m, blue triangles---Spitzer/MIPS 70~$\mu$m. The fits are made only to the data at wavelengths greater or equal to rest-frame 50~$\mu$m (points not included in the fits are marked with black squares). \\ The entire figure set (69 images) is available in the online journal. }
\label{sed_multiplot}
\end{figure}

In Figure~\ref{sed_multiplot}, we show the MBB SED fits for the 69 sources from the total sample that have redshifts.

%
%

\figsetstart
\figsetnum{12}
\figsettitle{MBB SED Fits}

\figsetgrpstart
\figsetgrpnum{12.1}
\figsetgrptitle{SED for ALMA 1}
\figsetplot{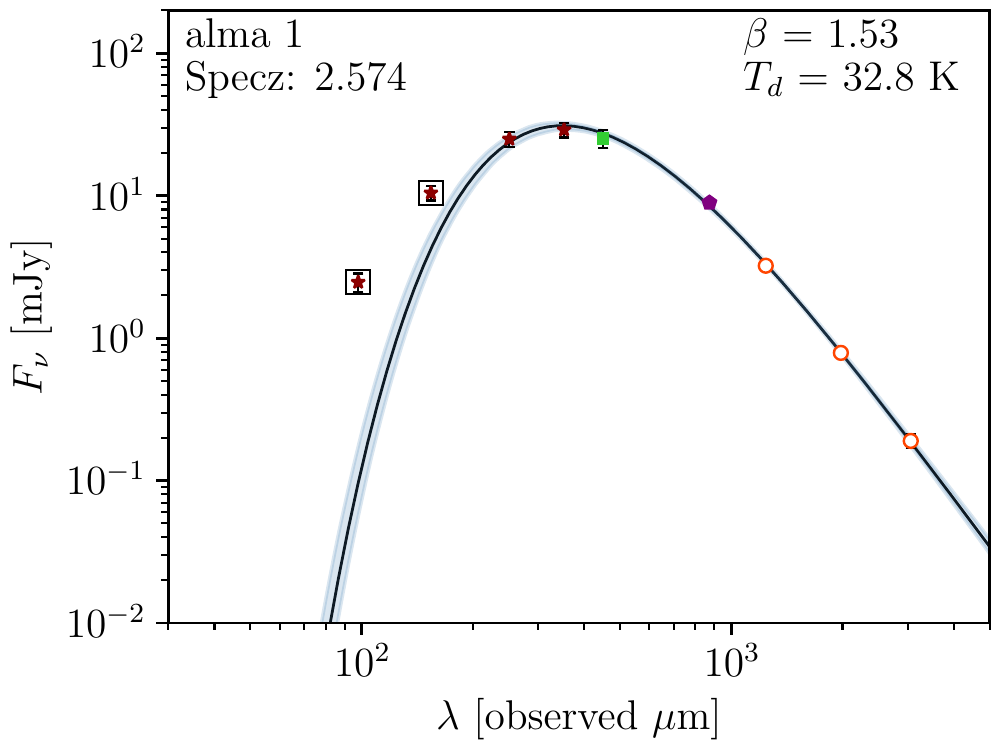}
\figsetgrpnote{Optically thin MBB SED fit (black curve) and 16th to 84th percentile range of the accepted MCMC models (blue shaded region). Photometry: Red circles---ALMA 1.1~mm, 1.2~mm, 2~mm, and 3~mm, maroon pentagon---ALMA 870~$\mu$m, green square---SCUBA-2 450~$\mu$m, dark red stars---Herschel/PACS 100 and 160~$\mu$m and SPIRE 250 and 350~$\mu$m, blue triangles---Spitzer/MIPS 70~$\mu$m. Points not included in the fits are marked with black squares.}
\figsetgrpend

\figsetgrpstart
\figsetgrpnum{12.2}
\figsetgrptitle{SED for ALMA 2}
\figsetplot{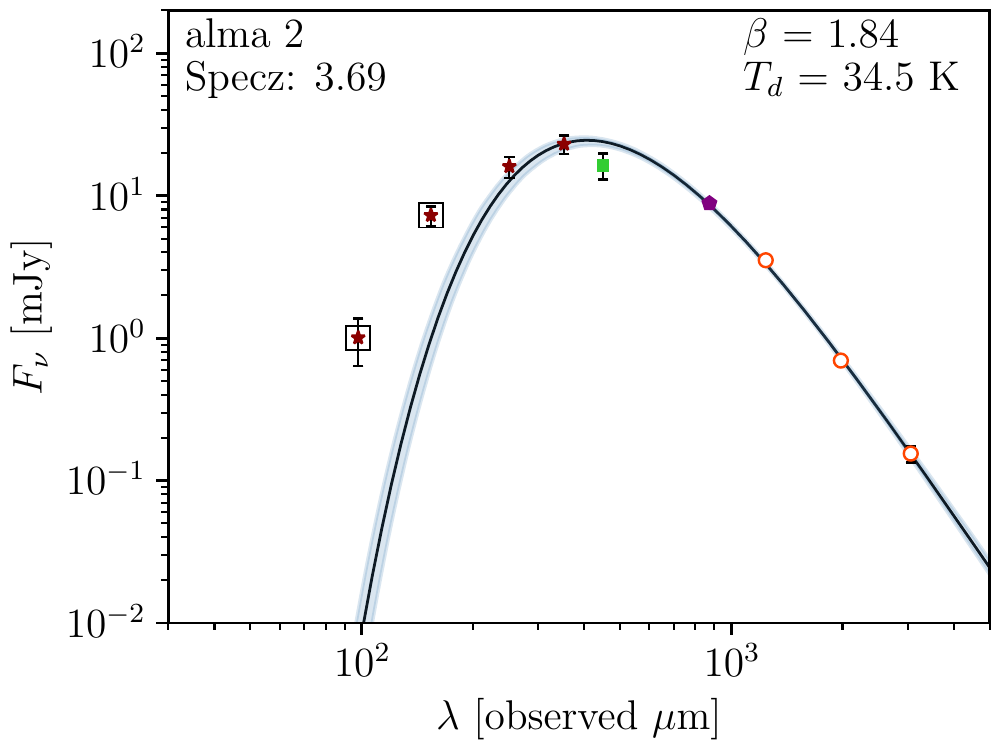}
\figsetgrpnote{Optically thin MBB SED fit (black curve) and 16th to 84th percentile range of the accepted MCMC models (blue shaded region). Photometry: Red circles---ALMA 1.1~mm, 1.2~mm, 2~mm, and 3~mm, maroon pentagon---ALMA 870~$\mu$m, green square---SCUBA-2 450~$\mu$m, dark red stars---Herschel/PACS 100 and 160~$\mu$m and SPIRE 250 and 350~$\mu$m, blue triangles---Spitzer/MIPS 70~$\mu$m. Points not included in the fits are marked with black squares.}
\figsetgrpend

\figsetgrpstart
\figsetgrpnum{12.3}
\figsetgrptitle{SED for ALMA 3}
\figsetplot{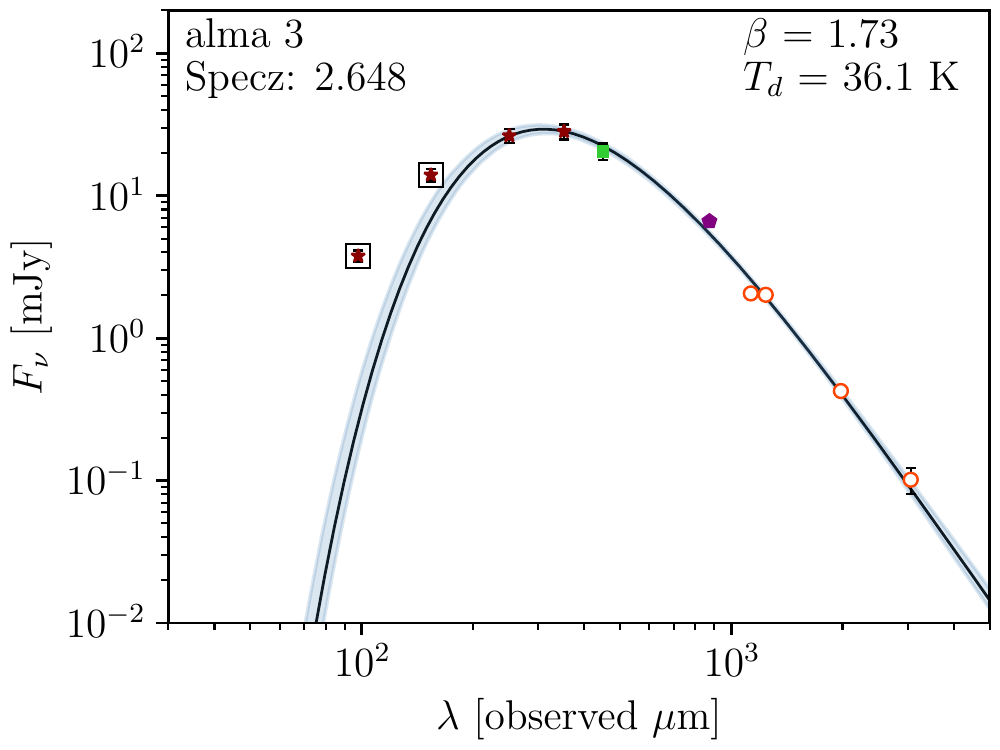}
\figsetgrpnote{Optically thin MBB SED fit (black curve) and 16th to 84th percentile range of the accepted MCMC models (blue shaded region). Photometry: Red circles---ALMA 1.1~mm, 1.2~mm, 2~mm, and 3~mm, maroon pentagon---ALMA 870~$\mu$m, green square---SCUBA-2 450~$\mu$m, dark red stars---Herschel/PACS 100 and 160~$\mu$m and SPIRE 250 and 350~$\mu$m, blue triangles---Spitzer/MIPS 70~$\mu$m. Points not included in the fits are marked with black squares.}
\figsetgrpend

\figsetgrpstart
\figsetgrpnum{12.4}
\figsetgrptitle{SED for ALMA 4}
\figsetplot{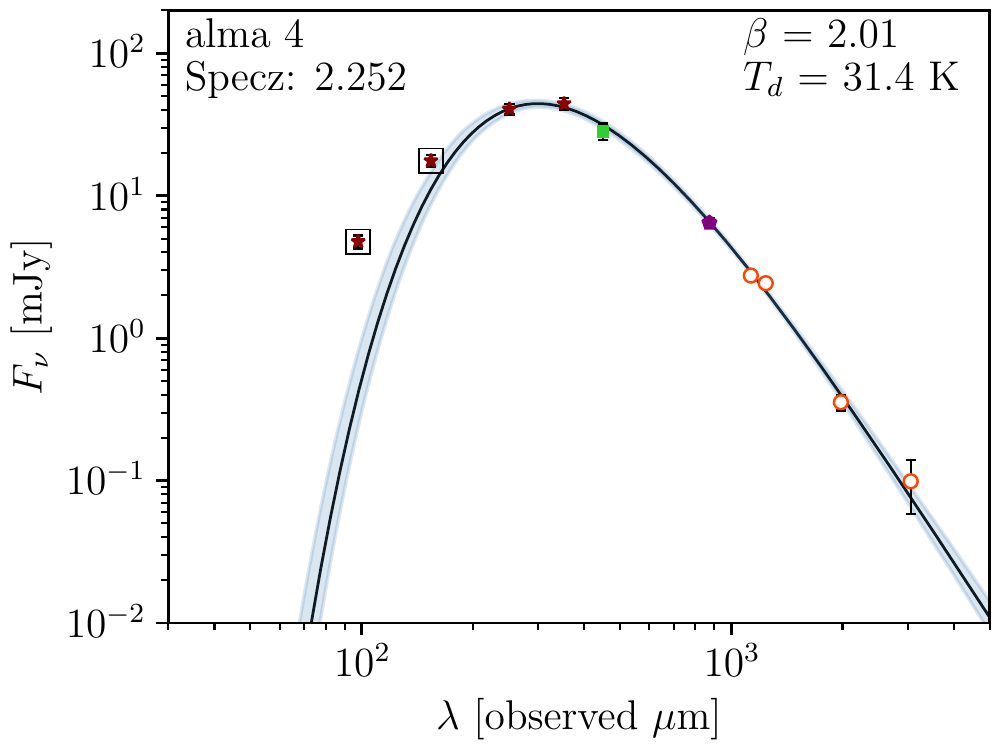}
\figsetgrpnote{Optically thin MBB SED fit (black curve) and 16th to 84th percentile range of the accepted MCMC models (blue shaded region). Photometry: Red circles---ALMA 1.1~mm, 1.2~mm, 2~mm, and 3~mm, maroon pentagon---ALMA 870~$\mu$m, green square---SCUBA-2 450~$\mu$m, dark red stars---Herschel/PACS 100 and 160~$\mu$m and SPIRE 250 and 350~$\mu$m, blue triangles---Spitzer/MIPS 70~$\mu$m. Points not included in the fits are marked with black squares.}
\figsetgrpend

\figsetgrpstart
\figsetgrpnum{12.5}
\figsetgrptitle{SED for ALMA 5}
\figsetplot{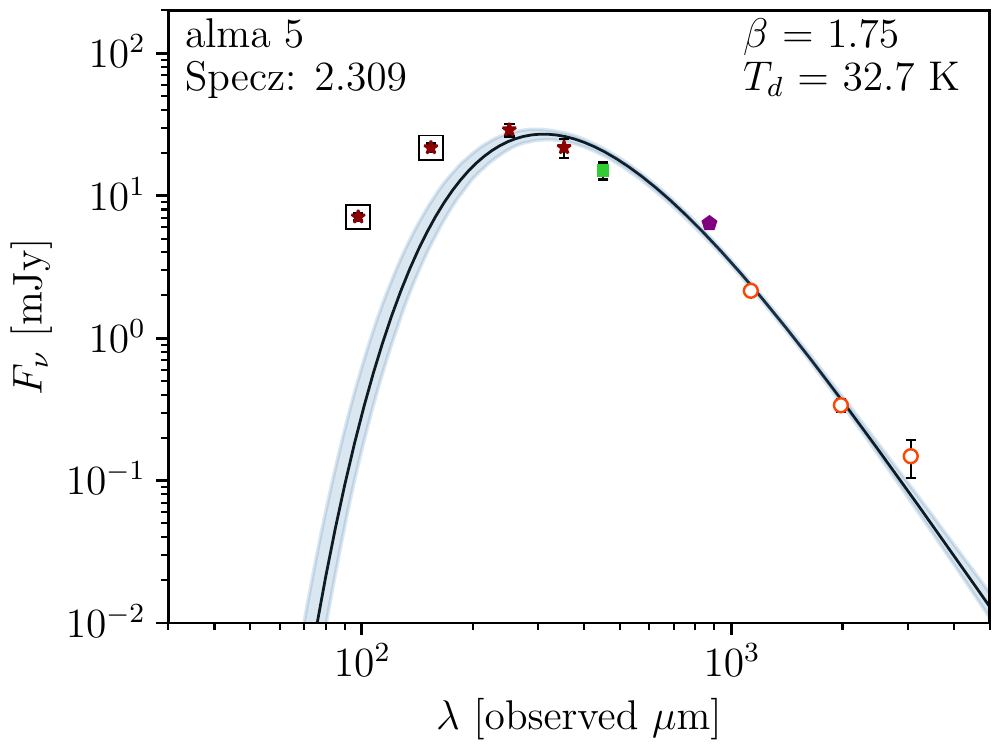}
\figsetgrpnote{Optically thin MBB SED fit (black curve) and 16th to 84th percentile range of the accepted MCMC models (blue shaded region). Photometry: Red circles---ALMA 1.1~mm, 1.2~mm, 2~mm, and 3~mm, maroon pentagon---ALMA 870~$\mu$m, green square---SCUBA-2 450~$\mu$m, dark red stars---Herschel/PACS 100 and 160~$\mu$m and SPIRE 250 and 350~$\mu$m, blue triangles---Spitzer/MIPS 70~$\mu$m. Points not included in the fits are marked with black squares.}
\figsetgrpend

\figsetgrpstart
\figsetgrpnum{12.6}
\figsetgrptitle{SED for ALMA 7}
\figsetplot{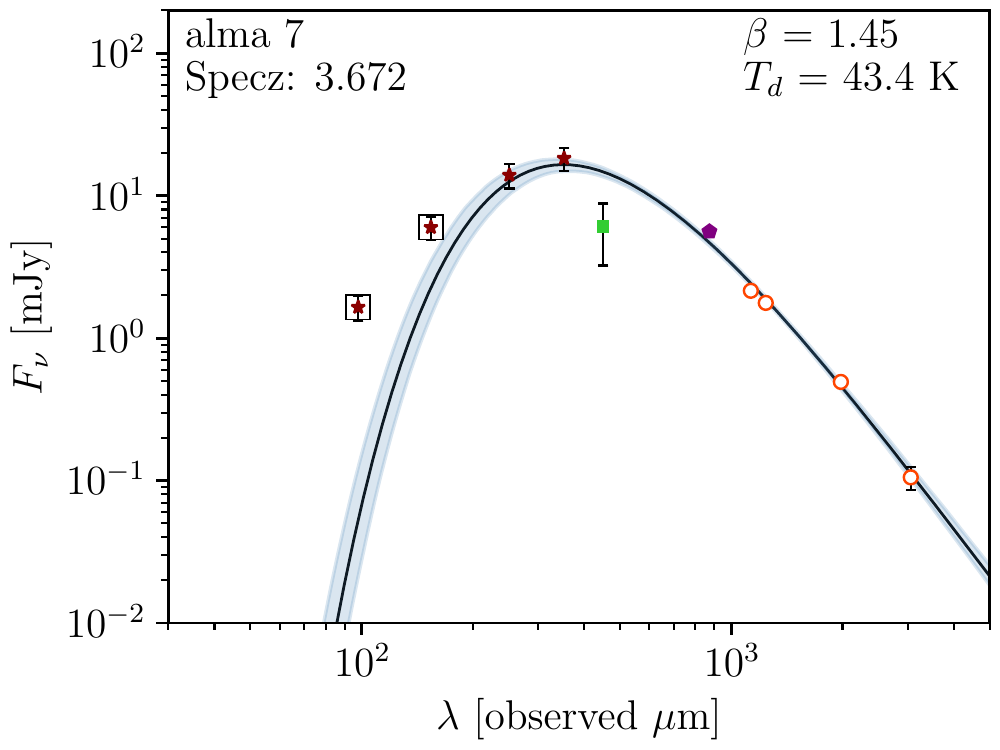}
\figsetgrpnote{Optically thin MBB SED fit (black curve) and 16th to 84th percentile range of the accepted MCMC models (blue shaded region). Photometry: Red circles---ALMA 1.1~mm, 1.2~mm, 2~mm, and 3~mm, maroon pentagon---ALMA 870~$\mu$m, green square---SCUBA-2 450~$\mu$m, dark red stars---Herschel/PACS 100 and 160~$\mu$m and SPIRE 250 and 350~$\mu$m, blue triangles---Spitzer/MIPS 70~$\mu$m. Points not included in the fits are marked with black squares.}
\figsetgrpend

\figsetgrpstart
\figsetgrpnum{12.7}
\figsetgrptitle{SED for ALMA 8}
\figsetplot{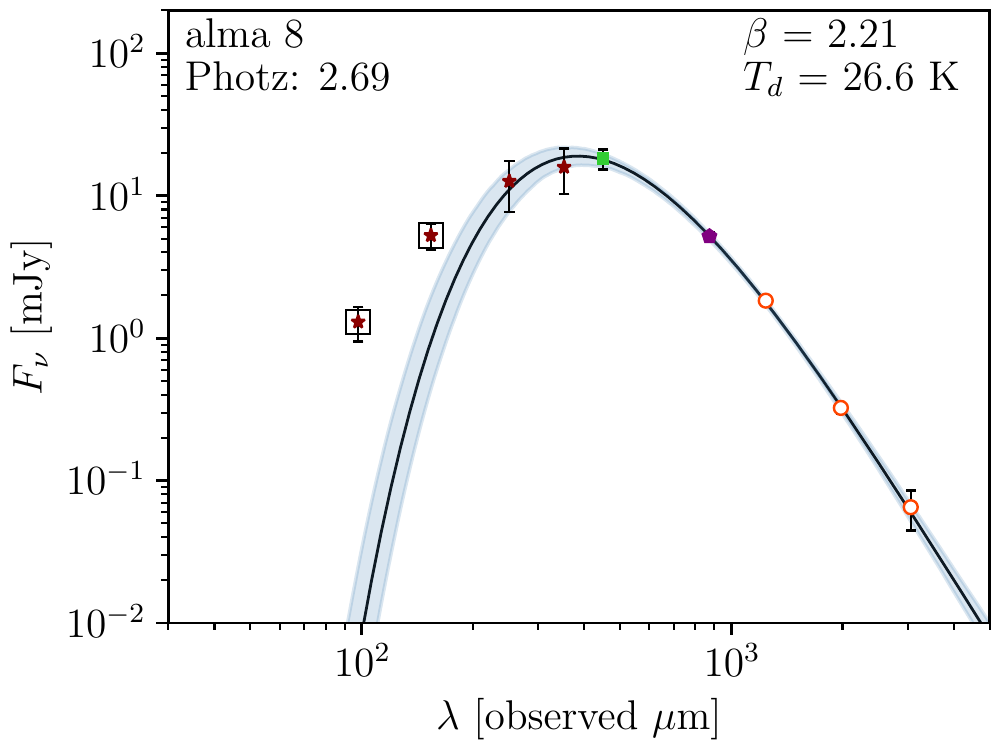}
\figsetgrpnote{Optically thin MBB SED fit (black curve) and 16th to 84th percentile range of the accepted MCMC models (blue shaded region). Photometry: Red circles---ALMA 1.1~mm, 1.2~mm, 2~mm, and 3~mm, maroon pentagon---ALMA 870~$\mu$m, green square---SCUBA-2 450~$\mu$m, dark red stars---Herschel/PACS 100 and 160~$\mu$m and SPIRE 250 and 350~$\mu$m, blue triangles---Spitzer/MIPS 70~$\mu$m. Points not included in the fits are marked with black squares.}
\figsetgrpend

\figsetgrpstart
\figsetgrpnum{12.8}
\figsetgrptitle{SED for ALMA 9}
\figsetplot{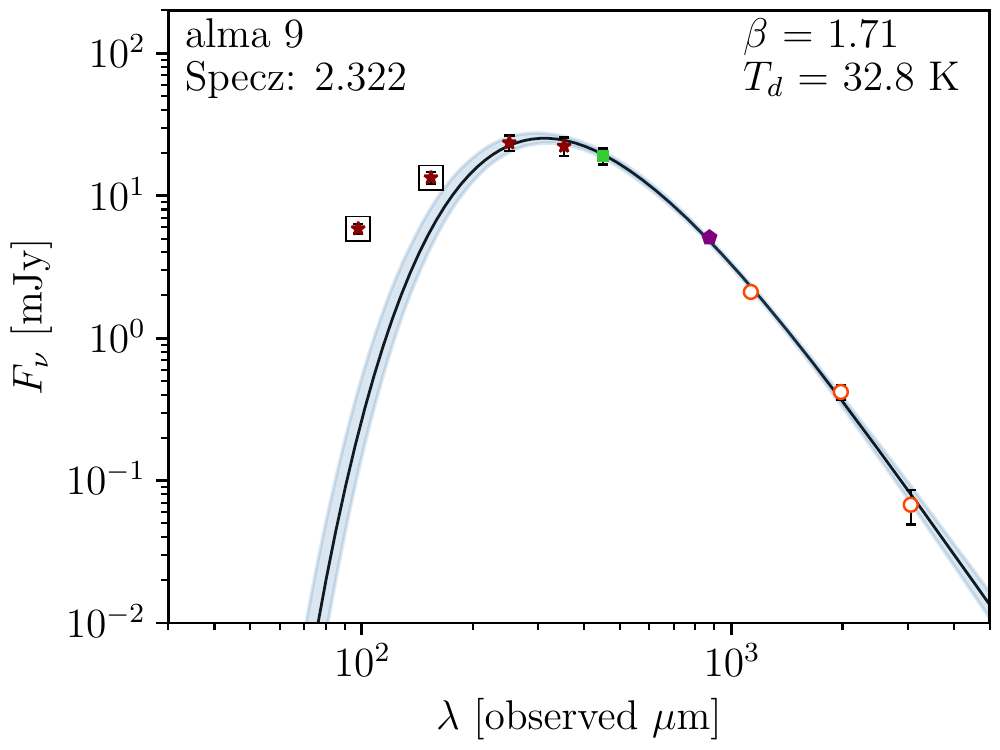}
\figsetgrpnote{Optically thin MBB SED fit (black curve) and 16th to 84th percentile range of the accepted MCMC models (blue shaded region). Photometry: Red circles---ALMA 1.1~mm, 1.2~mm, 2~mm, and 3~mm, maroon pentagon---ALMA 870~$\mu$m, green square---SCUBA-2 450~$\mu$m, dark red stars---Herschel/PACS 100 and 160~$\mu$m and SPIRE 250 and 350~$\mu$m, blue triangles---Spitzer/MIPS 70~$\mu$m. Points not included in the fits are marked with black squares.}
\figsetgrpend

\figsetgrpstart
\figsetgrpnum{12.9}
\figsetgrptitle{SED for ALMA 10}
\figsetplot{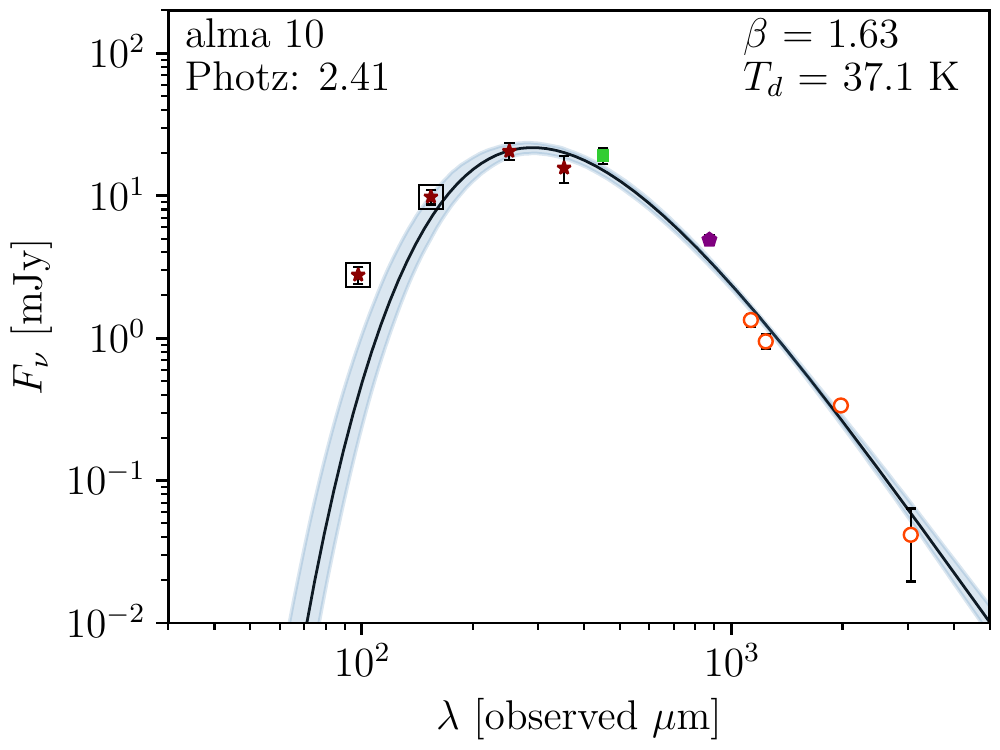}
\figsetgrpnote{Optically thin MBB SED fit (black curve) and 16th to 84th percentile range of the accepted MCMC models (blue shaded region). Photometry: Red circles---ALMA 1.1~mm, 1.2~mm, 2~mm, and 3~mm, maroon pentagon---ALMA 870~$\mu$m, green square---SCUBA-2 450~$\mu$m, dark red stars---Herschel/PACS 100 and 160~$\mu$m and SPIRE 250 and 350~$\mu$m, blue triangles---Spitzer/MIPS 70~$\mu$m. Points not included in the fits are marked with black squares.}
\figsetgrpend

\figsetgrpstart
\figsetgrpnum{12.10}
\figsetgrptitle{SED for ALMA 12}
\figsetplot{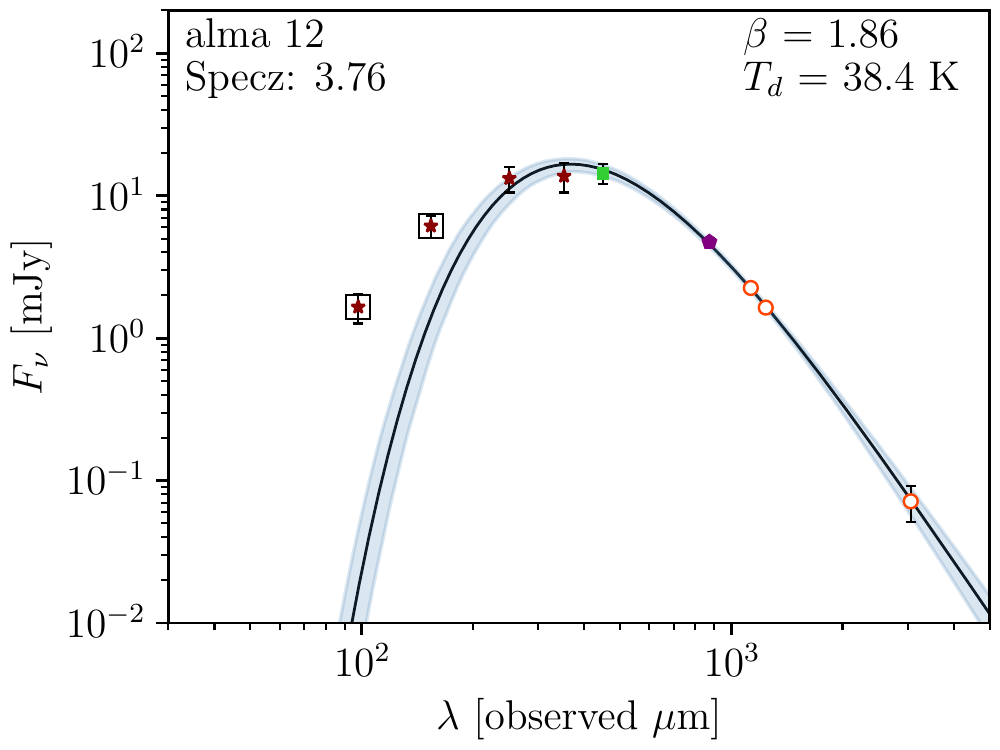}
\figsetgrpnote{Optically thin MBB SED fit (black curve) and 16th to 84th percentile range of the accepted MCMC models (blue shaded region). Photometry: Red circles---ALMA 1.1~mm, 1.2~mm, 2~mm, and 3~mm, maroon pentagon---ALMA 870~$\mu$m, green square---SCUBA-2 450~$\mu$m, dark red stars---Herschel/PACS 100 and 160~$\mu$m and SPIRE 250 and 350~$\mu$m, blue triangles---Spitzer/MIPS 70~$\mu$m. Points not included in the fits are marked with black squares.}
\figsetgrpend

\figsetgrpstart
\figsetgrpnum{12.11}
\figsetgrptitle{SED for ALMA 13}
\figsetplot{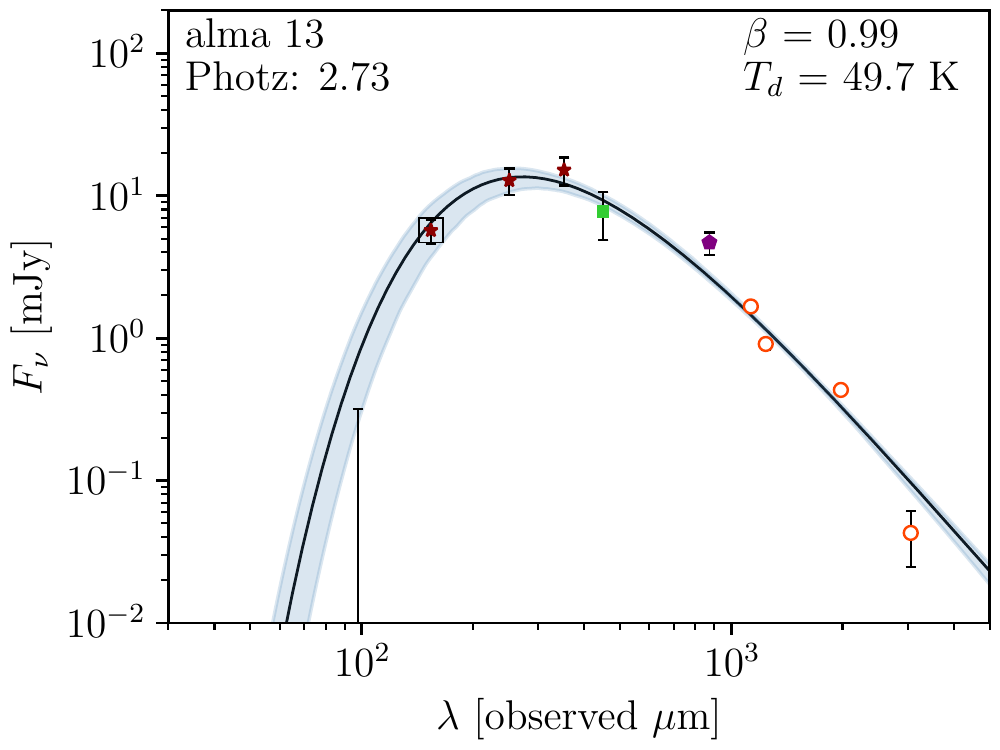}
\figsetgrpnote{Optically thin MBB SED fit (black curve) and 16th to 84th percentile range of the accepted MCMC models (blue shaded region). Photometry: Red circles---ALMA 1.1~mm, 1.2~mm, 2~mm, and 3~mm, maroon pentagon---ALMA 870~$\mu$m, green square---SCUBA-2 450~$\mu$m, dark red stars---Herschel/PACS 100 and 160~$\mu$m and SPIRE 250 and 350~$\mu$m, blue triangles---Spitzer/MIPS 70~$\mu$m. Points not included in the fits are marked with black squares.}
\figsetgrpend

\figsetgrpstart
\figsetgrpnum{12.12}
\figsetgrptitle{SED for ALMA 14}
\figsetplot{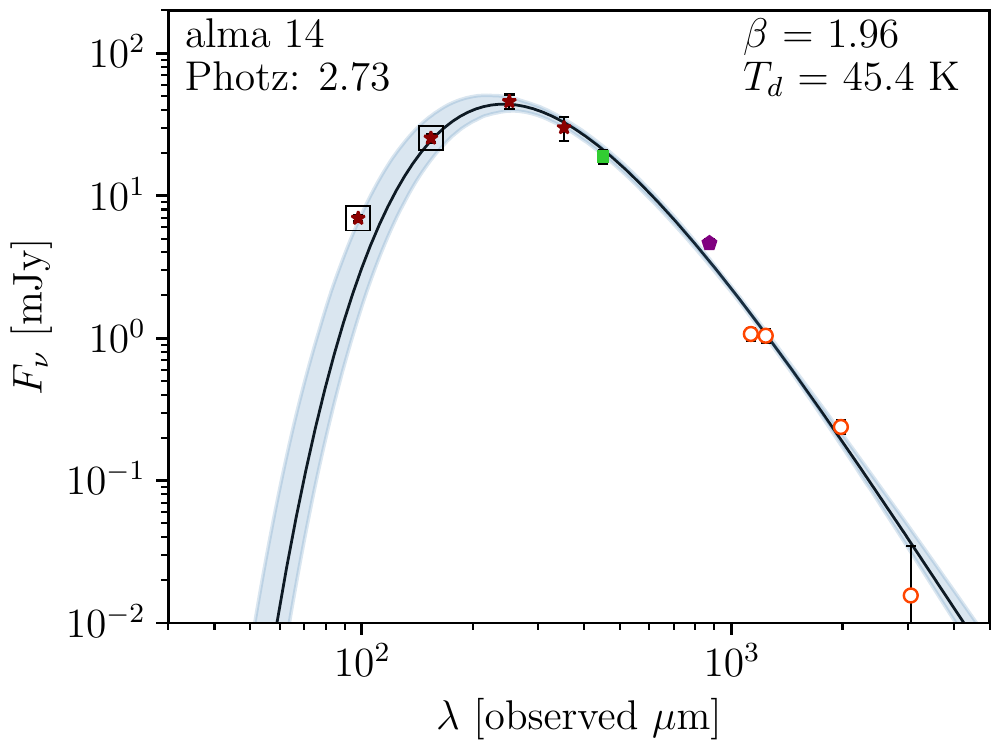}
\figsetgrpnote{Optically thin MBB SED fit (black curve) and 16th to 84th percentile range of the accepted MCMC models (blue shaded region). Photometry: Red circles---ALMA 1.1~mm, 1.2~mm, 2~mm, and 3~mm, maroon pentagon---ALMA 870~$\mu$m, green square---SCUBA-2 450~$\mu$m, dark red stars---Herschel/PACS 100 and 160~$\mu$m and SPIRE 250 and 350~$\mu$m, blue triangles---Spitzer/MIPS 70~$\mu$m. Points not included in the fits are marked with black squares.}
\figsetgrpend

\figsetgrpstart
\figsetgrpnum{12.13}
\figsetgrptitle{SED for ALMA 15}
\figsetplot{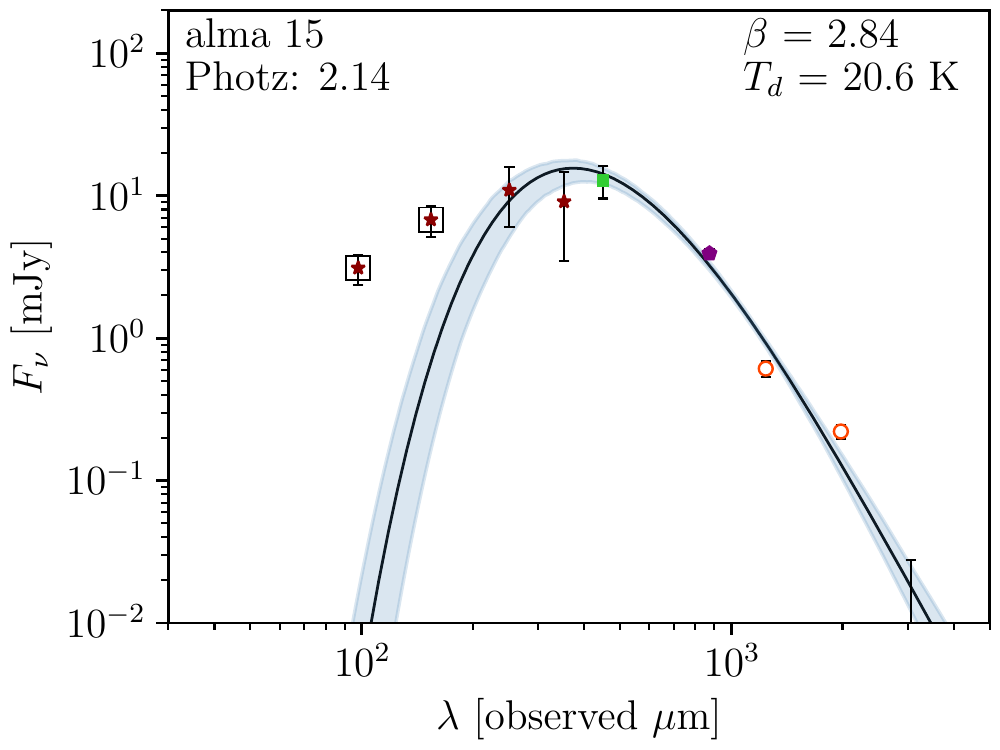}
\figsetgrpnote{Optically thin MBB SED fit (black curve) and 16th to 84th percentile range of the accepted MCMC models (blue shaded region). Photometry: Red circles---ALMA 1.1~mm, 1.2~mm, 2~mm, and 3~mm, maroon pentagon---ALMA 870~$\mu$m, green square---SCUBA-2 450~$\mu$m, dark red stars---Herschel/PACS 100 and 160~$\mu$m and SPIRE 250 and 350~$\mu$m, blue triangles---Spitzer/MIPS 70~$\mu$m. Points not included in the fits are marked with black squares.}
\figsetgrpend

\figsetgrpstart
\figsetgrpnum{12.14}
\figsetgrptitle{SED for ALMA 16}
\figsetplot{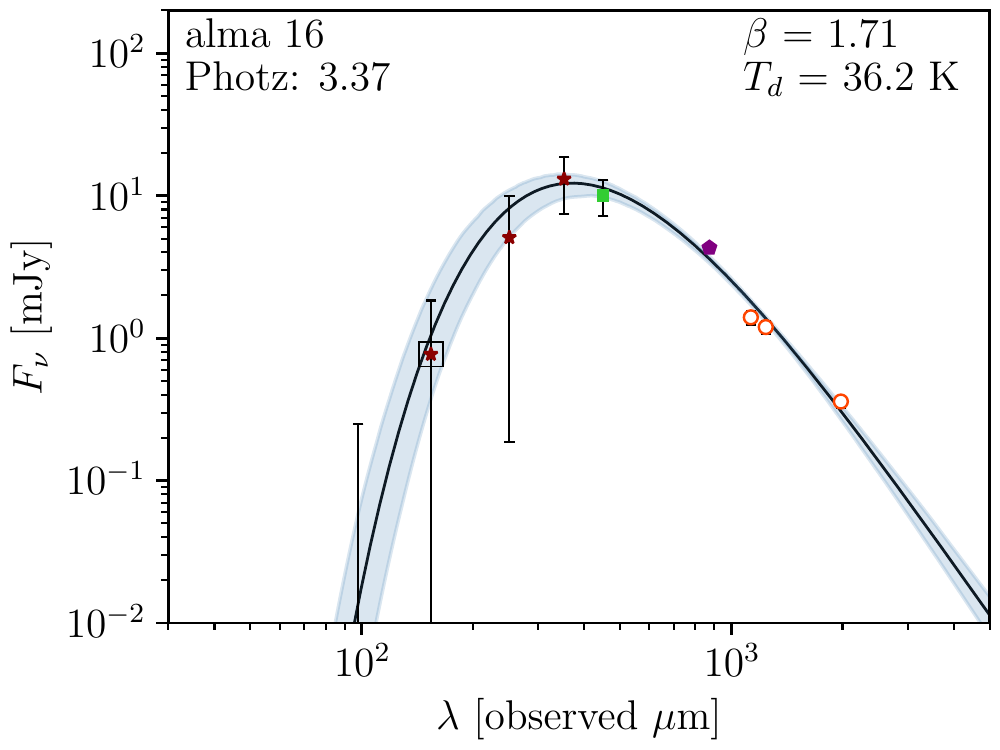}
\figsetgrpnote{Optically thin MBB SED fit (black curve) and 16th to 84th percentile range of the accepted MCMC models (blue shaded region). Photometry: Red circles---ALMA 1.1~mm, 1.2~mm, 2~mm, and 3~mm, maroon pentagon---ALMA 870~$\mu$m, green square---SCUBA-2 450~$\mu$m, dark red stars---Herschel/PACS 100 and 160~$\mu$m and SPIRE 250 and 350~$\mu$m, blue triangles---Spitzer/MIPS 70~$\mu$m. Points not included in the fits are marked with black squares.}
\figsetgrpend

\figsetgrpstart
\figsetgrpnum{12.15}
\figsetgrptitle{SED for ALMA 17}
\figsetplot{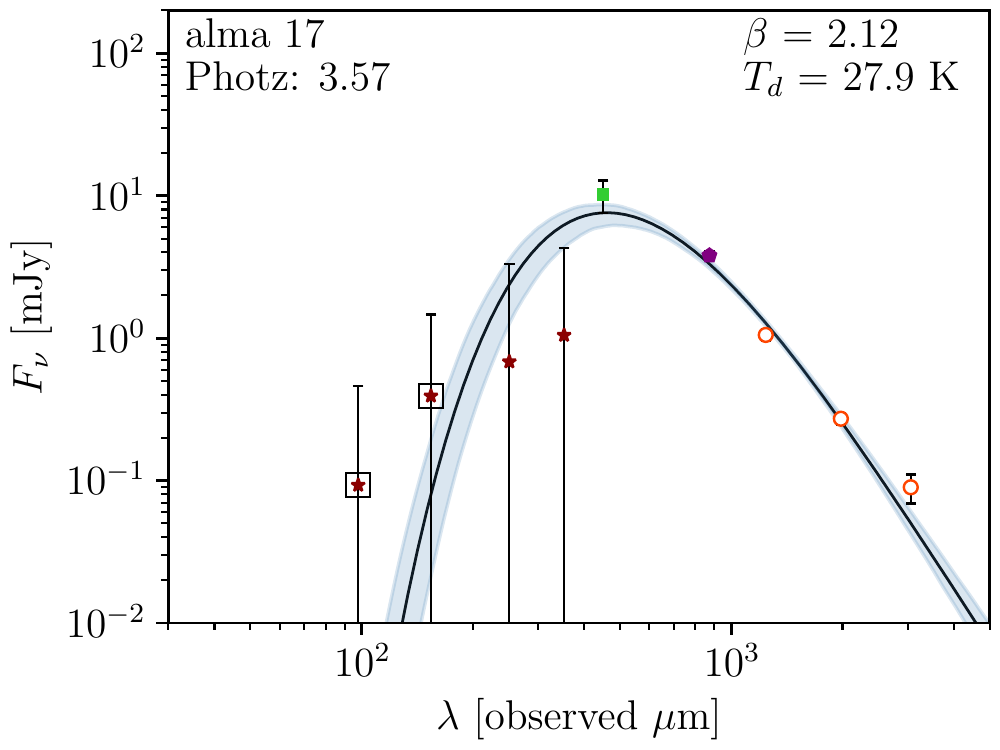}
\figsetgrpnote{Optically thin MBB SED fit (black curve) and 16th to 84th percentile range of the accepted MCMC models (blue shaded region). Photometry: Red circles---ALMA 1.1~mm, 1.2~mm, 2~mm, and 3~mm, maroon pentagon---ALMA 870~$\mu$m, green square---SCUBA-2 450~$\mu$m, dark red stars---Herschel/PACS 100 and 160~$\mu$m and SPIRE 250 and 350~$\mu$m, blue triangles---Spitzer/MIPS 70~$\mu$m. Points not included in the fits are marked with black squares.}
\figsetgrpend

\figsetgrpstart
\figsetgrpnum{12.16}
\figsetgrptitle{SED for ALMA 18}
\figsetplot{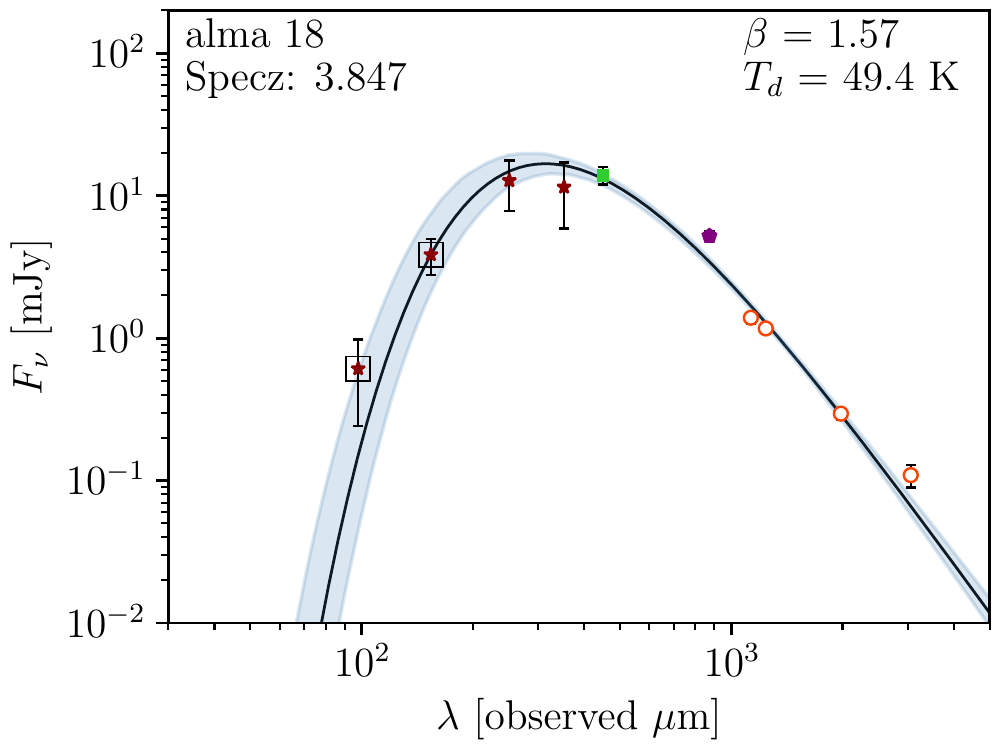}
\figsetgrpnote{Optically thin MBB SED fit (black curve) and 16th to 84th percentile range of the accepted MCMC models (blue shaded region). Photometry: Red circles---ALMA 1.1~mm, 1.2~mm, 2~mm, and 3~mm, maroon pentagon---ALMA 870~$\mu$m, green square---SCUBA-2 450~$\mu$m, dark red stars---Herschel/PACS 100 and 160~$\mu$m and SPIRE 250 and 350~$\mu$m, blue triangles---Spitzer/MIPS 70~$\mu$m. Points not included in the fits are marked with black squares.}
\figsetgrpend

\figsetgrpstart
\figsetgrpnum{12.17}
\figsetgrptitle{SED for ALMA 19}
\figsetplot{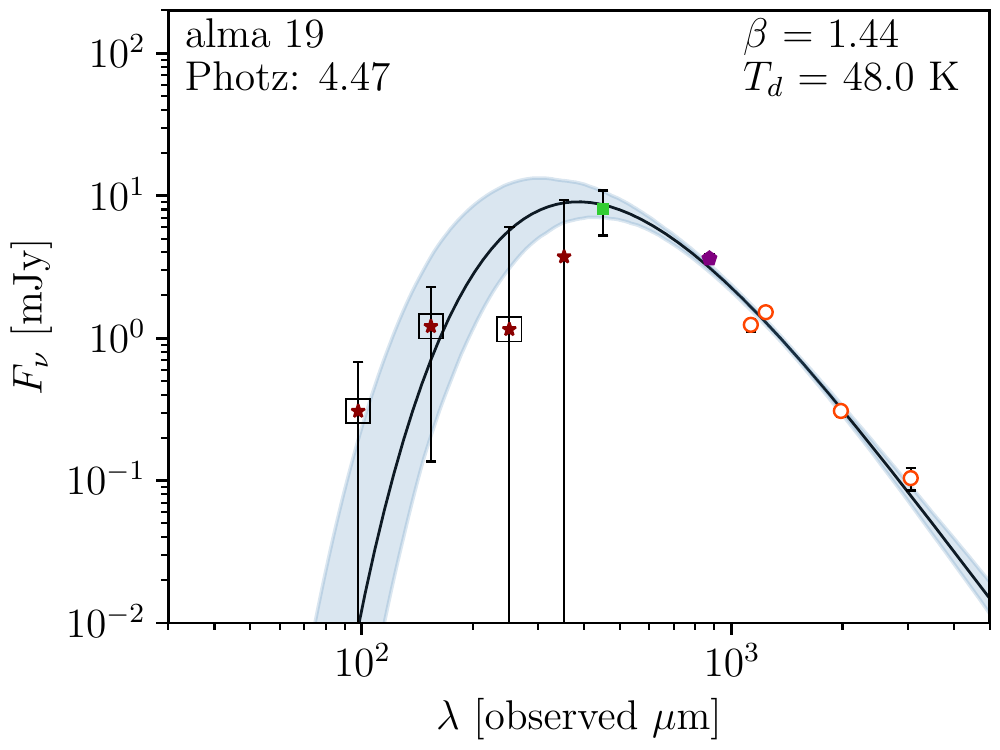}
\figsetgrpnote{Optically thin MBB SED fit (black curve) and 16th to 84th percentile range of the accepted MCMC models (blue shaded region). Photometry: Red circles---ALMA 1.1~mm, 1.2~mm, 2~mm, and 3~mm, maroon pentagon---ALMA 870~$\mu$m, green square---SCUBA-2 450~$\mu$m, dark red stars---Herschel/PACS 100 and 160~$\mu$m and SPIRE 250 and 350~$\mu$m, blue triangles---Spitzer/MIPS 70~$\mu$m. Points not included in the fits are marked with black squares.}
\figsetgrpend

\figsetgrpstart
\figsetgrpnum{12.18}
\figsetgrptitle{SED for ALMA 20}
\figsetplot{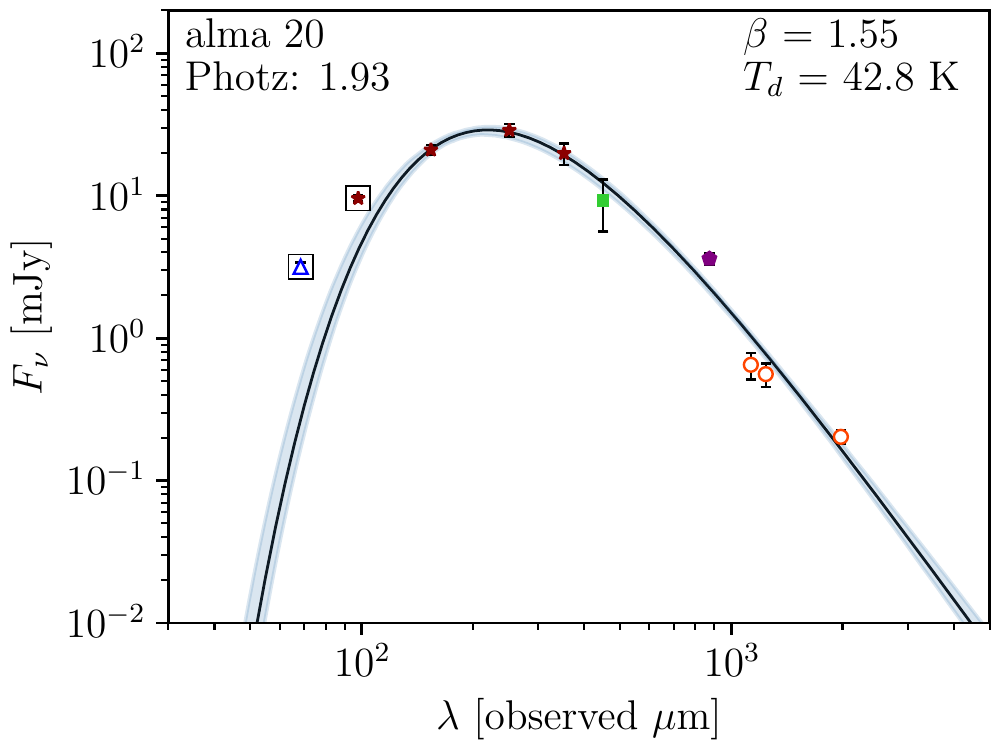}
\figsetgrpnote{Optically thin MBB SED fit (black curve) and 16th to 84th percentile range of the accepted MCMC models (blue shaded region). Photometry: Red circles---ALMA 1.1~mm, 1.2~mm, 2~mm, and 3~mm, maroon pentagon---ALMA 870~$\mu$m, green square---SCUBA-2 450~$\mu$m, dark red stars---Herschel/PACS 100 and 160~$\mu$m and SPIRE 250 and 350~$\mu$m, blue triangles---Spitzer/MIPS 70~$\mu$m. Points not included in the fits are marked with black squares.}
\figsetgrpend

\figsetgrpstart
\figsetgrpnum{12.19}
\figsetgrptitle{SED for ALMA 21}
\figsetplot{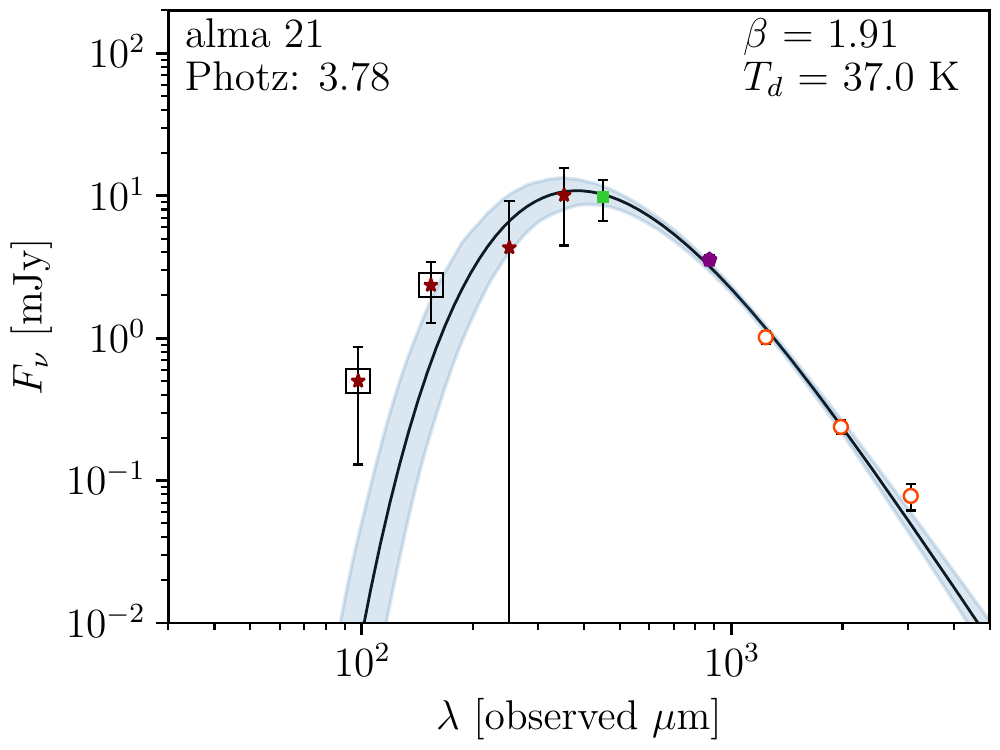}
\figsetgrpnote{Optically thin MBB SED fit (black curve) and 16th to 84th percentile range of the accepted MCMC models (blue shaded region). Photometry: Red circles---ALMA 1.1~mm, 1.2~mm, 2~mm, and 3~mm, maroon pentagon---ALMA 870~$\mu$m, green square---SCUBA-2 450~$\mu$m, dark red stars---Herschel/PACS 100 and 160~$\mu$m and SPIRE 250 and 350~$\mu$m, blue triangles---Spitzer/MIPS 70~$\mu$m. Points not included in the fits are marked with black squares.}
\figsetgrpend

\figsetgrpstart
\figsetgrpnum{12.20}
\figsetgrptitle{SED for ALMA 22}
\figsetplot{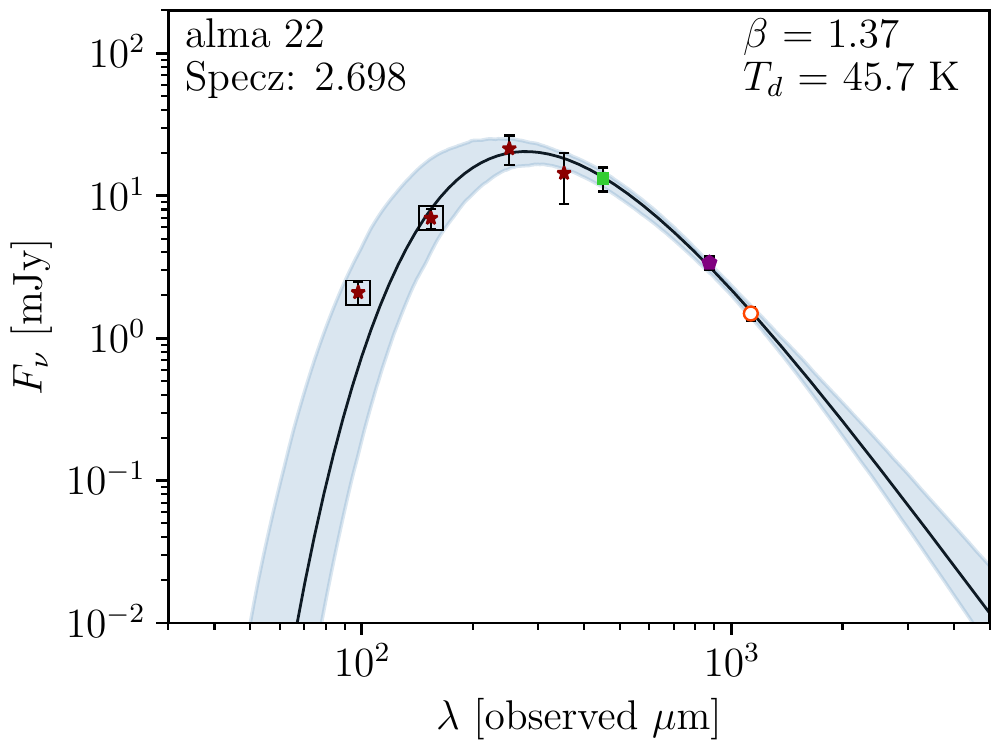}
\figsetgrpnote{Optically thin MBB SED fit (black curve) and 16th to 84th percentile range of the accepted MCMC models (blue shaded region). Photometry: Red circles---ALMA 1.1~mm, 1.2~mm, 2~mm, and 3~mm, maroon pentagon---ALMA 870~$\mu$m, green square---SCUBA-2 450~$\mu$m, dark red stars---Herschel/PACS 100 and 160~$\mu$m and SPIRE 250 and 350~$\mu$m, blue triangles---Spitzer/MIPS 70~$\mu$m. Points not included in the fits are marked with black squares.}
\figsetgrpend

\figsetgrpstart
\figsetgrpnum{12.21}
\figsetgrptitle{SED for ALMA 23}
\figsetplot{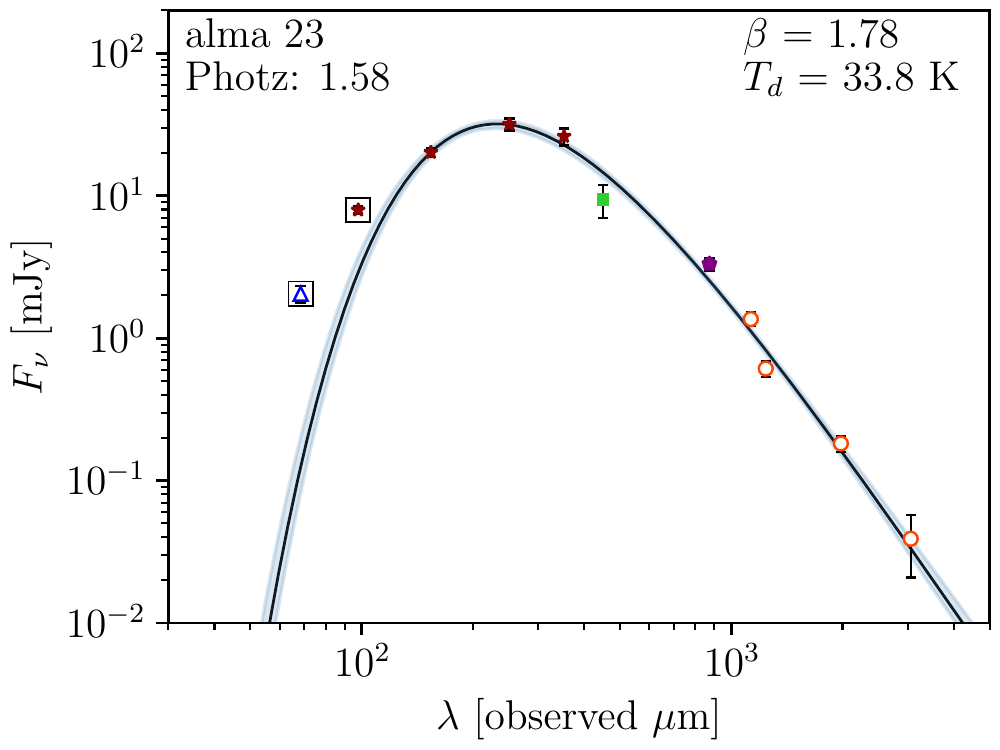}
\figsetgrpnote{Optically thin MBB SED fit (black curve) and 16th to 84th percentile range of the accepted MCMC models (blue shaded region). Photometry: Red circles---ALMA 1.1~mm, 1.2~mm, 2~mm, and 3~mm, maroon pentagon---ALMA 870~$\mu$m, green square---SCUBA-2 450~$\mu$m, dark red stars---Herschel/PACS 100 and 160~$\mu$m and SPIRE 250 and 350~$\mu$m, blue triangles---Spitzer/MIPS 70~$\mu$m. Points not included in the fits are marked with black squares.}
\figsetgrpend

\figsetgrpstart
\figsetgrpnum{12.22}
\figsetgrptitle{SED for ALMA 24}
\figsetplot{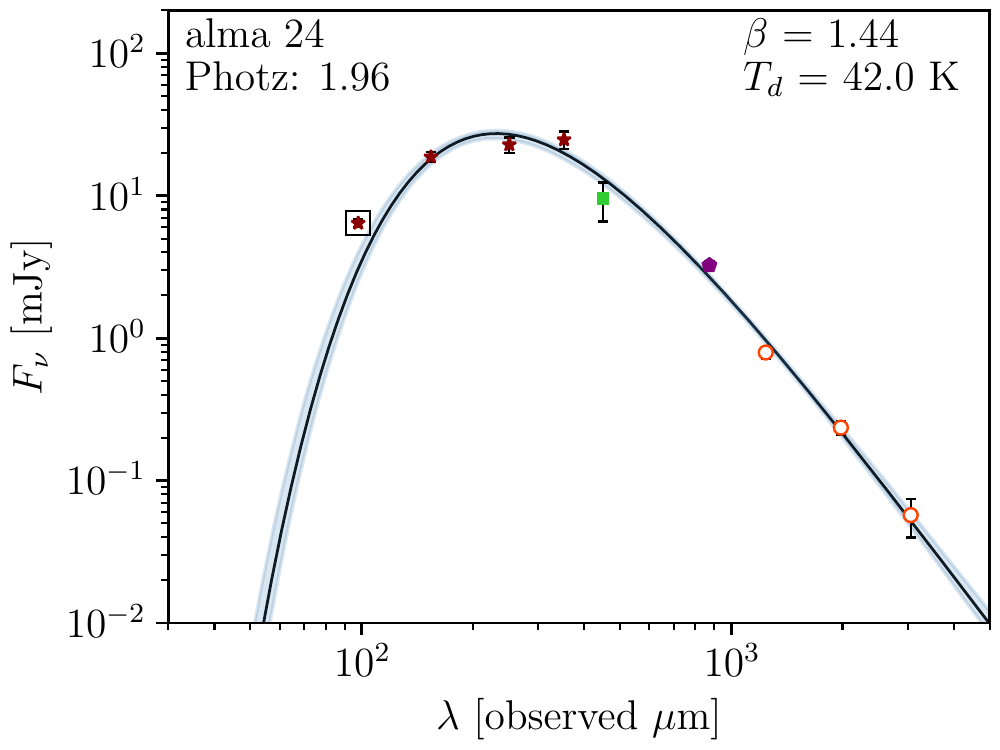}
\figsetgrpnote{Optically thin MBB SED fit (black curve) and 16th to 84th percentile range of the accepted MCMC models (blue shaded region). Photometry: Red circles---ALMA 1.1~mm, 1.2~mm, 2~mm, and 3~mm, maroon pentagon---ALMA 870~$\mu$m, green square---SCUBA-2 450~$\mu$m, dark red stars---Herschel/PACS 100 and 160~$\mu$m and SPIRE 250 and 350~$\mu$m, blue triangles---Spitzer/MIPS 70~$\mu$m. Points not included in the fits are marked with black squares.}
\figsetgrpend

\figsetgrpstart
\figsetgrpnum{12.23}
\figsetgrptitle{SED for ALMA 25}
\figsetplot{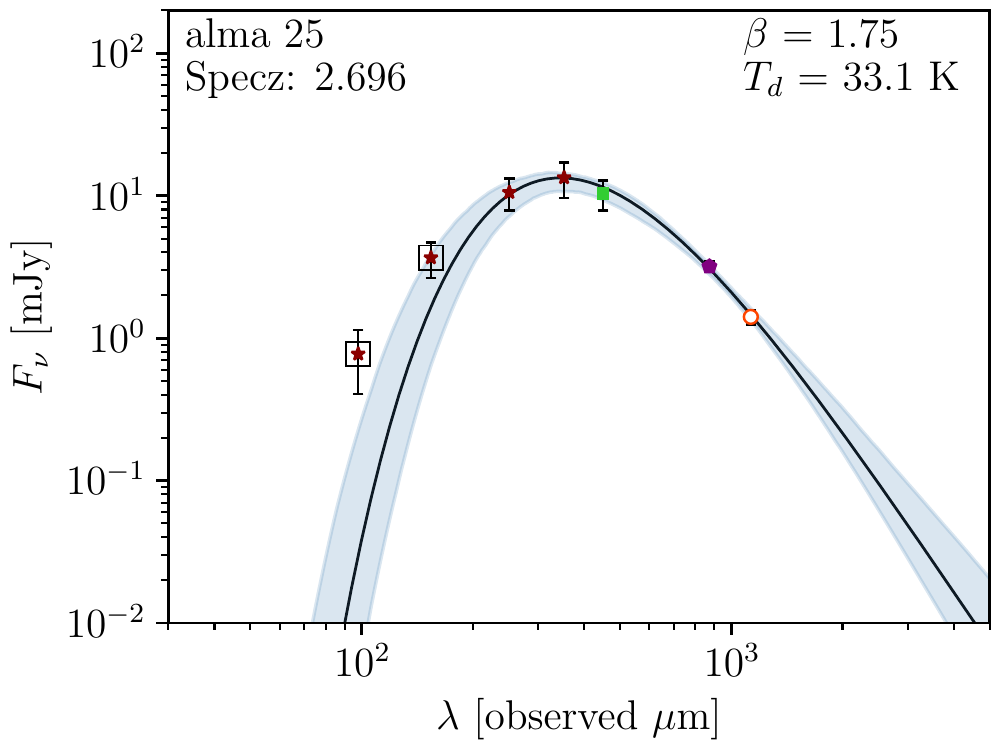}
\figsetgrpnote{Optically thin MBB SED fit (black curve) and 16th to 84th percentile range of the accepted MCMC models (blue shaded region). Photometry: Red circles---ALMA 1.1~mm, 1.2~mm, 2~mm, and 3~mm, maroon pentagon---ALMA 870~$\mu$m, green square---SCUBA-2 450~$\mu$m, dark red stars---Herschel/PACS 100 and 160~$\mu$m and SPIRE 250 and 350~$\mu$m, blue triangles---Spitzer/MIPS 70~$\mu$m. Points not included in the fits are marked with black squares.}
\figsetgrpend

\figsetgrpstart
\figsetgrpnum{12.24}
\figsetgrptitle{SED for ALMA 26}
\figsetplot{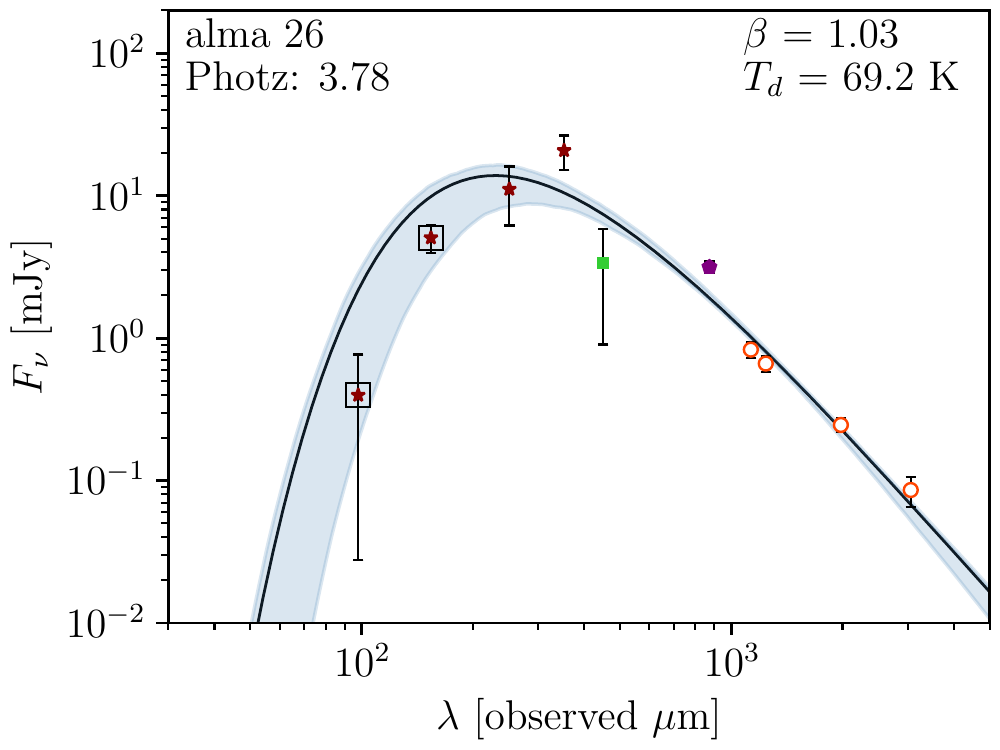}
\figsetgrpnote{Optically thin MBB SED fit (black curve) and 16th to 84th percentile range of the accepted MCMC models (blue shaded region). Photometry: Red circles---ALMA 1.1~mm, 1.2~mm, 2~mm, and 3~mm, maroon pentagon---ALMA 870~$\mu$m, green square---SCUBA-2 450~$\mu$m, dark red stars---Herschel/PACS 100 and 160~$\mu$m and SPIRE 250 and 350~$\mu$m, blue triangles---Spitzer/MIPS 70~$\mu$m. Points not included in the fits are marked with black squares.}
\figsetgrpend

\figsetgrpstart
\figsetgrpnum{12.25}
\figsetgrptitle{SED for ALMA 27}
\figsetplot{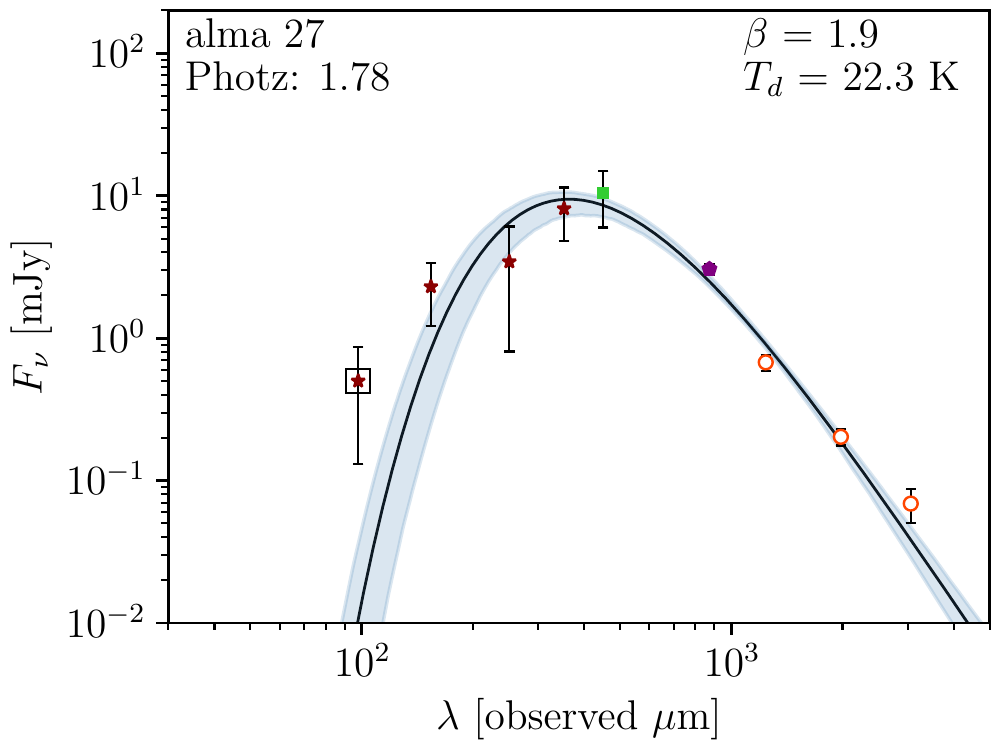}
\figsetgrpnote{Optically thin MBB SED fit (black curve) and 16th to 84th percentile range of the accepted MCMC models (blue shaded region). Photometry: Red circles---ALMA 1.1~mm, 1.2~mm, 2~mm, and 3~mm, maroon pentagon---ALMA 870~$\mu$m, green square---SCUBA-2 450~$\mu$m, dark red stars---Herschel/PACS 100 and 160~$\mu$m and SPIRE 250 and 350~$\mu$m, blue triangles---Spitzer/MIPS 70~$\mu$m. Points not included in the fits are marked with black squares.}
\figsetgrpend

\figsetgrpstart
\figsetgrpnum{12.26}
\figsetgrptitle{SED for ALMA 28}
\figsetplot{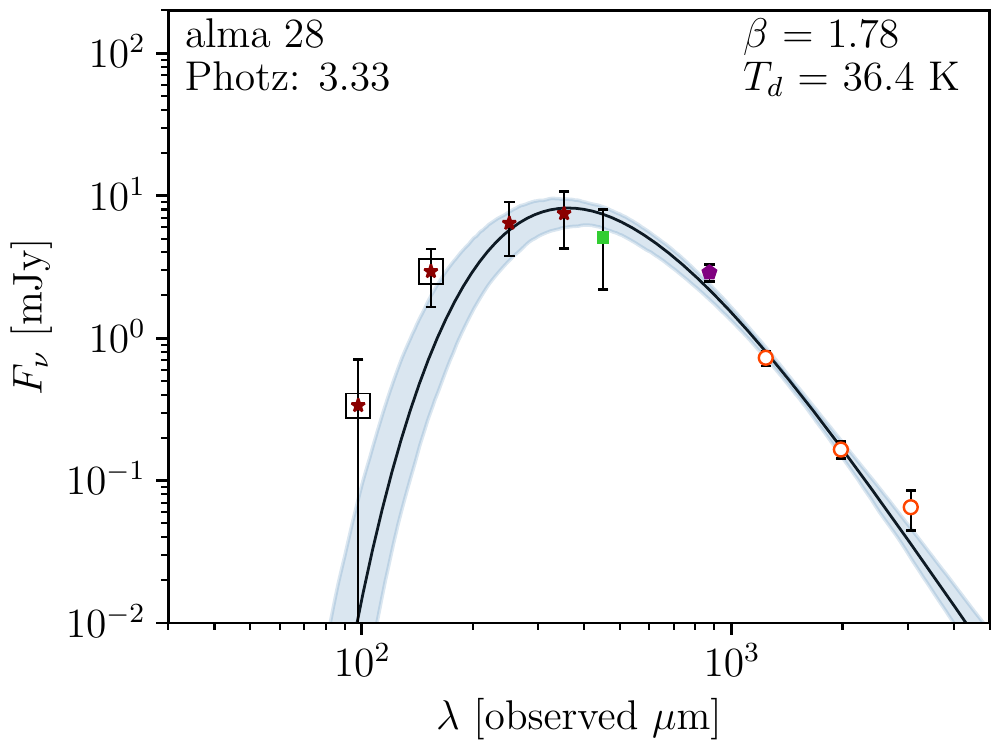}
\figsetgrpnote{Optically thin MBB SED fit (black curve) and 16th to 84th percentile range of the accepted MCMC models (blue shaded region). Photometry: Red circles---ALMA 1.1~mm, 1.2~mm, 2~mm, and 3~mm, maroon pentagon---ALMA 870~$\mu$m, green square---SCUBA-2 450~$\mu$m, dark red stars---Herschel/PACS 100 and 160~$\mu$m and SPIRE 250 and 350~$\mu$m, blue triangles---Spitzer/MIPS 70~$\mu$m. Points not included in the fits are marked with black squares.}
\figsetgrpend

\figsetgrpstart
\figsetgrpnum{12.27}
\figsetgrptitle{SED for ALMA 30}
\figsetplot{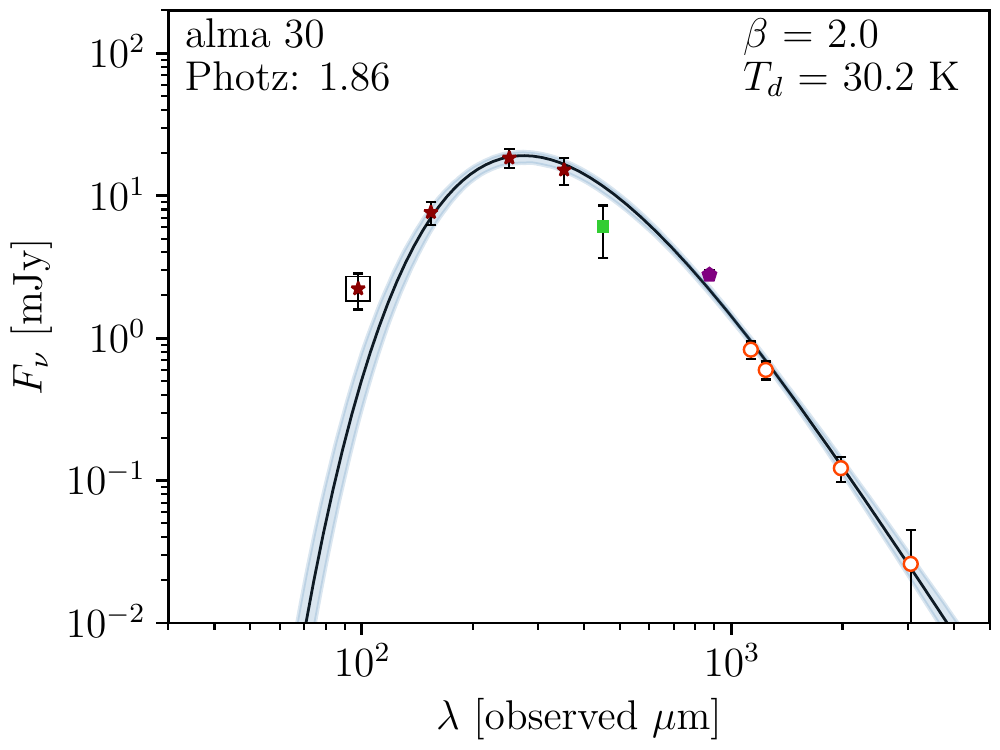}
\figsetgrpnote{Optically thin MBB SED fit (black curve) and 16th to 84th percentile range of the accepted MCMC models (blue shaded region). Photometry: Red circles---ALMA 1.1~mm, 1.2~mm, 2~mm, and 3~mm, maroon pentagon---ALMA 870~$\mu$m, green square---SCUBA-2 450~$\mu$m, dark red stars---Herschel/PACS 100 and 160~$\mu$m and SPIRE 250 and 350~$\mu$m, blue triangles---Spitzer/MIPS 70~$\mu$m. Points not included in the fits are marked with black squares.}
\figsetgrpend

\figsetgrpstart
\figsetgrpnum{12.28}
\figsetgrptitle{SED for ALMA 31}
\figsetplot{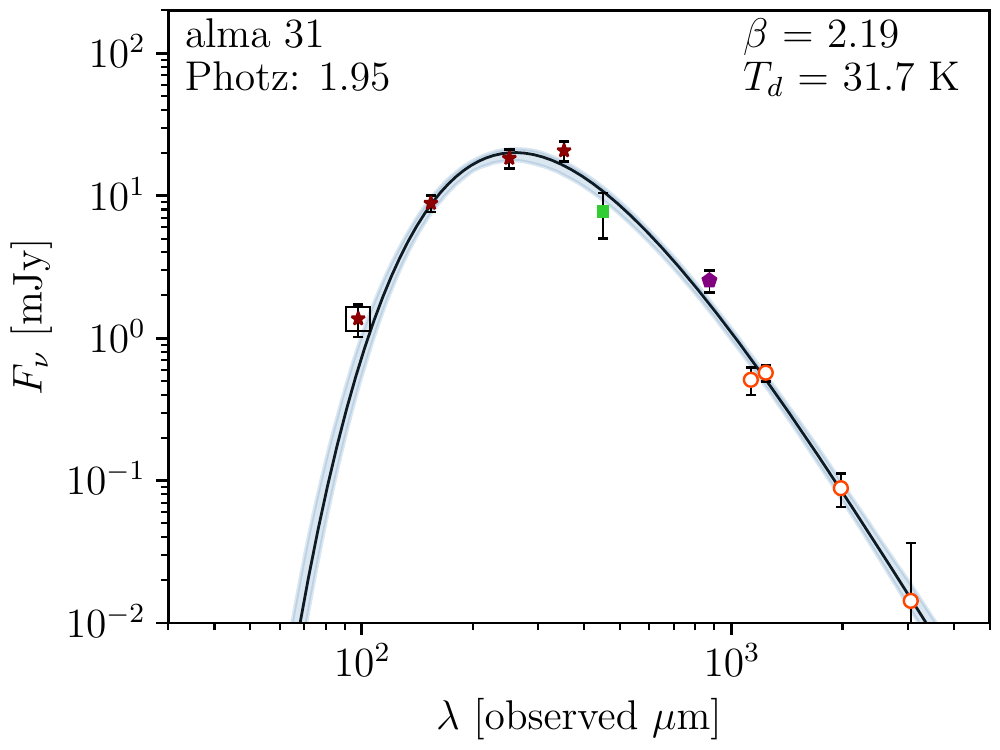}
\figsetgrpnote{Optically thin MBB SED fit (black curve) and 16th to 84th percentile range of the accepted MCMC models (blue shaded region). Photometry: Red circles---ALMA 1.1~mm, 1.2~mm, 2~mm, and 3~mm, maroon pentagon---ALMA 870~$\mu$m, green square---SCUBA-2 450~$\mu$m, dark red stars---Herschel/PACS 100 and 160~$\mu$m and SPIRE 250 and 350~$\mu$m, blue triangles---Spitzer/MIPS 70~$\mu$m. Points not included in the fits are marked with black squares.}
\figsetgrpend

\figsetgrpstart
\figsetgrpnum{12.29}
\figsetgrptitle{SED for ALMA 32}
\figsetplot{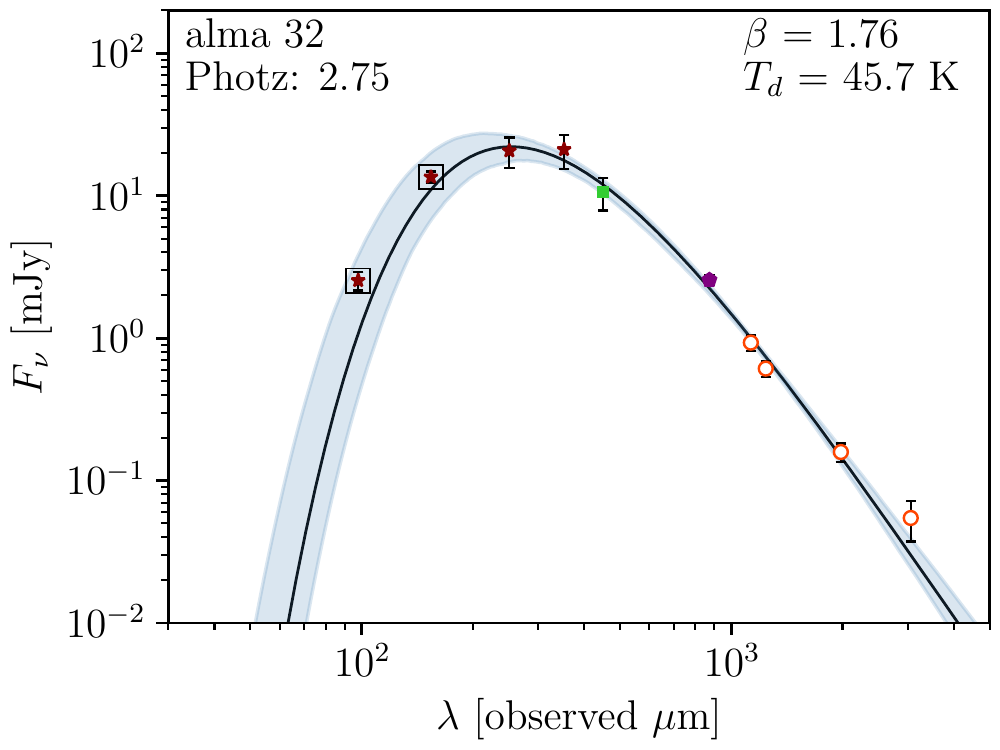}
\figsetgrpnote{Optically thin MBB SED fit (black curve) and 16th to 84th percentile range of the accepted MCMC models (blue shaded region). Photometry: Red circles---ALMA 1.1~mm, 1.2~mm, 2~mm, and 3~mm, maroon pentagon---ALMA 870~$\mu$m, green square---SCUBA-2 450~$\mu$m, dark red stars---Herschel/PACS 100 and 160~$\mu$m and SPIRE 250 and 350~$\mu$m, blue triangles---Spitzer/MIPS 70~$\mu$m. Points not included in the fits are marked with black squares.}
\figsetgrpend

\figsetgrpstart
\figsetgrpnum{12.30}
\figsetgrptitle{SED for ALMA 33}
\figsetplot{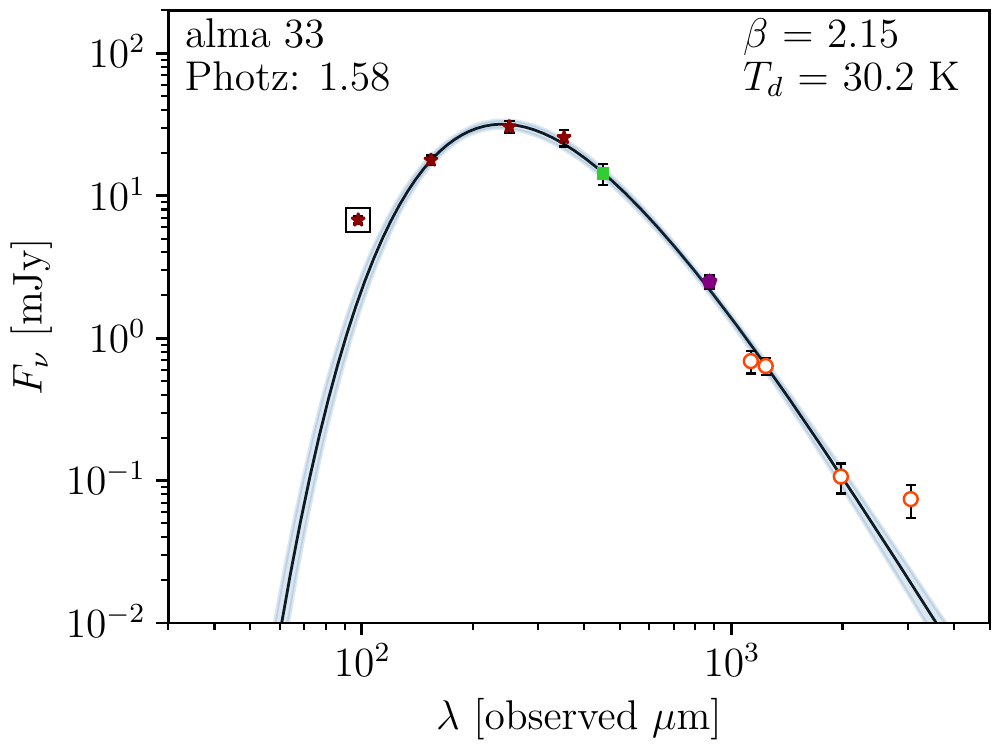}
\figsetgrpnote{Optically thin MBB SED fit (black curve) and 16th to 84th percentile range of the accepted MCMC models (blue shaded region). Photometry: Red circles---ALMA 1.1~mm, 1.2~mm, 2~mm, and 3~mm, maroon pentagon---ALMA 870~$\mu$m, green square---SCUBA-2 450~$\mu$m, dark red stars---Herschel/PACS 100 and 160~$\mu$m and SPIRE 250 and 350~$\mu$m, blue triangles---Spitzer/MIPS 70~$\mu$m. Points not included in the fits are marked with black squares.}
\figsetgrpend

\figsetgrpstart
\figsetgrpnum{12.31}
\figsetgrptitle{SED for ALMA 34}
\figsetplot{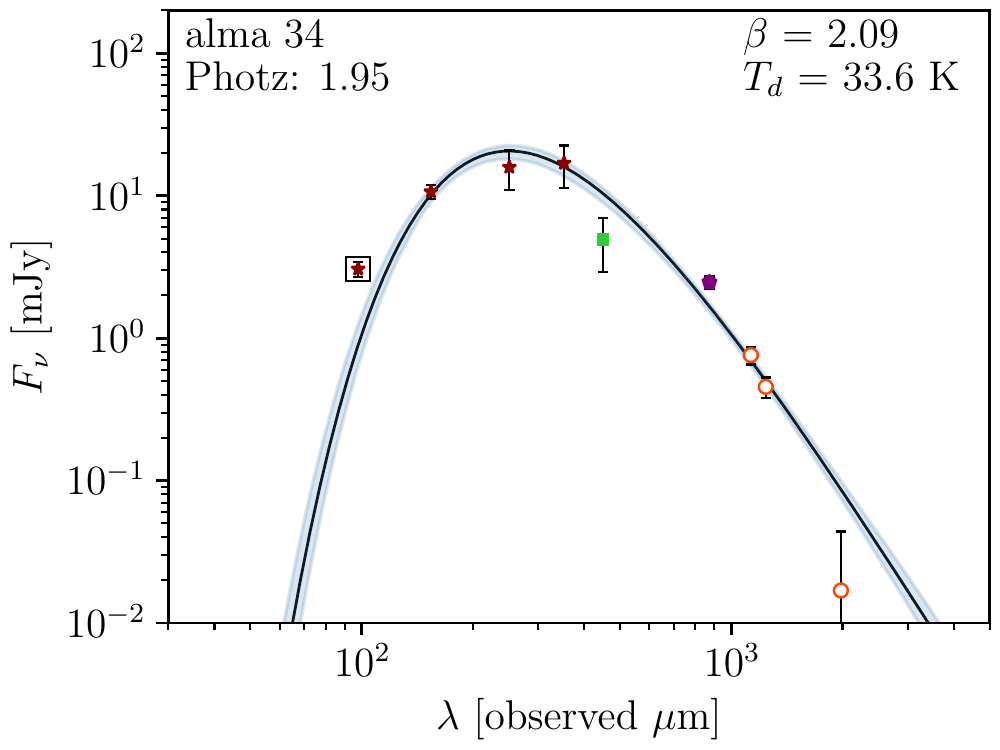}
\figsetgrpnote{Optically thin MBB SED fit (black curve) and 16th to 84th percentile range of the accepted MCMC models (blue shaded region). Photometry: Red circles---ALMA 1.1~mm, 1.2~mm, 2~mm, and 3~mm, maroon pentagon---ALMA 870~$\mu$m, green square---SCUBA-2 450~$\mu$m, dark red stars---Herschel/PACS 100 and 160~$\mu$m and SPIRE 250 and 350~$\mu$m, blue triangles---Spitzer/MIPS 70~$\mu$m. Points not included in the fits are marked with black squares.}
\figsetgrpend

\figsetgrpstart
\figsetgrpnum{12.32}
\figsetgrptitle{SED for ALMA 35}
\figsetplot{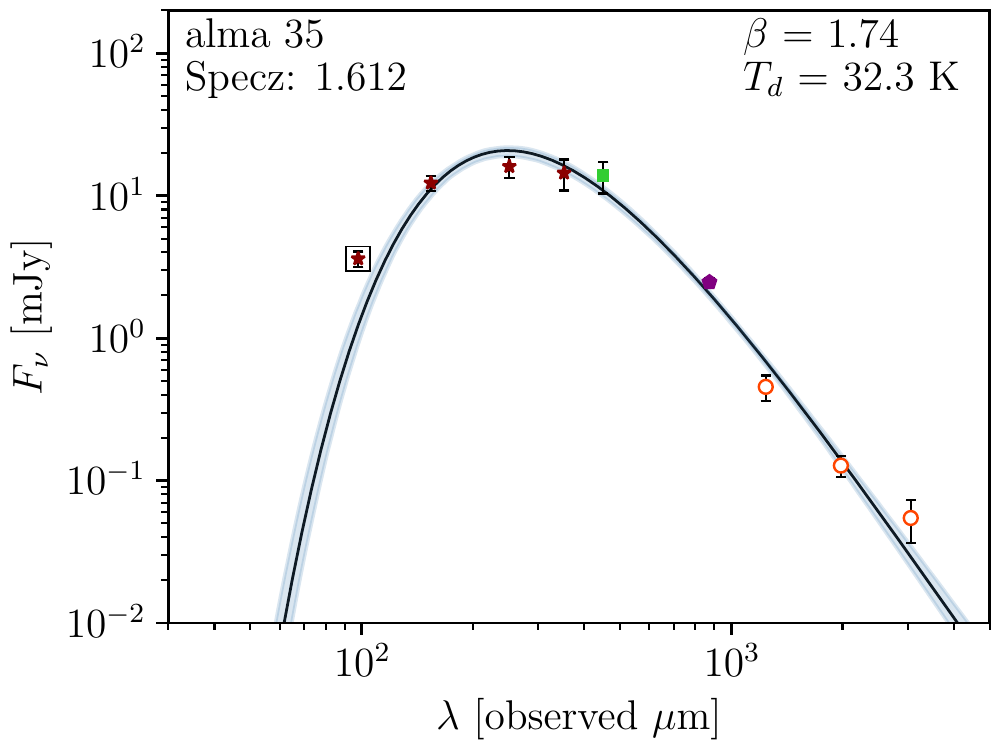}
\figsetgrpnote{Optically thin MBB SED fit (black curve) and 16th to 84th percentile range of the accepted MCMC models (blue shaded region). Photometry: Red circles---ALMA 1.1~mm, 1.2~mm, 2~mm, and 3~mm, maroon pentagon---ALMA 870~$\mu$m, green square---SCUBA-2 450~$\mu$m, dark red stars---Herschel/PACS 100 and 160~$\mu$m and SPIRE 250 and 350~$\mu$m, blue triangles---Spitzer/MIPS 70~$\mu$m. Points not included in the fits are marked with black squares.}
\figsetgrpend

\figsetgrpstart
\figsetgrpnum{12.33}
\figsetgrptitle{SED for ALMA 36}
\figsetplot{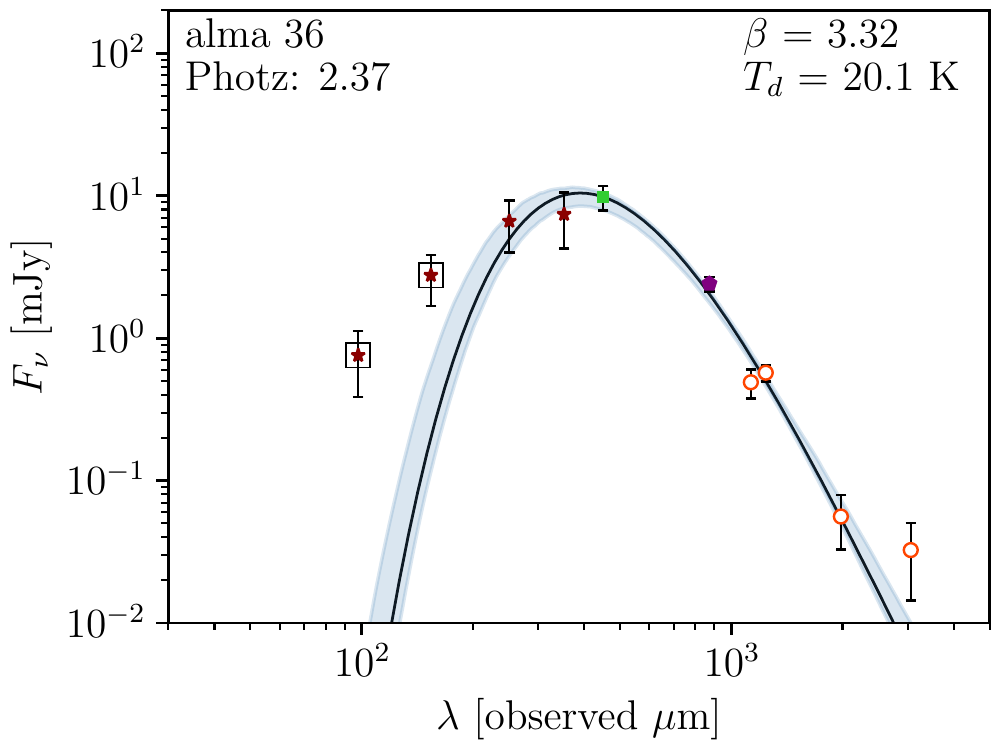}
\figsetgrpnote{Optically thin MBB SED fit (black curve) and 16th to 84th percentile range of the accepted MCMC models (blue shaded region). Photometry: Red circles---ALMA 1.1~mm, 1.2~mm, 2~mm, and 3~mm, maroon pentagon---ALMA 870~$\mu$m, green square---SCUBA-2 450~$\mu$m, dark red stars---Herschel/PACS 100 and 160~$\mu$m and SPIRE 250 and 350~$\mu$m, blue triangles---Spitzer/MIPS 70~$\mu$m. Points not included in the fits are marked with black squares.}
\figsetgrpend

\figsetgrpstart
\figsetgrpnum{12.34}
\figsetgrptitle{SED for ALMA 37}
\figsetplot{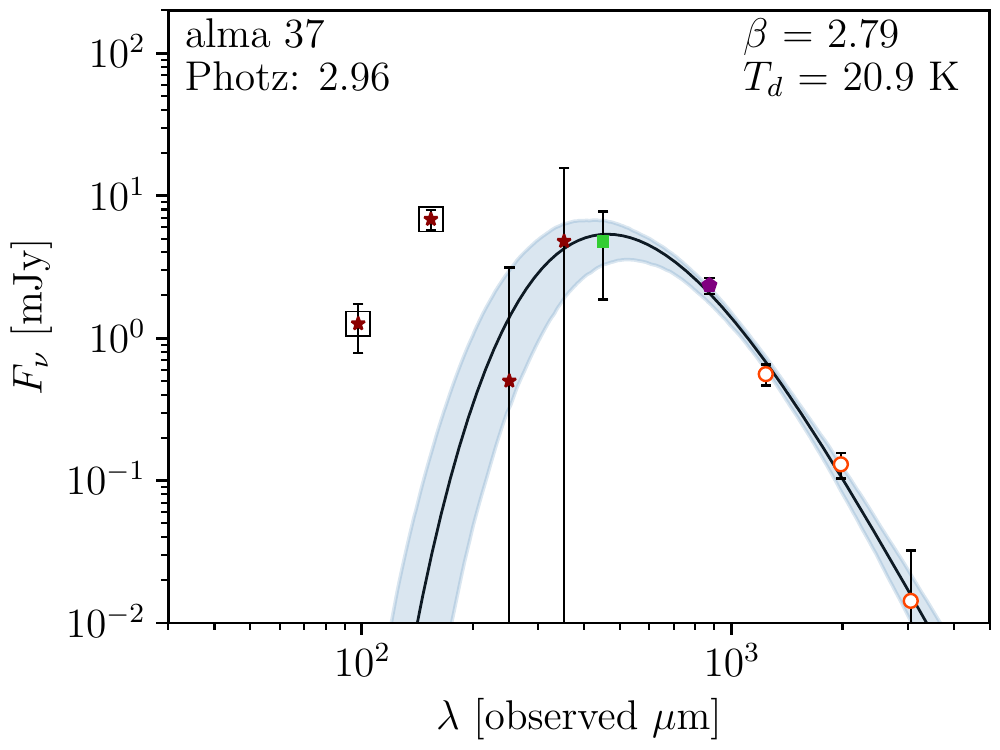}
\figsetgrpnote{Optically thin MBB SED fit (black curve) and 16th to 84th percentile range of the accepted MCMC models (blue shaded region). Photometry: Red circles---ALMA 1.1~mm, 1.2~mm, 2~mm, and 3~mm, maroon pentagon---ALMA 870~$\mu$m, green square---SCUBA-2 450~$\mu$m, dark red stars---Herschel/PACS 100 and 160~$\mu$m and SPIRE 250 and 350~$\mu$m, blue triangles---Spitzer/MIPS 70~$\mu$m. Points not included in the fits are marked with black squares.}
\figsetgrpend

\figsetgrpstart
\figsetgrpnum{12.35}
\figsetgrptitle{SED for ALMA 38}
\figsetplot{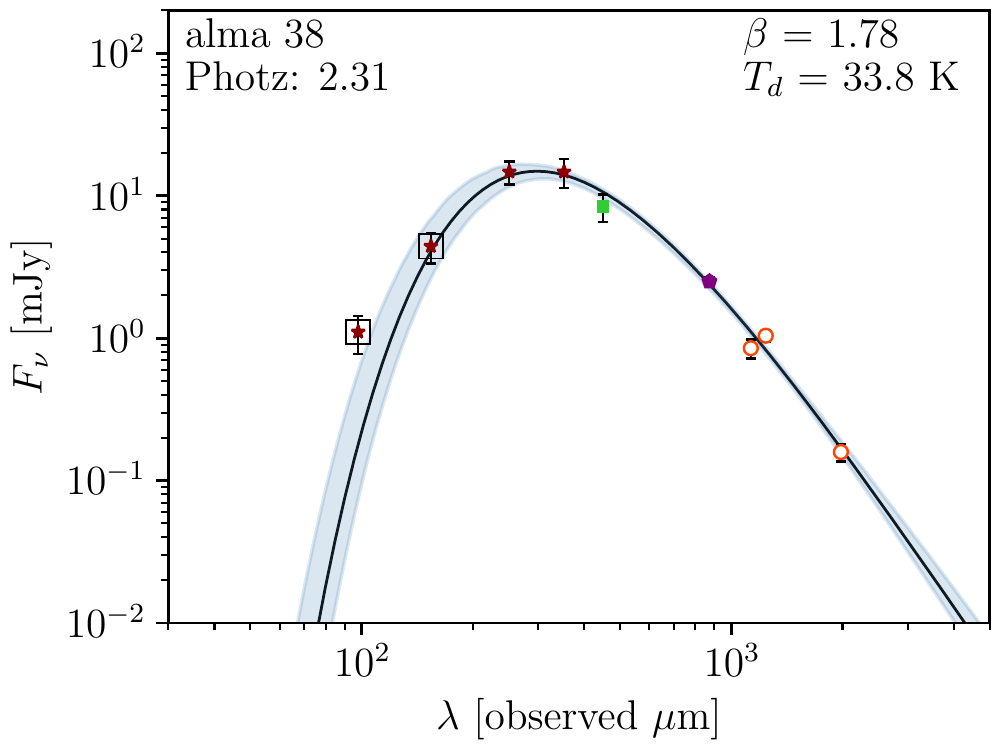}
\figsetgrpnote{Optically thin MBB SED fit (black curve) and 16th to 84th percentile range of the accepted MCMC models (blue shaded region). Photometry: Red circles---ALMA 1.1~mm, 1.2~mm, 2~mm, and 3~mm, maroon pentagon---ALMA 870~$\mu$m, green square---SCUBA-2 450~$\mu$m, dark red stars---Herschel/PACS 100 and 160~$\mu$m and SPIRE 250 and 350~$\mu$m, blue triangles---Spitzer/MIPS 70~$\mu$m. Points not included in the fits are marked with black squares.}
\figsetgrpend

\figsetgrpstart
\figsetgrpnum{12.36}
\figsetgrptitle{SED for ALMA 39}
\figsetplot{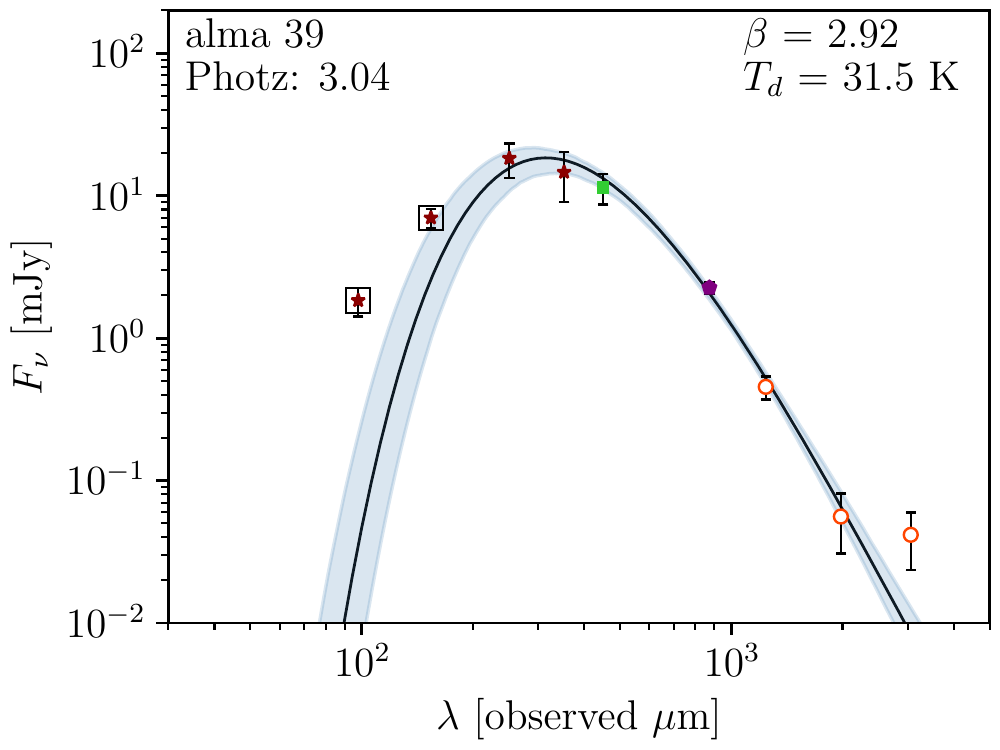}
\figsetgrpnote{Optically thin MBB SED fit (black curve) and 16th to 84th percentile range of the accepted MCMC models (blue shaded region). Photometry: Red circles---ALMA 1.1~mm, 1.2~mm, 2~mm, and 3~mm, maroon pentagon---ALMA 870~$\mu$m, green square---SCUBA-2 450~$\mu$m, dark red stars---Herschel/PACS 100 and 160~$\mu$m and SPIRE 250 and 350~$\mu$m, blue triangles---Spitzer/MIPS 70~$\mu$m. Points not included in the fits are marked with black squares.}
\figsetgrpend

\figsetgrpstart
\figsetgrpnum{12.37}
\figsetgrptitle{SED for ALMA 40}
\figsetplot{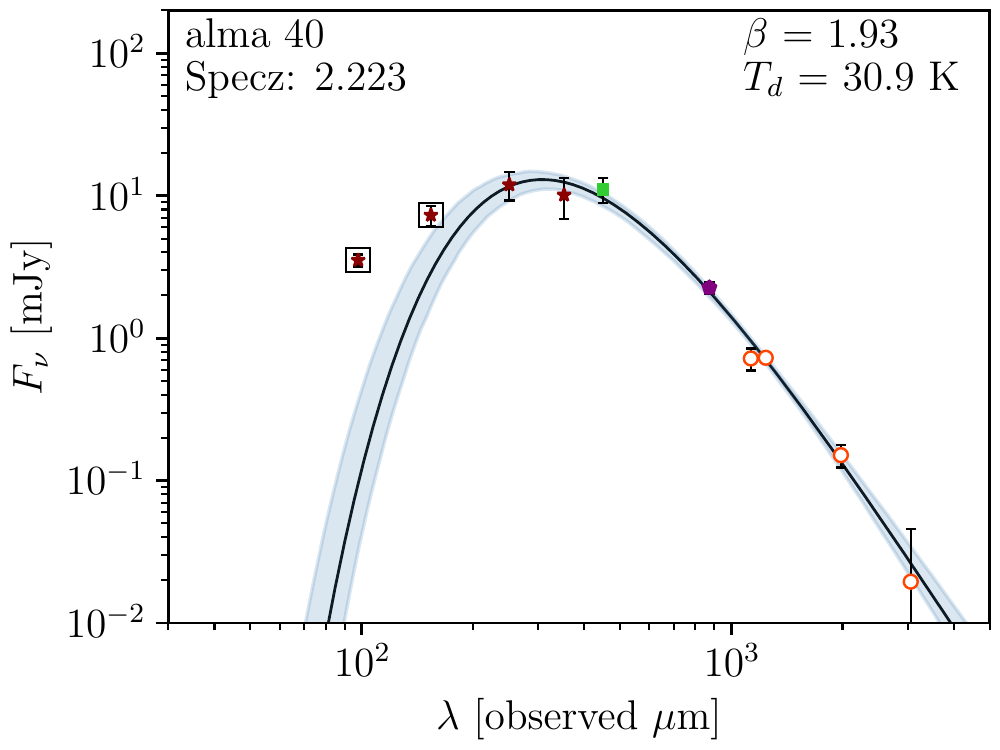}
\figsetgrpnote{Optically thin MBB SED fit (black curve) and 16th to 84th percentile range of the accepted MCMC models (blue shaded region). Photometry: Red circles---ALMA 1.1~mm, 1.2~mm, 2~mm, and 3~mm, maroon pentagon---ALMA 870~$\mu$m, green square---SCUBA-2 450~$\mu$m, dark red stars---Herschel/PACS 100 and 160~$\mu$m and SPIRE 250 and 350~$\mu$m, blue triangles---Spitzer/MIPS 70~$\mu$m. Points not included in the fits are marked with black squares.}
\figsetgrpend

\figsetgrpstart
\figsetgrpnum{12.38}
\figsetgrptitle{SED for ALMA 41}
\figsetplot{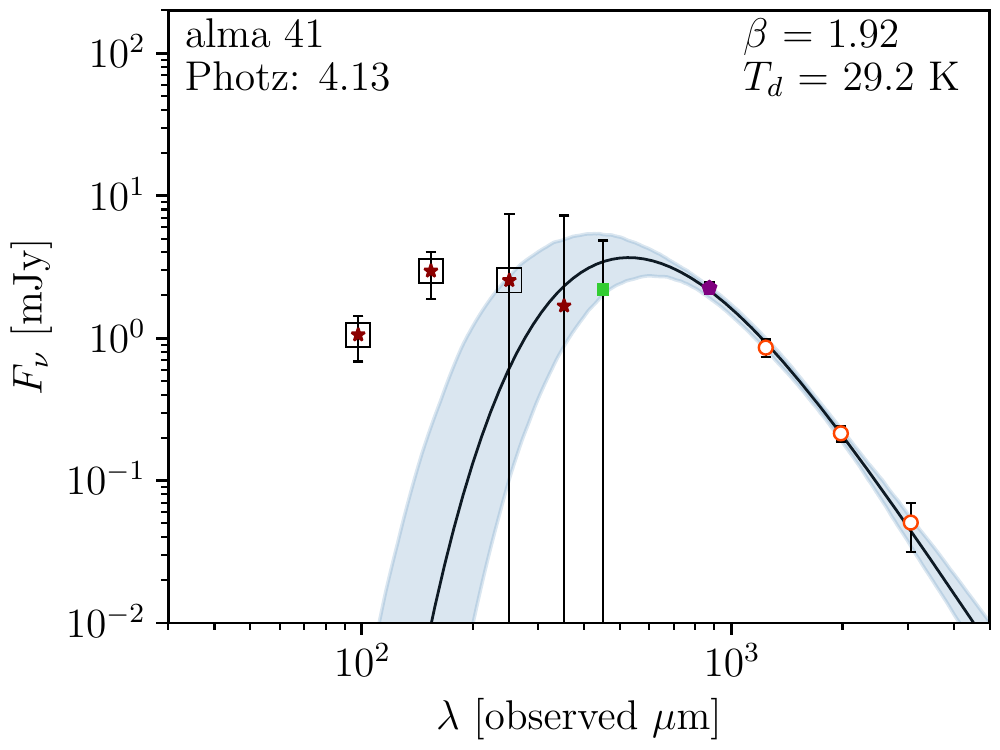}
\figsetgrpnote{Optically thin MBB SED fit (black curve) and 16th to 84th percentile range of the accepted MCMC models (blue shaded region). Photometry: Red circles---ALMA 1.1~mm, 1.2~mm, 2~mm, and 3~mm, maroon pentagon---ALMA 870~$\mu$m, green square---SCUBA-2 450~$\mu$m, dark red stars---Herschel/PACS 100 and 160~$\mu$m and SPIRE 250 and 350~$\mu$m, blue triangles---Spitzer/MIPS 70~$\mu$m. Points not included in the fits are marked with black squares.}
\figsetgrpend

\figsetgrpstart
\figsetgrpnum{12.39}
\figsetgrptitle{SED for ALMA 42}
\figsetplot{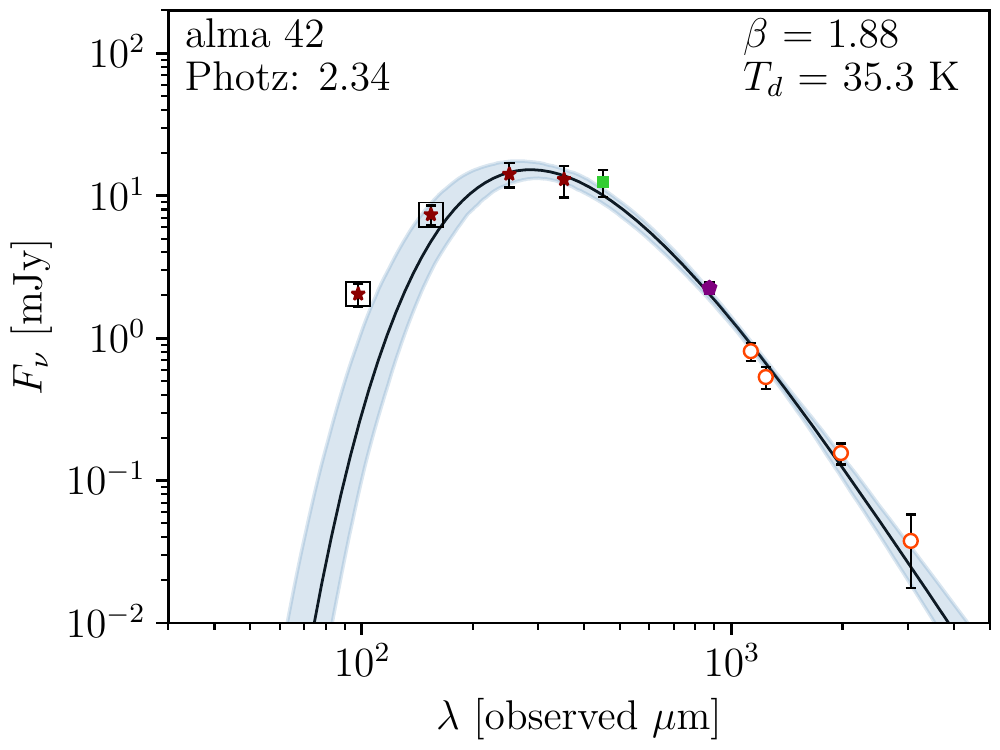}
\figsetgrpnote{Optically thin MBB SED fit (black curve) and 16th to 84th percentile range of the accepted MCMC models (blue shaded region). Photometry: Red circles---ALMA 1.1~mm, 1.2~mm, 2~mm, and 3~mm, maroon pentagon---ALMA 870~$\mu$m, green square---SCUBA-2 450~$\mu$m, dark red stars---Herschel/PACS 100 and 160~$\mu$m and SPIRE 250 and 350~$\mu$m, blue triangles---Spitzer/MIPS 70~$\mu$m. Points not included in the fits are marked with black squares.}
\figsetgrpend

\figsetgrpstart
\figsetgrpnum{12.40}
\figsetgrptitle{SED for ALMA 43}
\figsetplot{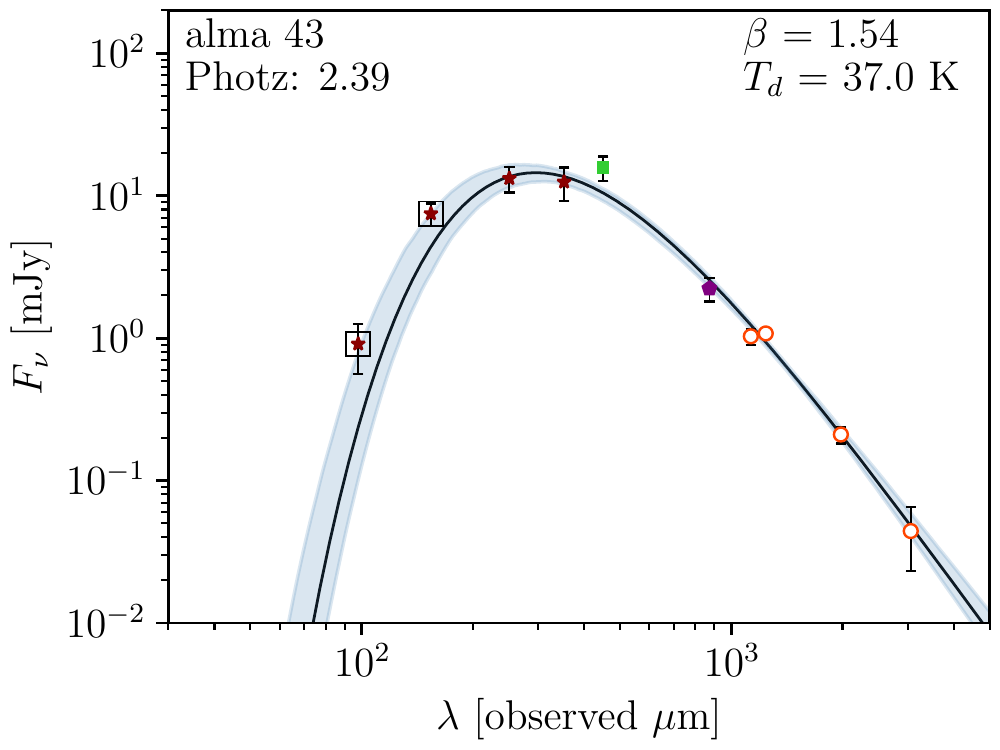}
\figsetgrpnote{Optically thin MBB SED fit (black curve) and 16th to 84th percentile range of the accepted MCMC models (blue shaded region). Photometry: Red circles---ALMA 1.1~mm, 1.2~mm, 2~mm, and 3~mm, maroon pentagon---ALMA 870~$\mu$m, green square---SCUBA-2 450~$\mu$m, dark red stars---Herschel/PACS 100 and 160~$\mu$m and SPIRE 250 and 350~$\mu$m, blue triangles---Spitzer/MIPS 70~$\mu$m. Points not included in the fits are marked with black squares.}
\figsetgrpend

\figsetgrpstart
\figsetgrpnum{12.41}
\figsetgrptitle{SED for ALMA 45}
\figsetplot{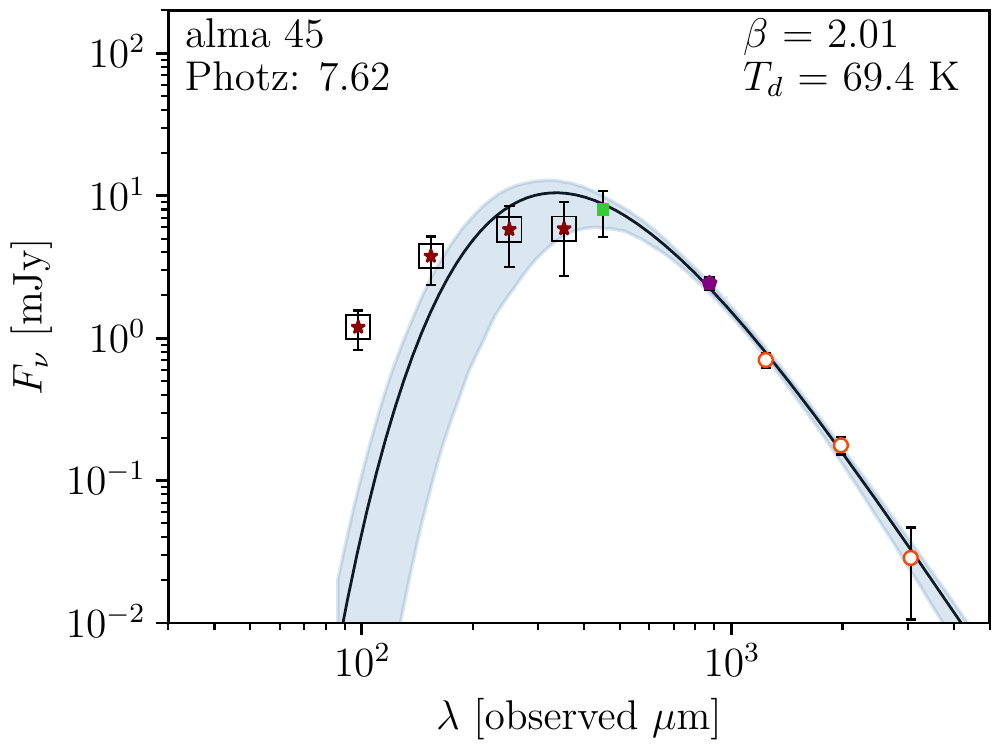}
\figsetgrpnote{Optically thin MBB SED fit (black curve) and 16th to 84th percentile range of the accepted MCMC models (blue shaded region). Photometry: Red circles---ALMA 1.1~mm, 1.2~mm, 2~mm, and 3~mm, maroon pentagon---ALMA 870~$\mu$m, green square---SCUBA-2 450~$\mu$m, dark red stars---Herschel/PACS 100 and 160~$\mu$m and SPIRE 250 and 350~$\mu$m, blue triangles---Spitzer/MIPS 70~$\mu$m. Points not included in the fits are marked with black squares.}
\figsetgrpend

\figsetgrpstart
\figsetgrpnum{12.42}
\figsetgrptitle{SED for ALMA 46}
\figsetplot{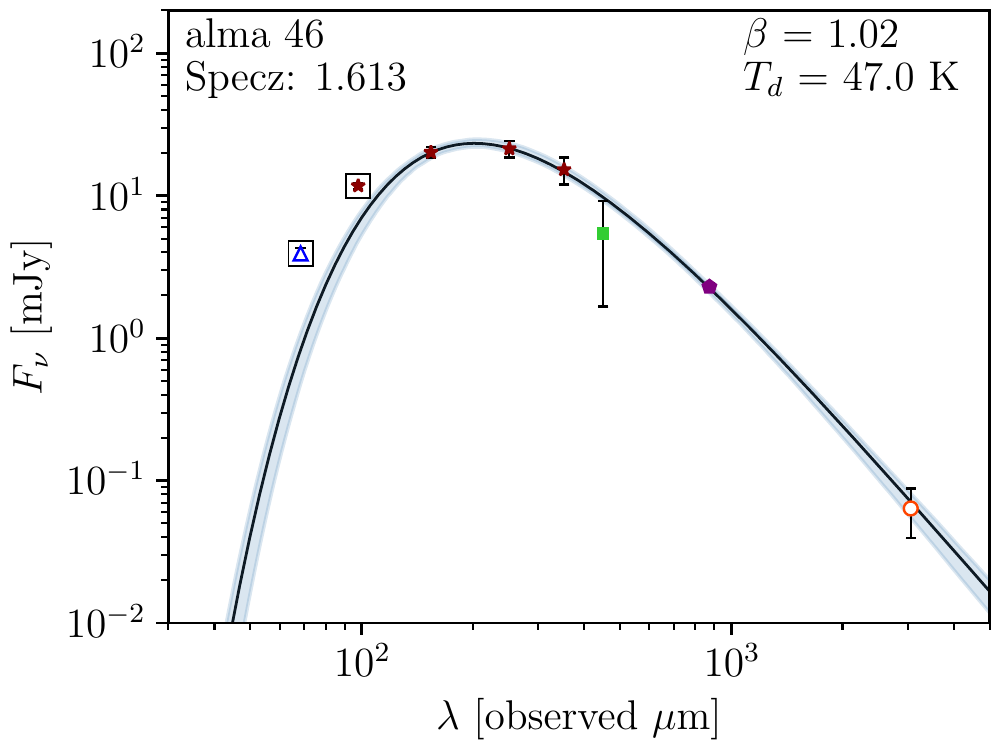}
\figsetgrpnote{Optically thin MBB SED fit (black curve) and 16th to 84th percentile range of the accepted MCMC models (blue shaded region). Photometry: Red circles---ALMA 1.1~mm, 1.2~mm, 2~mm, and 3~mm, maroon pentagon---ALMA 870~$\mu$m, green square---SCUBA-2 450~$\mu$m, dark red stars---Herschel/PACS 100 and 160~$\mu$m and SPIRE 250 and 350~$\mu$m, blue triangles---Spitzer/MIPS 70~$\mu$m. Points not included in the fits are marked with black squares.}
\figsetgrpend

\figsetgrpstart
\figsetgrpnum{12.43}
\figsetgrptitle{SED for ALMA 47}
\figsetplot{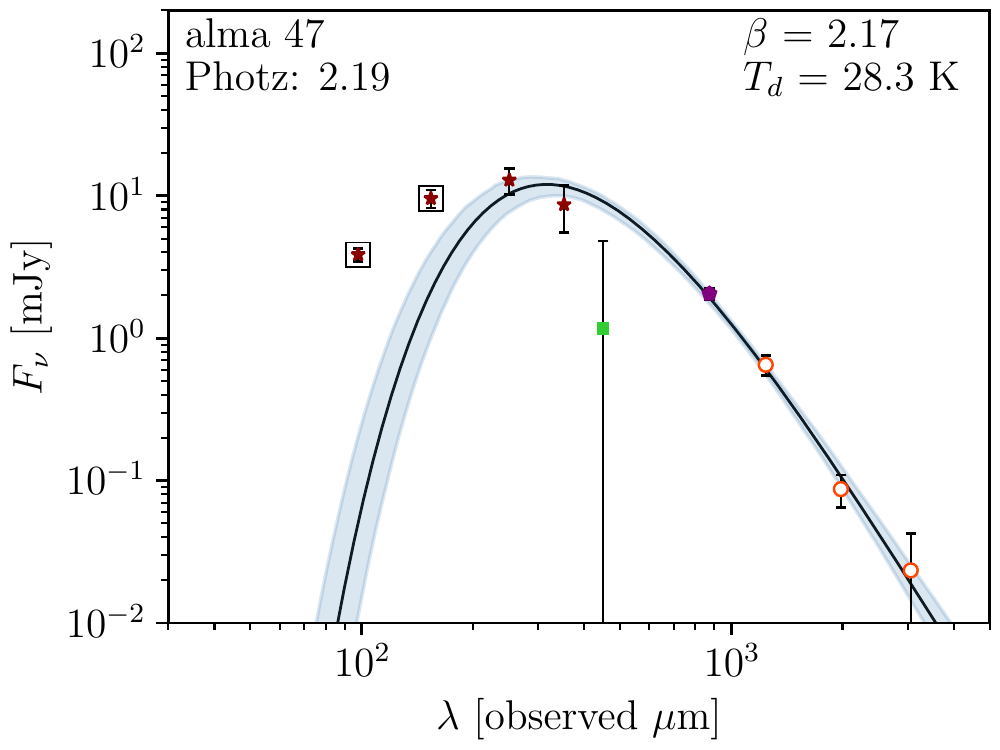}
\figsetgrpnote{Optically thin MBB SED fit (black curve) and 16th to 84th percentile range of the accepted MCMC models (blue shaded region). Photometry: Red circles---ALMA 1.1~mm, 1.2~mm, 2~mm, and 3~mm, maroon pentagon---ALMA 870~$\mu$m, green square---SCUBA-2 450~$\mu$m, dark red stars---Herschel/PACS 100 and 160~$\mu$m and SPIRE 250 and 350~$\mu$m, blue triangles---Spitzer/MIPS 70~$\mu$m. Points not included in the fits are marked with black squares.}
\figsetgrpend

\figsetgrpstart
\figsetgrpnum{12.44}
\figsetgrptitle{SED for ALMA 48}
\figsetplot{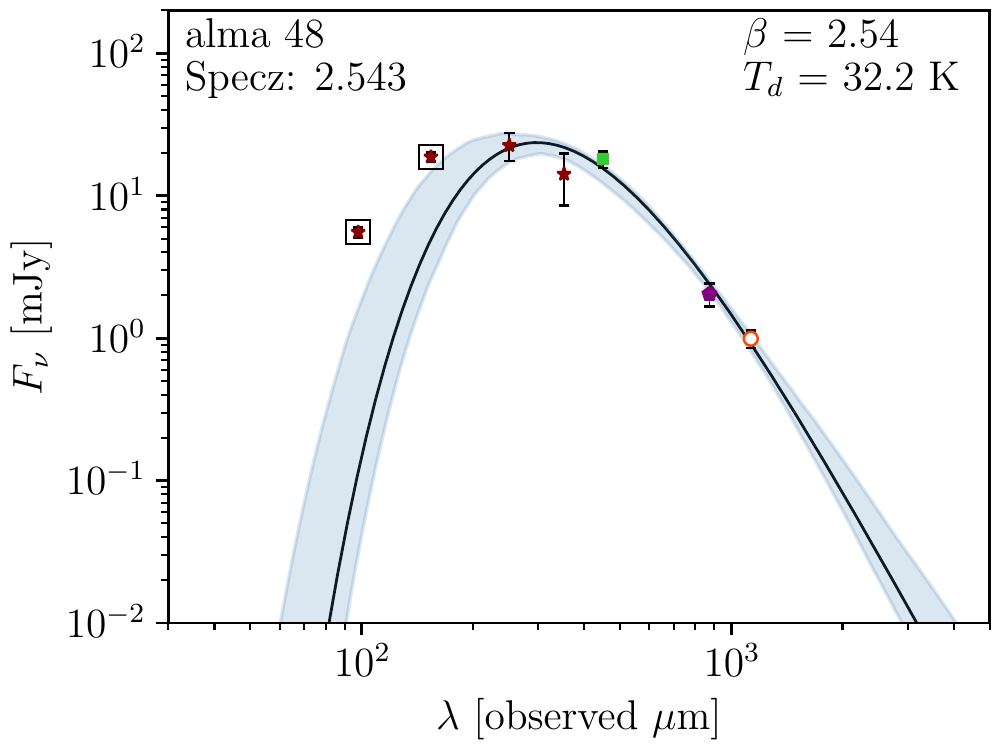}
\figsetgrpnote{Optically thin MBB SED fit (black curve) and 16th to 84th percentile range of the accepted MCMC models (blue shaded region). Photometry: Red circles---ALMA 1.1~mm, 1.2~mm, 2~mm, and 3~mm, maroon pentagon---ALMA 870~$\mu$m, green square---SCUBA-2 450~$\mu$m, dark red stars---Herschel/PACS 100 and 160~$\mu$m and SPIRE 250 and 350~$\mu$m, blue triangles---Spitzer/MIPS 70~$\mu$m. Points not included in the fits are marked with black squares.}
\figsetgrpend

\figsetgrpstart
\figsetgrpnum{12.45}
\figsetgrptitle{SED for ALMA 49}
\figsetplot{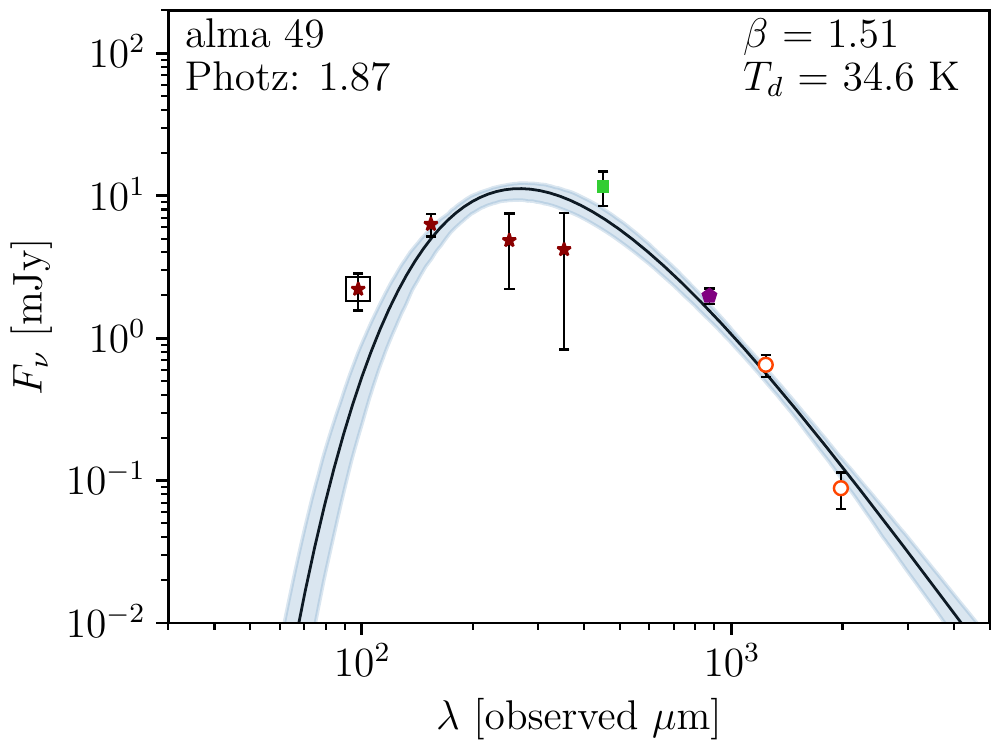}
\figsetgrpnote{Optically thin MBB SED fit (black curve) and 16th to 84th percentile range of the accepted MCMC models (blue shaded region). Photometry: Red circles---ALMA 1.1~mm, 1.2~mm, 2~mm, and 3~mm, maroon pentagon---ALMA 870~$\mu$m, green square---SCUBA-2 450~$\mu$m, dark red stars---Herschel/PACS 100 and 160~$\mu$m and SPIRE 250 and 350~$\mu$m, blue triangles---Spitzer/MIPS 70~$\mu$m. Points not included in the fits are marked with black squares.}
\figsetgrpend

\figsetgrpstart
\figsetgrpnum{12.46}
\figsetgrptitle{SED for ALMA 50}
\figsetplot{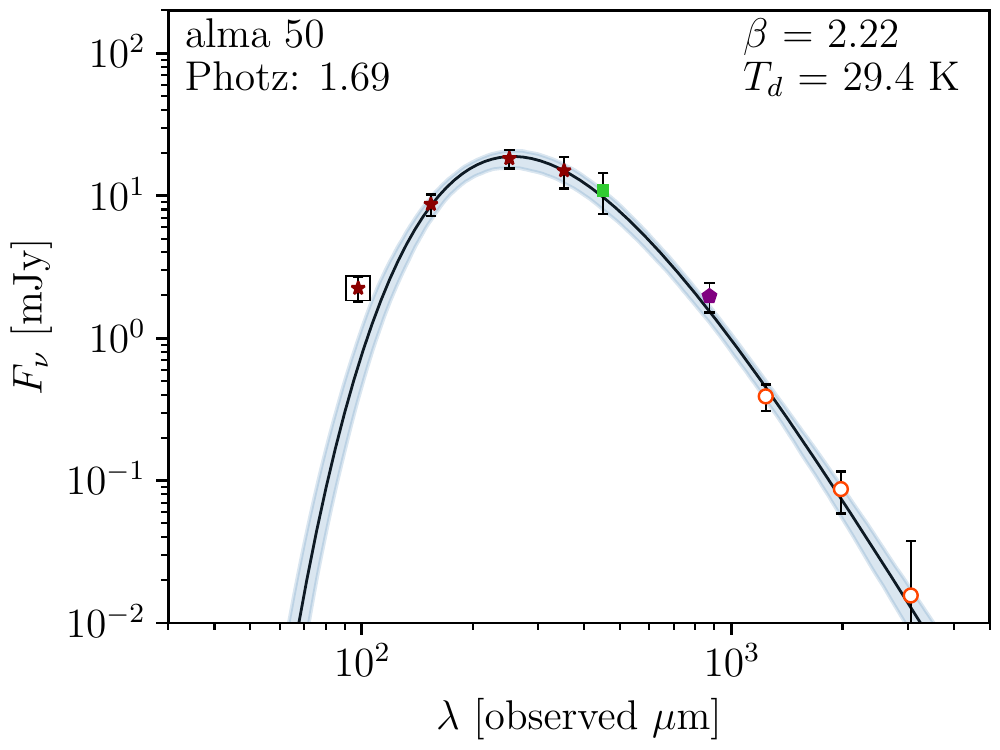}
\figsetgrpnote{Optically thin MBB SED fit (black curve) and 16th to 84th percentile range of the accepted MCMC models (blue shaded region). Photometry: Red circles---ALMA 1.1~mm, 1.2~mm, 2~mm, and 3~mm, maroon pentagon---ALMA 870~$\mu$m, green square---SCUBA-2 450~$\mu$m, dark red stars---Herschel/PACS 100 and 160~$\mu$m and SPIRE 250 and 350~$\mu$m, blue triangles---Spitzer/MIPS 70~$\mu$m. Points not included in the fits are marked with black squares.}
\figsetgrpend

\figsetgrpstart
\figsetgrpnum{12.47}
\figsetgrptitle{SED for ALMA 51}
\figsetplot{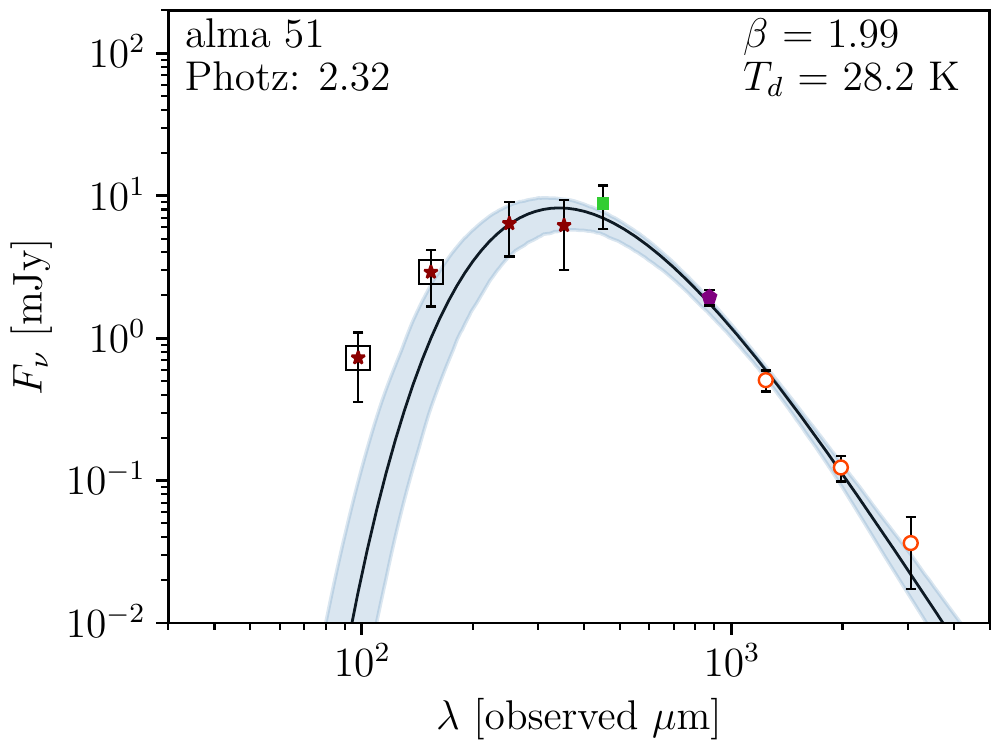}
\figsetgrpnote{Optically thin MBB SED fit (black curve) and 16th to 84th percentile range of the accepted MCMC models (blue shaded region). Photometry: Red circles---ALMA 1.1~mm, 1.2~mm, 2~mm, and 3~mm, maroon pentagon---ALMA 870~$\mu$m, green square---SCUBA-2 450~$\mu$m, dark red stars---Herschel/PACS 100 and 160~$\mu$m and SPIRE 250 and 350~$\mu$m, blue triangles---Spitzer/MIPS 70~$\mu$m. Points not included in the fits are marked with black squares.}
\figsetgrpend

\figsetgrpstart
\figsetgrpnum{12.48}
\figsetgrptitle{SED for ALMA 52}
\figsetplot{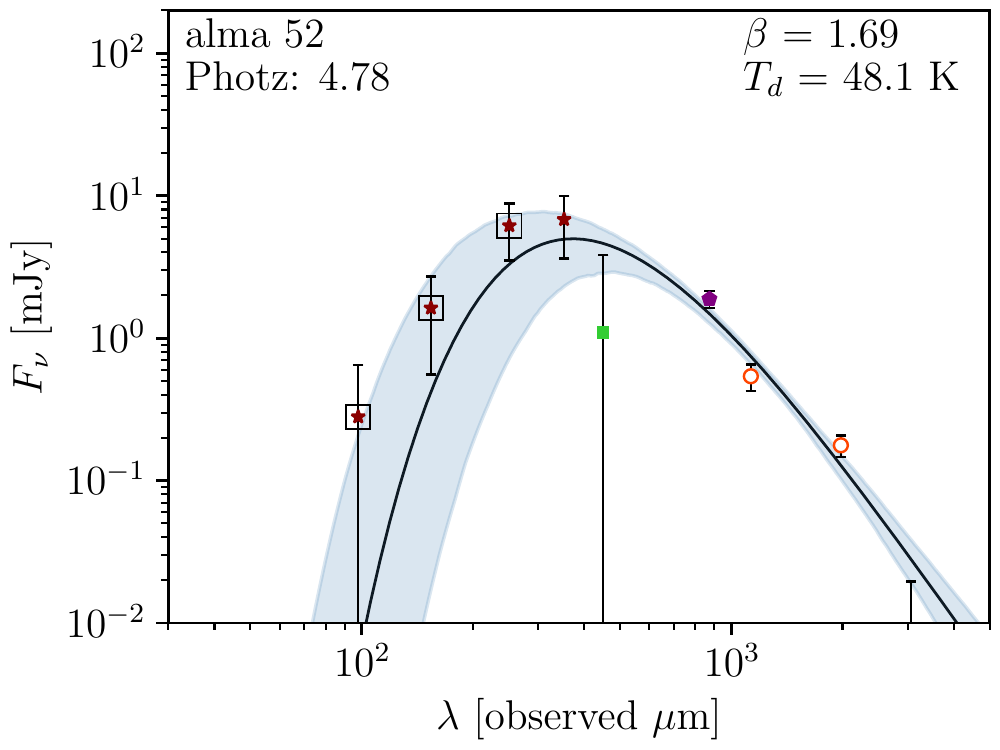}
\figsetgrpnote{Optically thin MBB SED fit (black curve) and 16th to 84th percentile range of the accepted MCMC models (blue shaded region). Photometry: Red circles---ALMA 1.1~mm, 1.2~mm, 2~mm, and 3~mm, maroon pentagon---ALMA 870~$\mu$m, green square---SCUBA-2 450~$\mu$m, dark red stars---Herschel/PACS 100 and 160~$\mu$m and SPIRE 250 and 350~$\mu$m, blue triangles---Spitzer/MIPS 70~$\mu$m. Points not included in the fits are marked with black squares.}
\figsetgrpend

\figsetgrpstart
\figsetgrpnum{12.49}
\figsetgrptitle{SED for ALMA 53}
\figsetplot{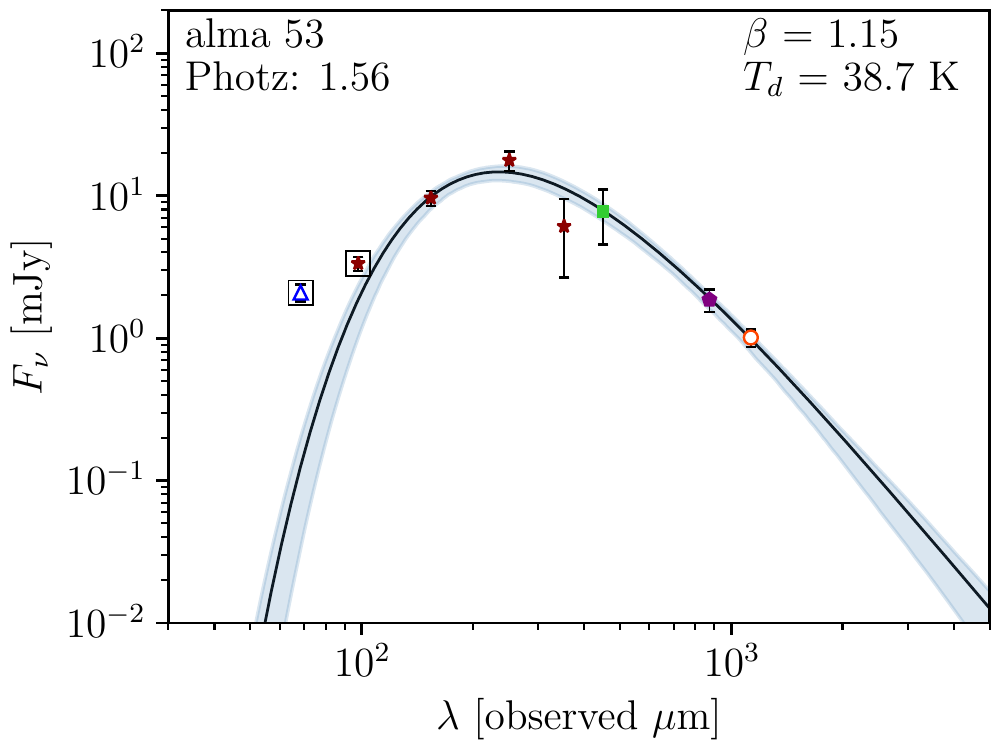}
\figsetgrpnote{Optically thin MBB SED fit (black curve) and 16th to 84th percentile range of the accepted MCMC models (blue shaded region). Photometry: Red circles---ALMA 1.1~mm, 1.2~mm, 2~mm, and 3~mm, maroon pentagon---ALMA 870~$\mu$m, green square---SCUBA-2 450~$\mu$m, dark red stars---Herschel/PACS 100 and 160~$\mu$m and SPIRE 250 and 350~$\mu$m, blue triangles---Spitzer/MIPS 70~$\mu$m. Points not included in the fits are marked with black squares.}
\figsetgrpend

\figsetgrpstart
\figsetgrpnum{12.50}
\figsetgrptitle{SED for ALMA 54}
\figsetplot{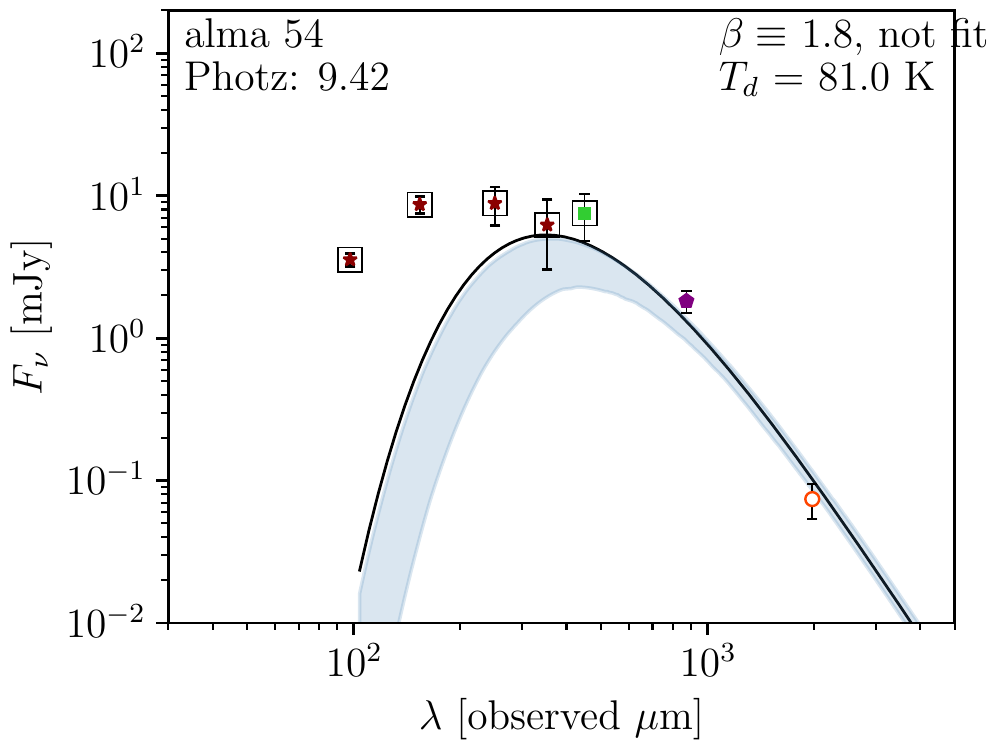}
\figsetgrpnote{Optically thin MBB SED fit (black curve) and 16th to 84th percentile range of the accepted MCMC models (blue shaded region). Photometry: Red circles---ALMA 1.1~mm, 1.2~mm, 2~mm, and 3~mm, maroon pentagon---ALMA 870~$\mu$m, green square---SCUBA-2 450~$\mu$m, dark red stars---Herschel/PACS 100 and 160~$\mu$m and SPIRE 250 and 350~$\mu$m, blue triangles---Spitzer/MIPS 70~$\mu$m. Points not included in the fits are marked with black squares.}
\figsetgrpend

\figsetgrpstart
\figsetgrpnum{12.51}
\figsetgrptitle{SED for ALMA 56}
\figsetplot{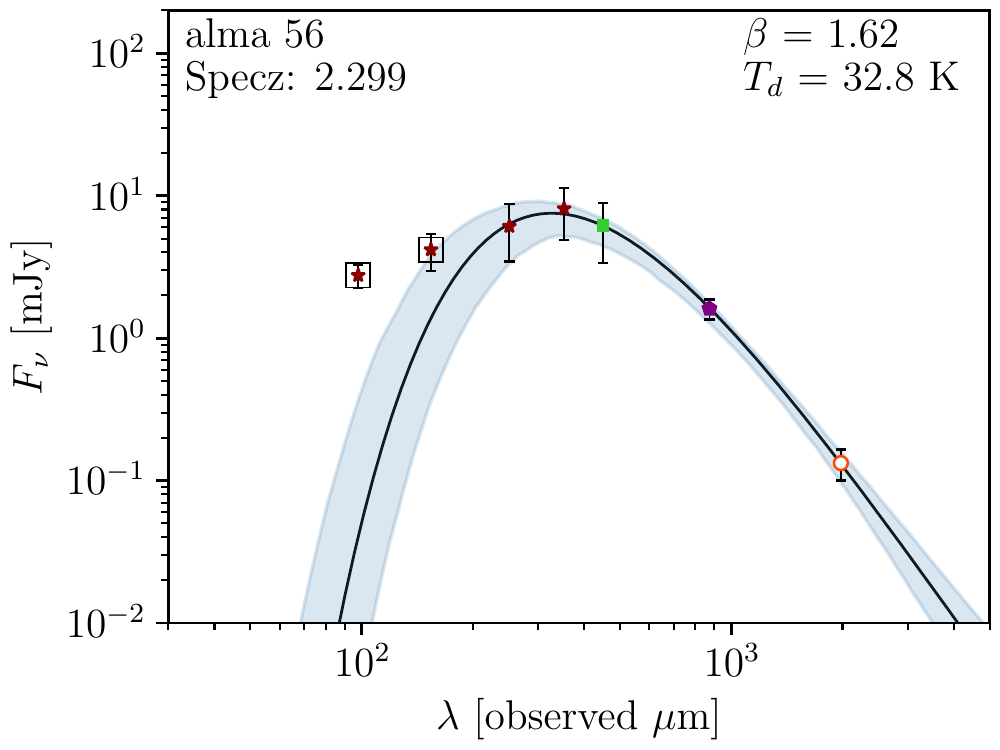}
\figsetgrpnote{Optically thin MBB SED fit (black curve) and 16th to 84th percentile range of the accepted MCMC models (blue shaded region). Photometry: Red circles---ALMA 1.1~mm, 1.2~mm, 2~mm, and 3~mm, maroon pentagon---ALMA 870~$\mu$m, green square---SCUBA-2 450~$\mu$m, dark red stars---Herschel/PACS 100 and 160~$\mu$m and SPIRE 250 and 350~$\mu$m, blue triangles---Spitzer/MIPS 70~$\mu$m. Points not included in the fits are marked with black squares.}
\figsetgrpend

\figsetgrpstart
\figsetgrpnum{12.52}
\figsetgrptitle{SED for ALMA 57}
\figsetplot{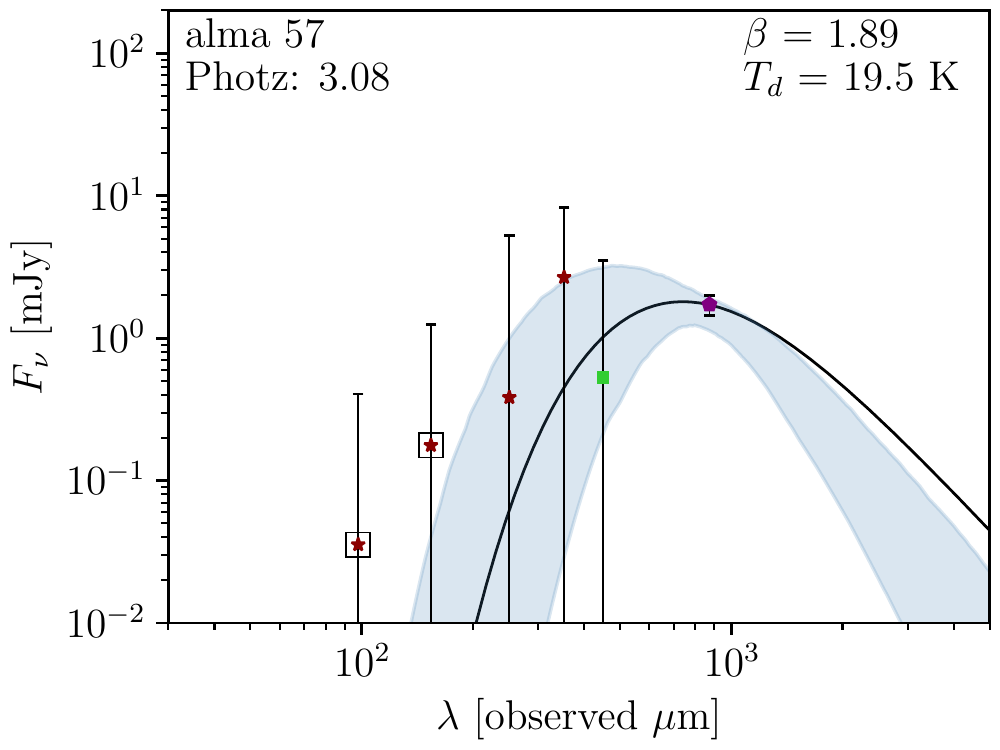}
\figsetgrpnote{Optically thin MBB SED fit (black curve) and 16th to 84th percentile range of the accepted MCMC models (blue shaded region). Photometry: Red circles---ALMA 1.1~mm, 1.2~mm, 2~mm, and 3~mm, maroon pentagon---ALMA 870~$\mu$m, green square---SCUBA-2 450~$\mu$m, dark red stars---Herschel/PACS 100 and 160~$\mu$m and SPIRE 250 and 350~$\mu$m, blue triangles---Spitzer/MIPS 70~$\mu$m. Points not included in the fits are marked with black squares.}
\figsetgrpend

\figsetgrpstart
\figsetgrpnum{12.53}
\figsetgrptitle{SED for ALMA 58}
\figsetplot{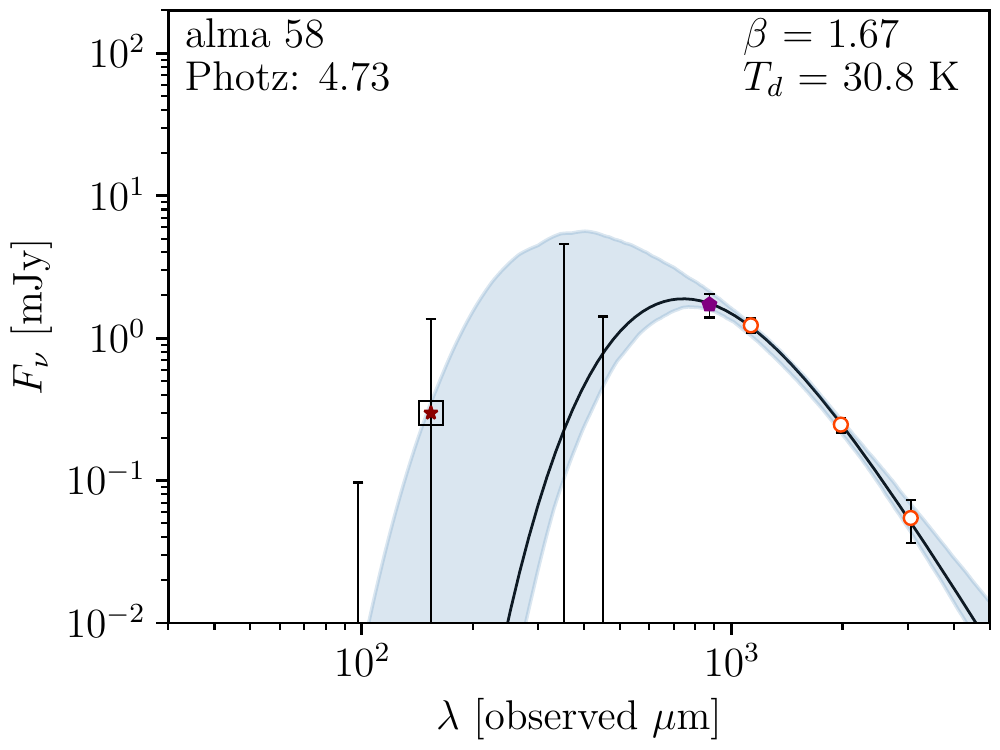}
\figsetgrpnote{Optically thin MBB SED fit (black curve) and 16th to 84th percentile range of the accepted MCMC models (blue shaded region). Photometry: Red circles---ALMA 1.1~mm, 1.2~mm, 2~mm, and 3~mm, maroon pentagon---ALMA 870~$\mu$m, green square---SCUBA-2 450~$\mu$m, dark red stars---Herschel/PACS 100 and 160~$\mu$m and SPIRE 250 and 350~$\mu$m, blue triangles---Spitzer/MIPS 70~$\mu$m. Points not included in the fits are marked with black squares.}
\figsetgrpend

\figsetgrpstart
\figsetgrpnum{12.54}
\figsetgrptitle{SED for ALMA 59}
\figsetplot{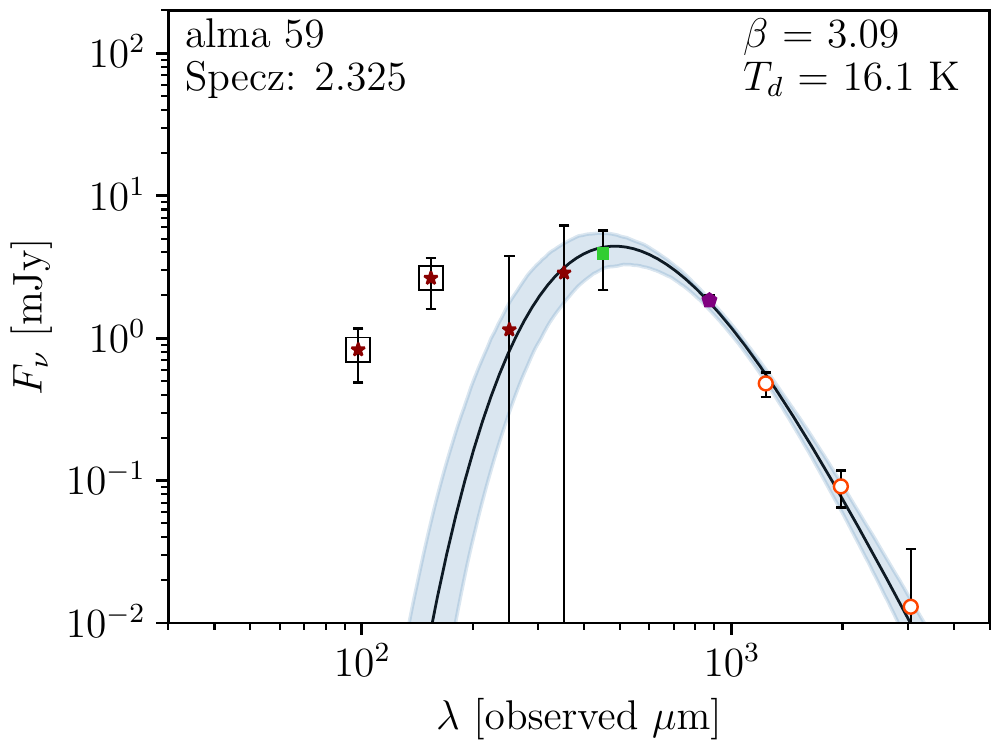}
\figsetgrpnote{Optically thin MBB SED fit (black curve) and 16th to 84th percentile range of the accepted MCMC models (blue shaded region). Photometry: Red circles---ALMA 1.1~mm, 1.2~mm, 2~mm, and 3~mm, maroon pentagon---ALMA 870~$\mu$m, green square---SCUBA-2 450~$\mu$m, dark red stars---Herschel/PACS 100 and 160~$\mu$m and SPIRE 250 and 350~$\mu$m, blue triangles---Spitzer/MIPS 70~$\mu$m. Points not included in the fits are marked with black squares.}
\figsetgrpend

\figsetgrpstart
\figsetgrpnum{12.55}
\figsetgrptitle{SED for ALMA 60}
\figsetplot{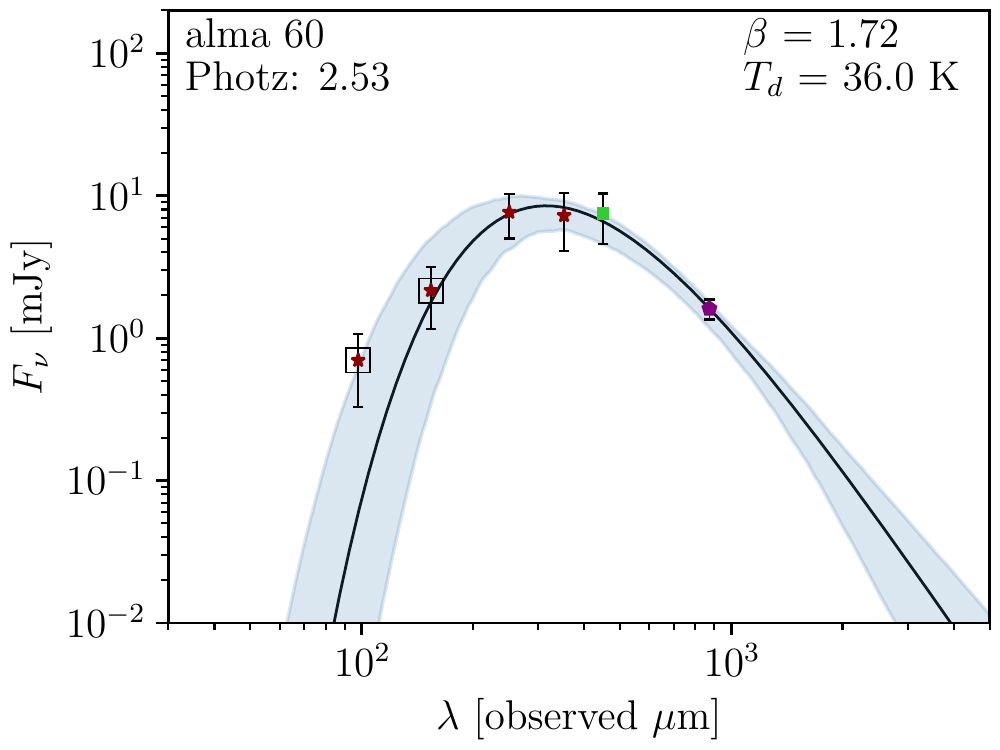}
\figsetgrpnote{Optically thin MBB SED fit (black curve) and 16th to 84th percentile range of the accepted MCMC models (blue shaded region). Photometry: Red circles---ALMA 1.1~mm, 1.2~mm, 2~mm, and 3~mm, maroon pentagon---ALMA 870~$\mu$m, green square---SCUBA-2 450~$\mu$m, dark red stars---Herschel/PACS 100 and 160~$\mu$m and SPIRE 250 and 350~$\mu$m, blue triangles---Spitzer/MIPS 70~$\mu$m. Points not included in the fits are marked with black squares.}
\figsetgrpend

\figsetgrpstart
\figsetgrpnum{12.56}
\figsetgrptitle{SED for ALMA 61}
\figsetplot{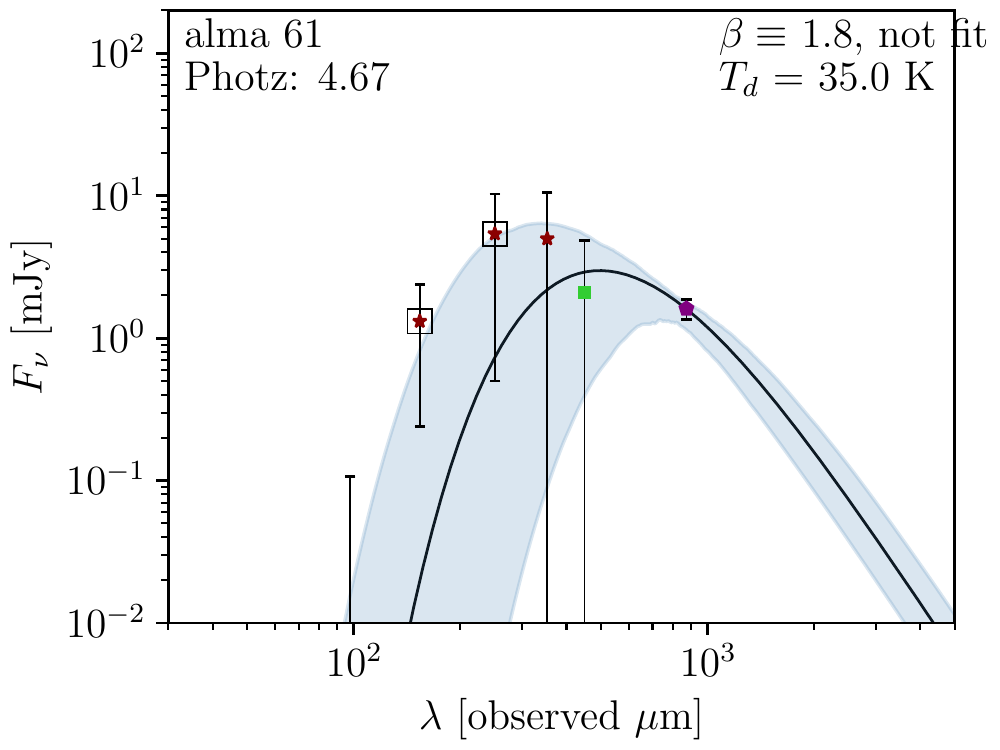}
\figsetgrpnote{Optically thin MBB SED fit (black curve) and 16th to 84th percentile range of the accepted MCMC models (blue shaded region). Photometry: Red circles---ALMA 1.1~mm, 1.2~mm, 2~mm, and 3~mm, maroon pentagon---ALMA 870~$\mu$m, green square---SCUBA-2 450~$\mu$m, dark red stars---Herschel/PACS 100 and 160~$\mu$m and SPIRE 250 and 350~$\mu$m, blue triangles---Spitzer/MIPS 70~$\mu$m. Points not included in the fits are marked with black squares.}
\figsetgrpend

\figsetgrpstart
\figsetgrpnum{12.57}
\figsetgrptitle{SED for ALMA 62}
\figsetplot{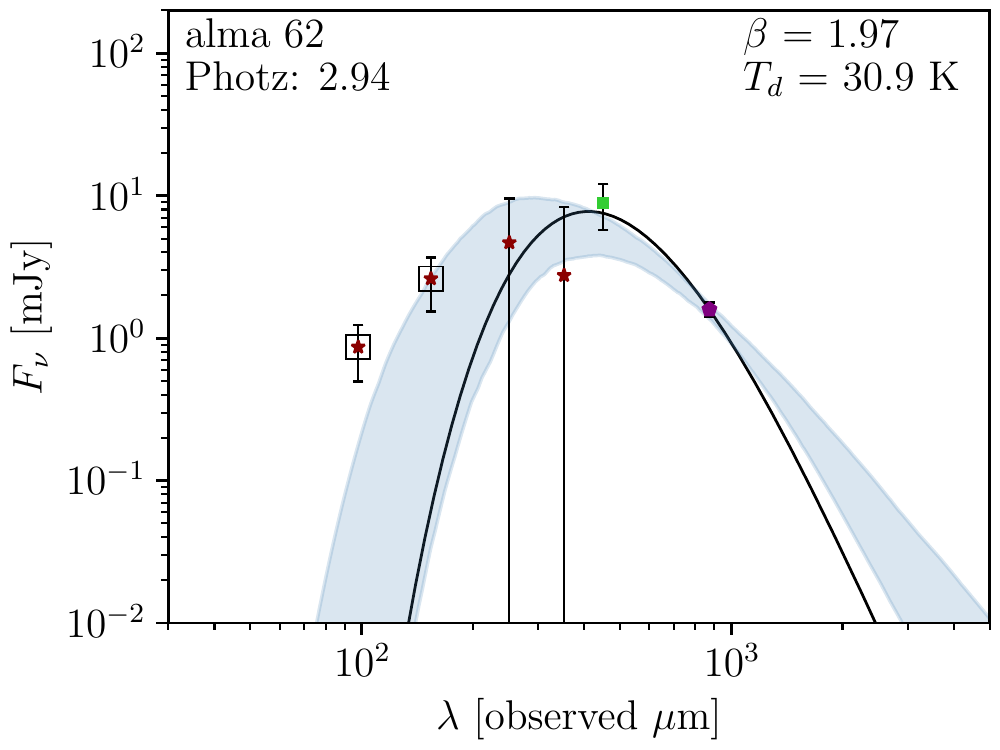}
\figsetgrpnote{Optically thin MBB SED fit (black curve) and 16th to 84th percentile range of the accepted MCMC models (blue shaded region). Photometry: Red circles---ALMA 1.1~mm, 1.2~mm, 2~mm, and 3~mm, maroon pentagon---ALMA 870~$\mu$m, green square---SCUBA-2 450~$\mu$m, dark red stars---Herschel/PACS 100 and 160~$\mu$m and SPIRE 250 and 350~$\mu$m, blue triangles---Spitzer/MIPS 70~$\mu$m. Points not included in the fits are marked with black squares.}
\figsetgrpend

\figsetgrpstart
\figsetgrpnum{12.58}
\figsetgrptitle{SED for ALMA 63}
\figsetplot{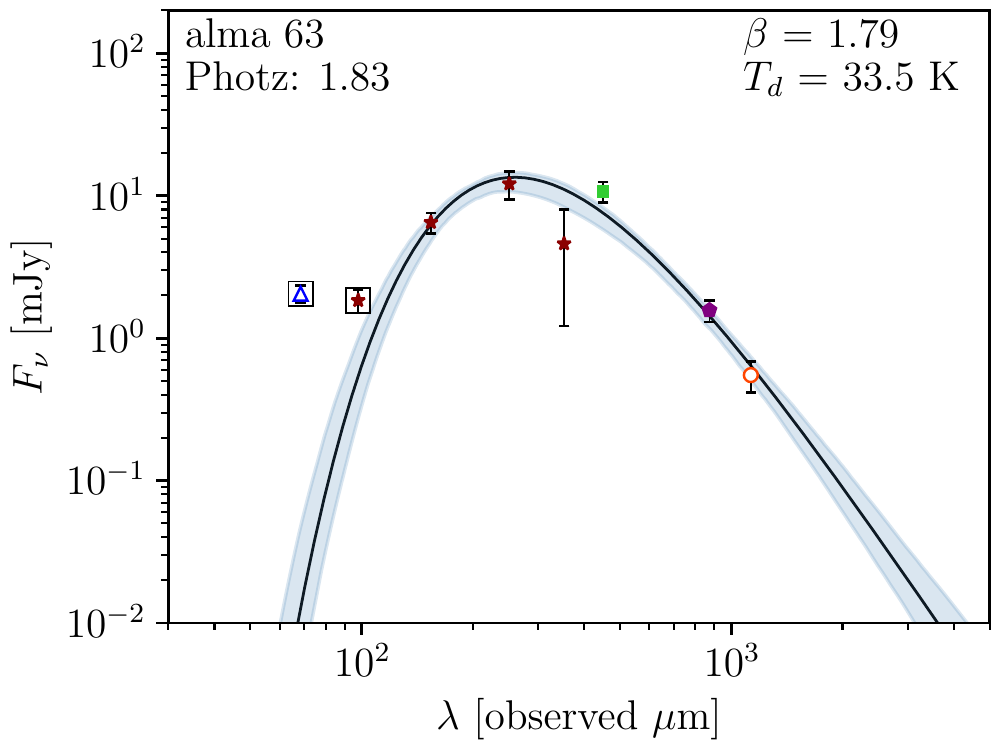}
\figsetgrpnote{Optically thin MBB SED fit (black curve) and 16th to 84th percentile range of the accepted MCMC models (blue shaded region). Photometry: Red circles---ALMA 1.1~mm, 1.2~mm, 2~mm, and 3~mm, maroon pentagon---ALMA 870~$\mu$m, green square---SCUBA-2 450~$\mu$m, dark red stars---Herschel/PACS 100 and 160~$\mu$m and SPIRE 250 and 350~$\mu$m, blue triangles---Spitzer/MIPS 70~$\mu$m. Points not included in the fits are marked with black squares.}
\figsetgrpend

\figsetgrpstart
\figsetgrpnum{12.59}
\figsetgrptitle{SED for ALMA 64}
\figsetplot{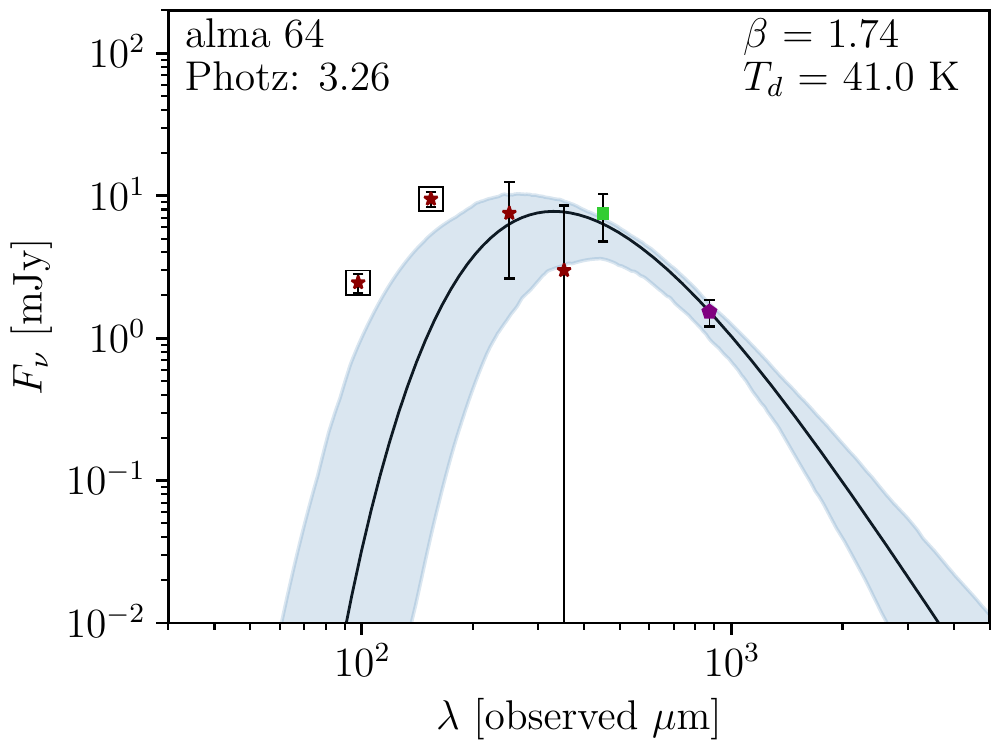}
\figsetgrpnote{Optically thin MBB SED fit (black curve) and 16th to 84th percentile range of the accepted MCMC models (blue shaded region). Photometry: Red circles---ALMA 1.1~mm, 1.2~mm, 2~mm, and 3~mm, maroon pentagon---ALMA 870~$\mu$m, green square---SCUBA-2 450~$\mu$m, dark red stars---Herschel/PACS 100 and 160~$\mu$m and SPIRE 250 and 350~$\mu$m, blue triangles---Spitzer/MIPS 70~$\mu$m. Points not included in the fits are marked with black squares.}
\figsetgrpend

\figsetgrpstart
\figsetgrpnum{12.60}
\figsetgrptitle{SED for ALMA 65}
\figsetplot{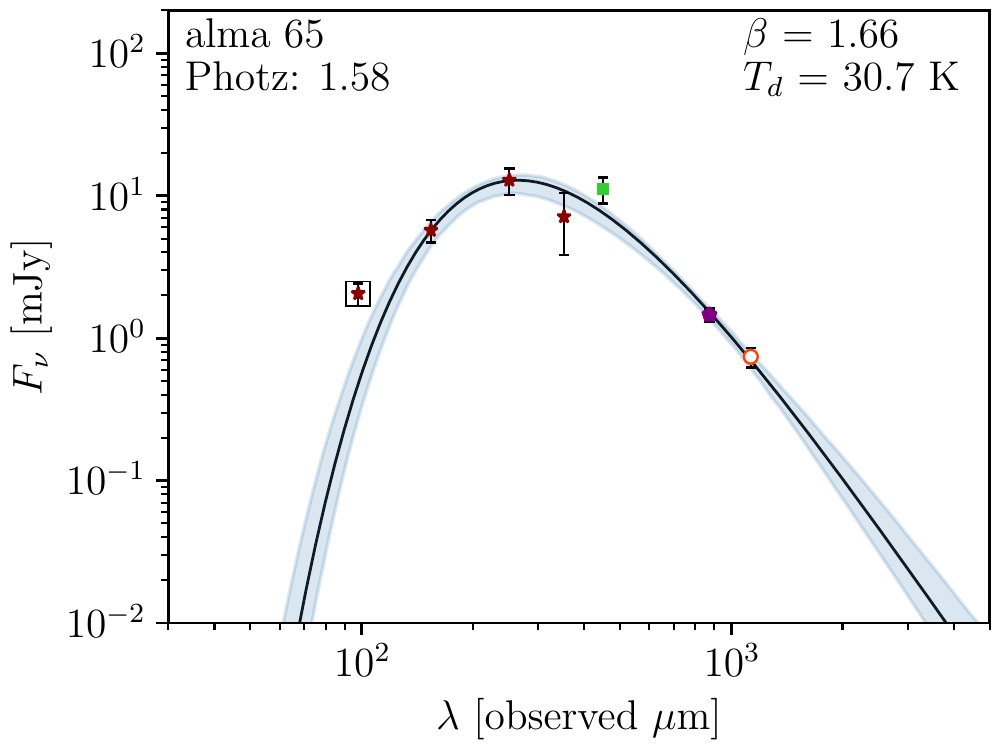}
\figsetgrpnote{Optically thin MBB SED fit (black curve) and 16th to 84th percentile range of the accepted MCMC models (blue shaded region). Photometry: Red circles---ALMA 1.1~mm, 1.2~mm, 2~mm, and 3~mm, maroon pentagon---ALMA 870~$\mu$m, green square---SCUBA-2 450~$\mu$m, dark red stars---Herschel/PACS 100 and 160~$\mu$m and SPIRE 250 and 350~$\mu$m, blue triangles---Spitzer/MIPS 70~$\mu$m. Points not included in the fits are marked with black squares.}
\figsetgrpend

\figsetgrpstart
\figsetgrpnum{12.61}
\figsetgrptitle{SED for ALMA 66}
\figsetplot{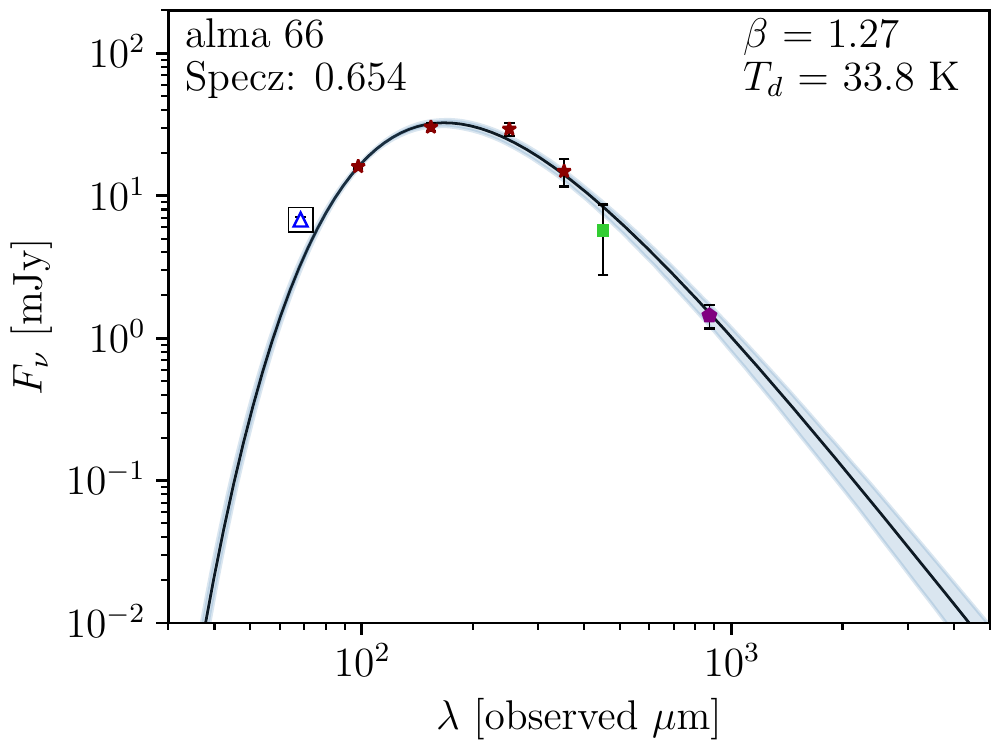}
\figsetgrpnote{Optically thin MBB SED fit (black curve) and 16th to 84th percentile range of the accepted MCMC models (blue shaded region). Photometry: Red circles---ALMA 1.1~mm, 1.2~mm, 2~mm, and 3~mm, maroon pentagon---ALMA 870~$\mu$m, green square---SCUBA-2 450~$\mu$m, dark red stars---Herschel/PACS 100 and 160~$\mu$m and SPIRE 250 and 350~$\mu$m, blue triangles---Spitzer/MIPS 70~$\mu$m. Points not included in the fits are marked with black squares.}
\figsetgrpend

\figsetgrpstart
\figsetgrpnum{12.62}
\figsetgrptitle{SED for ALMA 67}
\figsetplot{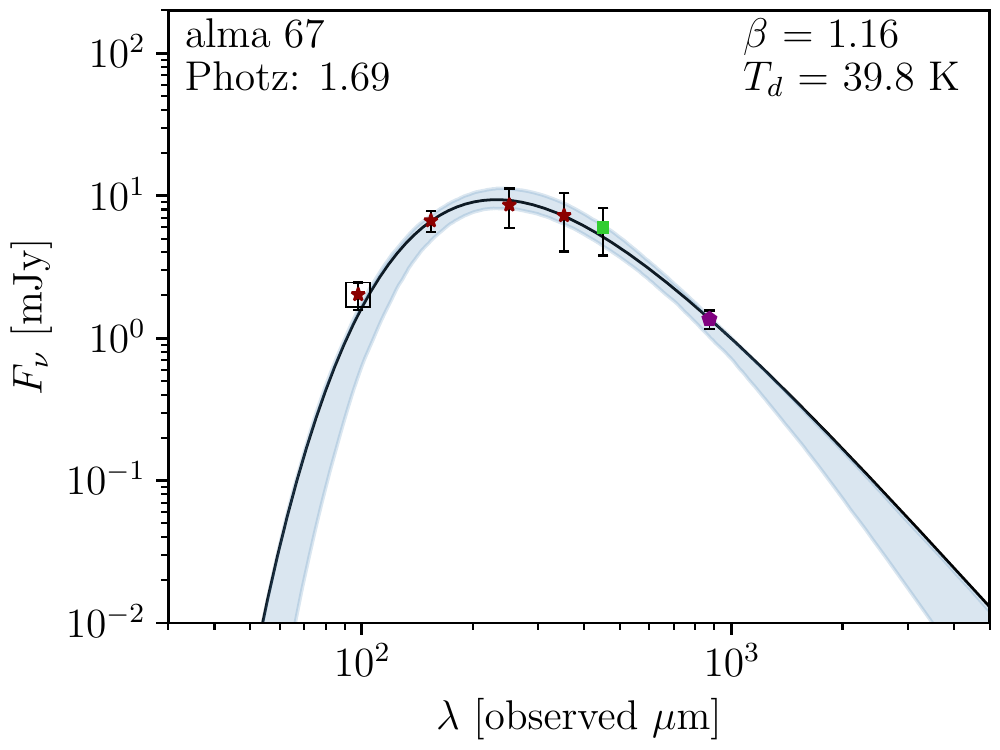}
\figsetgrpnote{Optically thin MBB SED fit (black curve) and 16th to 84th percentile range of the accepted MCMC models (blue shaded region). Photometry: Red circles---ALMA 1.1~mm, 1.2~mm, 2~mm, and 3~mm, maroon pentagon---ALMA 870~$\mu$m, green square---SCUBA-2 450~$\mu$m, dark red stars---Herschel/PACS 100 and 160~$\mu$m and SPIRE 250 and 350~$\mu$m, blue triangles---Spitzer/MIPS 70~$\mu$m. Points not included in the fits are marked with black squares.}
\figsetgrpend

\figsetgrpstart
\figsetgrpnum{12.63}
\figsetgrptitle{SED for ALMA 68}
\figsetplot{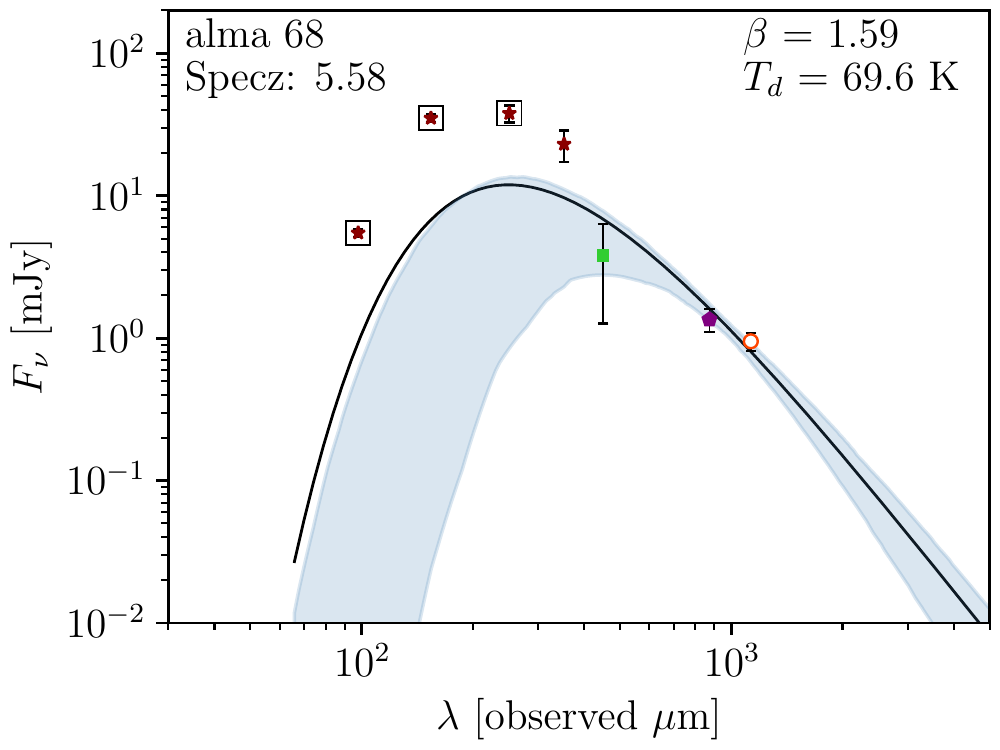}
\figsetgrpnote{Optically thin MBB SED fit (black curve) and 16th to 84th percentile range of the accepted MCMC models (blue shaded region). Photometry: Red circles---ALMA 1.1~mm, 1.2~mm, 2~mm, and 3~mm, maroon pentagon---ALMA 870~$\mu$m, green square---SCUBA-2 450~$\mu$m, dark red stars---Herschel/PACS 100 and 160~$\mu$m and SPIRE 250 and 350~$\mu$m, blue triangles---Spitzer/MIPS 70~$\mu$m. Points not included in the fits are marked with black squares.}
\figsetgrpend

\figsetgrpstart
\figsetgrpnum{12.64}
\figsetgrptitle{SED for ALMA 69}
\figsetplot{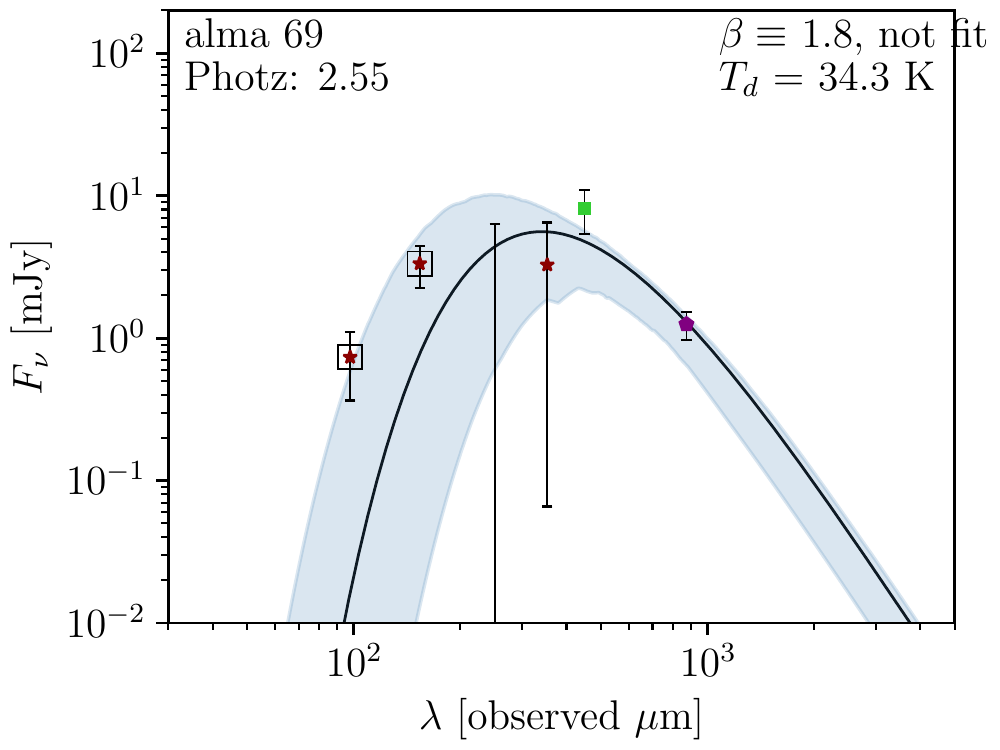}
\figsetgrpnote{Optically thin MBB SED fit (black curve) and 16th to 84th percentile range of the accepted MCMC models (blue shaded region). Photometry: Red circles---ALMA 1.1~mm, 1.2~mm, 2~mm, and 3~mm, maroon pentagon---ALMA 870~$\mu$m, green square---SCUBA-2 450~$\mu$m, dark red stars---Herschel/PACS 100 and 160~$\mu$m and SPIRE 250 and 350~$\mu$m, blue triangles---Spitzer/MIPS 70~$\mu$m. Points not included in the fits are marked with black squares.}
\figsetgrpend

\figsetgrpstart
\figsetgrpnum{12.65}
\figsetgrptitle{SED for ALMA 70}
\figsetplot{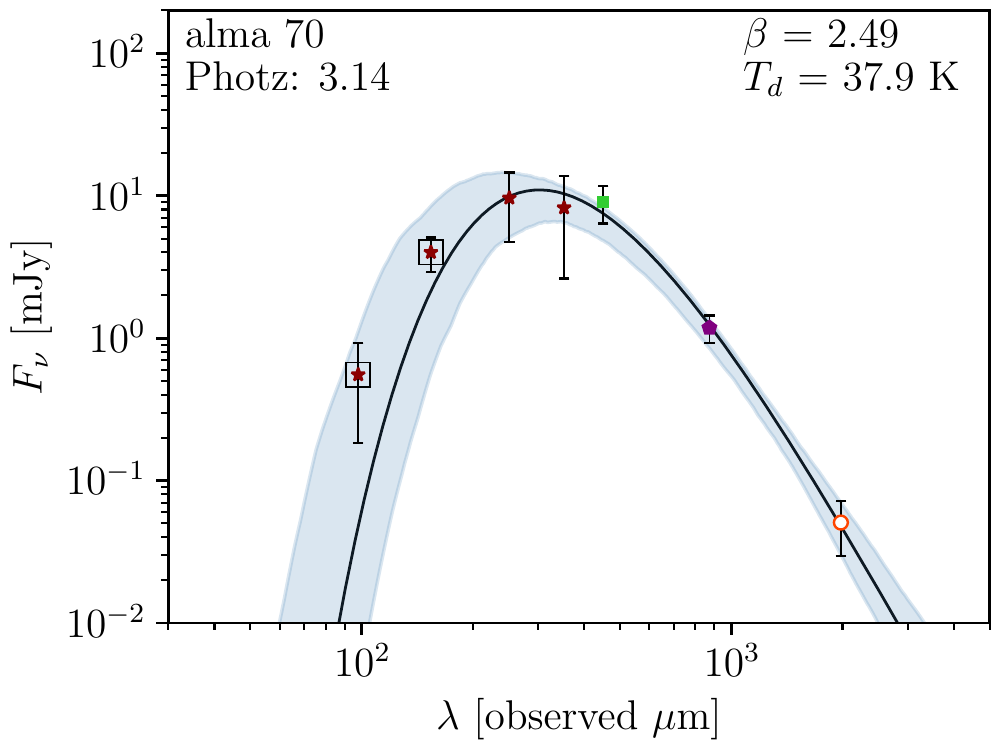}
\figsetgrpnote{Optically thin MBB SED fit (black curve) and 16th to 84th percentile range of the accepted MCMC models (blue shaded region). Photometry: Red circles---ALMA 1.1~mm, 1.2~mm, 2~mm, and 3~mm, maroon pentagon---ALMA 870~$\mu$m, green square---SCUBA-2 450~$\mu$m, dark red stars---Herschel/PACS 100 and 160~$\mu$m and SPIRE 250 and 350~$\mu$m, blue triangles---Spitzer/MIPS 70~$\mu$m. Points not included in the fits are marked with black squares.}
\figsetgrpend

\figsetgrpstart
\figsetgrpnum{12.66}
\figsetgrptitle{SED for ALMA 71}
\figsetplot{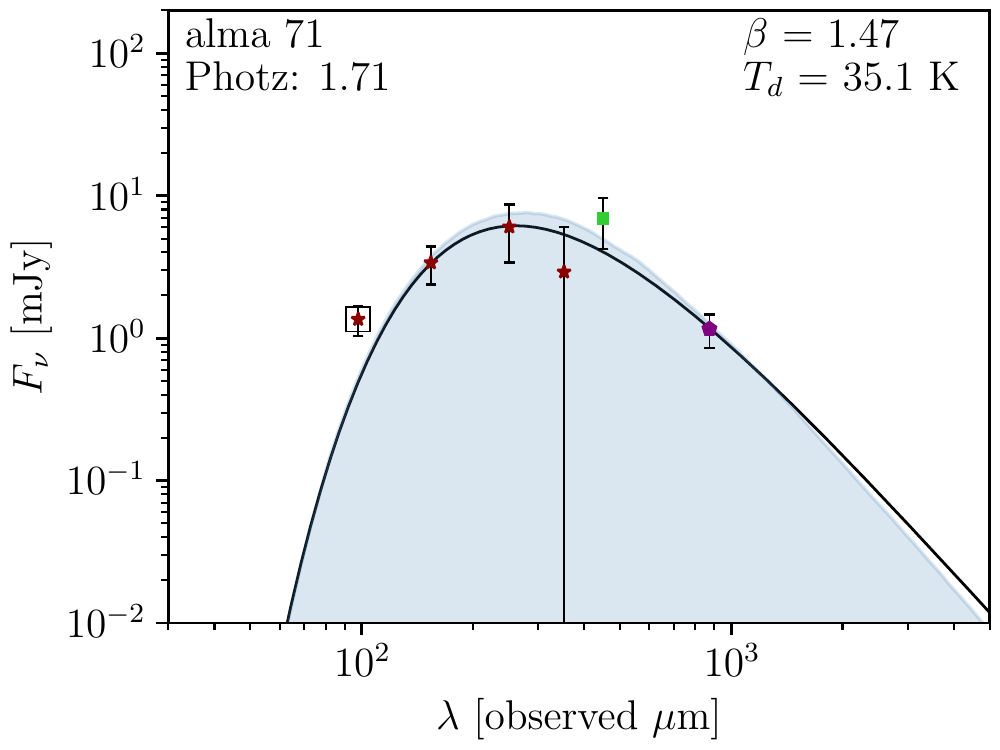}
\figsetgrpnote{Optically thin MBB SED fit (black curve) and 16th to 84th percentile range of the accepted MCMC models (blue shaded region). Photometry: Red circles---ALMA 1.1~mm, 1.2~mm, 2~mm, and 3~mm, maroon pentagon---ALMA 870~$\mu$m, green square---SCUBA-2 450~$\mu$m, dark red stars---Herschel/PACS 100 and 160~$\mu$m and SPIRE 250 and 350~$\mu$m, blue triangles---Spitzer/MIPS 70~$\mu$m. Points not included in the fits are marked with black squares.}
\figsetgrpend

\figsetgrpstart
\figsetgrpnum{12.67}
\figsetgrptitle{SED for ALMA 72}
\figsetplot{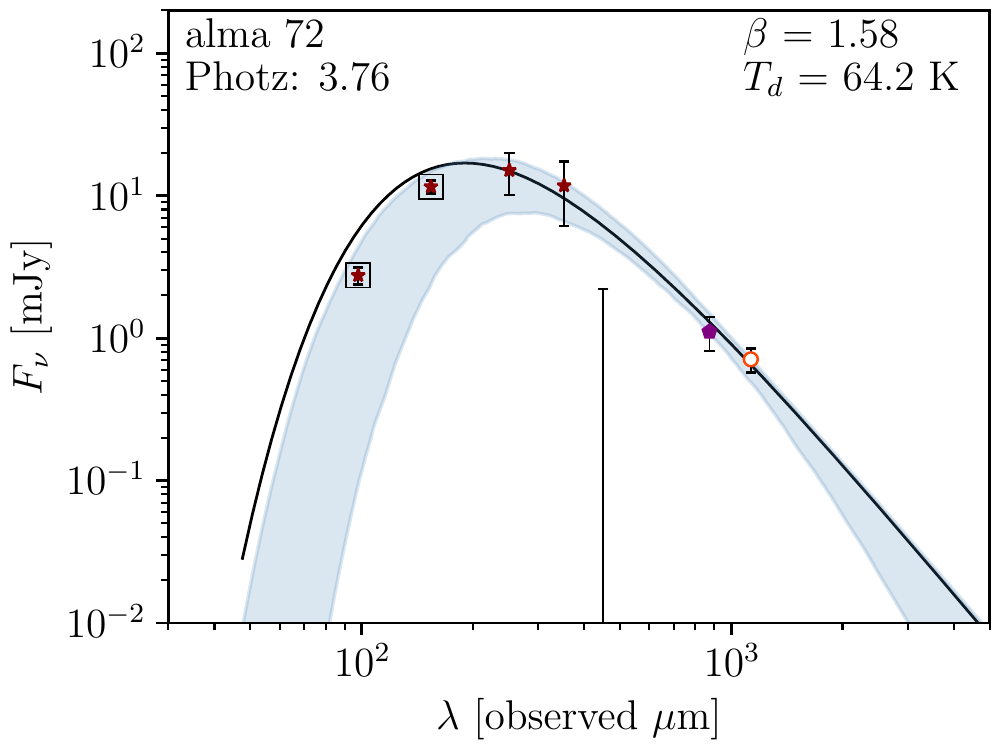}
\figsetgrpnote{Optically thin MBB SED fit (black curve) and 16th to 84th percentile range of the accepted MCMC models (blue shaded region). Photometry: Red circles---ALMA 1.1~mm, 1.2~mm, 2~mm, and 3~mm, maroon pentagon---ALMA 870~$\mu$m, green square---SCUBA-2 450~$\mu$m, dark red stars---Herschel/PACS 100 and 160~$\mu$m and SPIRE 250 and 350~$\mu$m, blue triangles---Spitzer/MIPS 70~$\mu$m. Points not included in the fits are marked with black squares.}
\figsetgrpend

\figsetgrpstart
\figsetgrpnum{12.68}
\figsetgrptitle{SED for ALMA 73}
\figsetplot{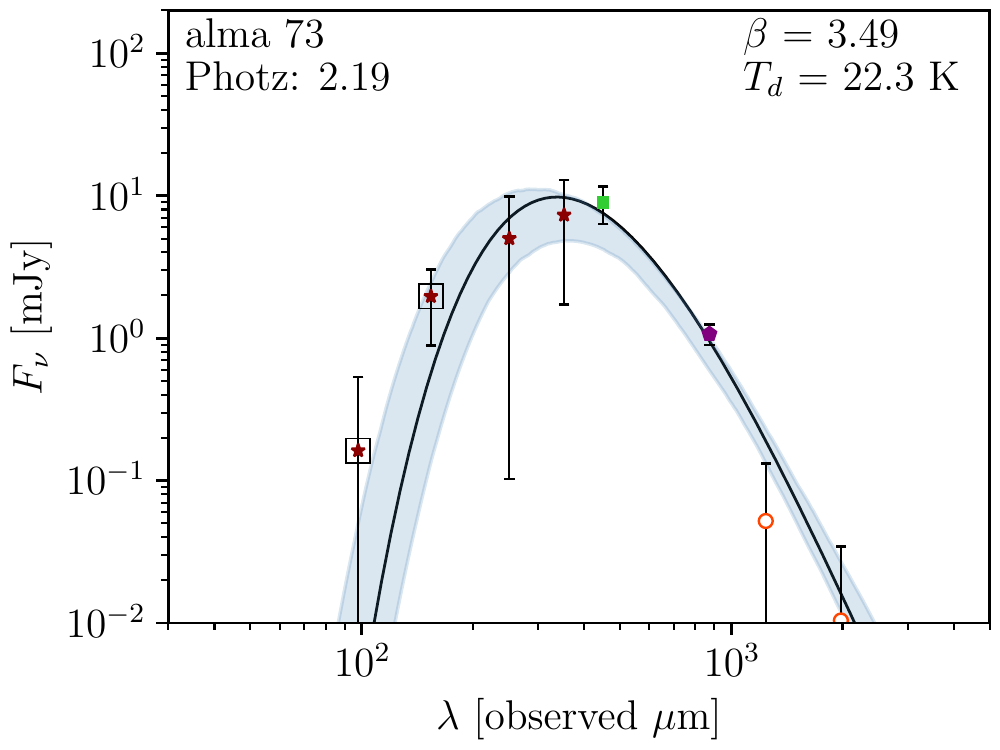}
\figsetgrpnote{Optically thin MBB SED fit (black curve) and 16th to 84th percentile range of the accepted MCMC models (blue shaded region). Photometry: Red circles---ALMA 1.1~mm, 1.2~mm, 2~mm, and 3~mm, maroon pentagon---ALMA 870~$\mu$m, green square---SCUBA-2 450~$\mu$m, dark red stars---Herschel/PACS 100 and 160~$\mu$m and SPIRE 250 and 350~$\mu$m, blue triangles---Spitzer/MIPS 70~$\mu$m. Points not included in the fits are marked with black squares.}
\figsetgrpend

\figsetgrpstart
\figsetgrpnum{12.69}
\figsetgrptitle{SED for ALMA 74}
\figsetplot{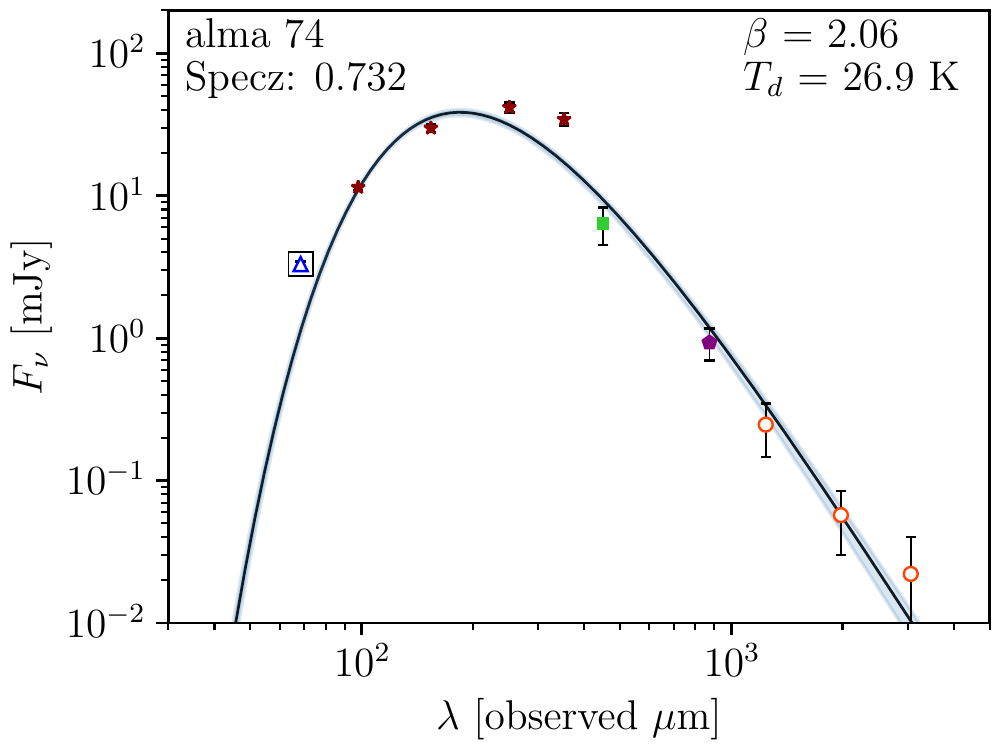}
\figsetgrpnote{Optically thin MBB SED fit (black curve) and 16th to 84th percentile range of the accepted MCMC models (blue shaded region). Photometry: Red circles---ALMA 1.1~mm, 1.2~mm, 2~mm, and 3~mm, maroon pentagon---ALMA 870~$\mu$m, green square---SCUBA-2 450~$\mu$m, dark red stars---Herschel/PACS 100 and 160~$\mu$m and SPIRE 250 and 350~$\mu$m, blue triangles---Spitzer/MIPS 70~$\mu$m. Points not included in the fits are marked with black squares.}
\figsetgrpend

\figsetend

\bibliography{references}{}
\bibliographystyle{aasjournal}

\end{document}